\documentclass[prx, letterpaper, superscriptaddress, amsmath, amssymb, reprint, floatfix, nofootinbib]{revtex4-2}

\usepackage{graphicx}
\usepackage{dcolumn}
\usepackage{bm}
\usepackage{xcolor} 
\usepackage[normalem]{ulem}
\usepackage{braket}

\usepackage{upgreek}
\usepackage{placeins}

\newcommand{\parens}[1]{\left({#1}\right)}
\newcommand{\eqr}[1]{Eq.~(\ref{#1})}
\newcommand{\figr}[1]{Fig.~\ref{#1}}
\newcommand{\tr}[1]{\mathbf{Tr}\parens{#1}}
\newcommand{\registeredmark}{\textsuperscript{\tiny\textregistered}}

\begin{document}

\preprint{APS/123-QED}

\title{Compact Quantum Dot Models for Analog Microwave co-Simulation}

\author{Lorenzo Peri}
\email{lp586@cam.ac.uk}
\affiliation{Quantum Motion, 9 Sterling Way, London, N7 9HJ, United Kingdom}
\affiliation{Cavendish Laboratory, University of Cambridge, JJ Thomson Ave, Cambridge CB3 0HE, United Kingdom}
 
\author{Alberto Gomez-Saiz}
\affiliation{Quantum Motion, 9 Sterling Way, London, N7 9HJ, United Kingdom}
\affiliation{Department of Electrical and Electronic Engineering, Imperial College London, London SW7 2AZ, United Kingdom}

\author{Christopher J. B. Ford}
\affiliation{Cavendish Laboratory, University of Cambridge, JJ Thomson Ave, Cambridge CB3 0HE, United Kingdom}

\author{M. Fernando Gonzalez-Zalba}
\email{fernando@quantummotion.tech}
\affiliation{Quantum Motion, 9 Sterling Way, London, N7 9HJ, United Kingdom}

\date{\today}

\begin{abstract}

    Scalable solid-state quantum computers will require integration with analog and digital electronics. Efficiently simulating the quantum-classical electronic interface is hence of paramount importance. Here, we present Verilog-A compact models with a focus on quantum-dot-based systems, relevant to semiconductor- and Majorana-based quantum computing. Our models are capable of faithfully reproducing coherent quantum behavior within a standard electronic circuit simulator, enabling compromise-free co-simulation of hybrid quantum devices. In particular, we present results from co-simulations performed in Cadence Spectre\registeredmark, showcasing coherent quantum phenomena in circuits with both quantum and classical components using an industry-standard electronic design and automation tool.
    Our work paves the way for a new paradigm in the design of quantum systems, which leverages the many decades of development of electronic computer-aided design and automation tools in the semiconductor industry to now simulate and optimize quantum processing units, quantum-classical interfaces, and hybrid quantum-analog circuits.
\end{abstract}

\maketitle

\section{Introduction}

Quantum computation has achieved remarkable progress over the past decade, demonstrating unprecedented computational capabilities and potential \cite{Arute_2019,Preskill_2023,Bluvstein2024,Bravyi2024, Acharya2024,paetznick2024, putterman2024}. 
However, many challenges still stand in the way of fulfilling Feynman's proposition of performing tasks outside the classical reach \cite{Feynman_1960}, one of which is undoubtedly scaling quantum systems to the likely millions of physical qubits required to tackle problems of scientific, commercial, and societal impact \cite{Fowler_2012,Campbell_Terhal_Vuillot_2017,chan2023algorithmic,Huang_Kueng_Preskill_2020}, as well as managing the classical signals necessary for their initialization, control, and readout \cite{Reilly2015, Reilly2019, Gonzalez-Zalba_2021,Pauka_2021}.
To perform operations on a quantum processor, each qubit needs to interface with classical hardware, potentially operating at cryogenic temperatures, with demanding specifications in terms of accuracy, noise, and power budget, which require careful design \cite{Hornibrook_2015,Charbon_2016,Sebastiano_2017,Park_Subramanian_2021}. 
Moreover, transients and non-idealities in both the quantum and classical layers may have unexpected and difficult-to-predict effects and interactions, ultimately degrading the performance of the quantum processing \cite{vanDijk_2019,Eggli_2024}. In addition, there has been a growing interest in leveraging the properties of quantum devices to create implementations of traditional analog circuits and sensors with low power dissipation, a nanoscale footprint, and capable of working at cryogenic temperatures \cite{Sliwa2015,Ahmed_2018,Aumentado2020,Cochrane_2022,Oakes_Peri_2023,Hogg_2023,Phan_2023,Navarathna2023}.

To meet the stringent specifications and efficiently design quantum-enabled devices at scale, efficient co-simulation of the quantum and classical layers is of paramount importance. 
Until now, seminal efforts have shown the great potential of co-simulation \cite{Csurgay_Porod_2001,Csurgay_2007}, showing how the quantum dynamics may be recast and expressed in a compatible way with standard analog circuit simulators \cite{spine_2019,Acharya_2021,Gys_2021,Pesic_2024}. In particular, the system's quantum dynamics can be expressed in Verilog-A, the industry-standard hardware description language for behavioral modeling of analog and mixed-signal systems, which is supported by most modern circuit simulators \cite{McAndrew_2015}. 
In this work, we leverage the expressiveness of Verilog-A to map with no approximations the equations describing the quantum dynamics to equivalent electrical circuits and integrate them into a Verilog-A compact model of the quantum device. Our models retain all coherent behavior, while being able to include decoherence effects that lead to important phenomena such as the finite lifetime of quantum states and dynamical power dissipation \cite{Persson_2010}.
The framework we discuss is based on the Lindblad master equation and is thus applicable to any arbitrary multilevel quantum system, although here we focus on systems based on quantum dots (QDs), which are particularly relevant to semiconductor- and Majorana-based quantum computing~\cite{Xue_2022,Noiri_2022,Burkard_2023,Aghaee2024}. Overall, our framework enables quantum information processing technologies to leverage the many decades of development in electronic computer-aided design and automation responsible for the very large-scale integration achieved by the semiconductor industry for the purpose of simulating and optimizing hybrid quantum systems. 

\begin{figure}[htb!]
    \centering
    \includegraphics[width=0.9\linewidth]{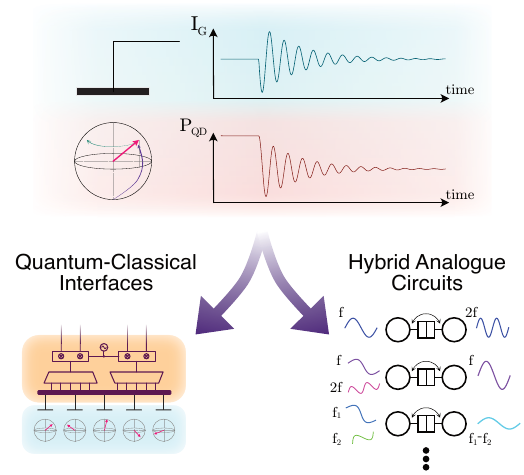}
    \caption{\textbf{Applications of quantum-classical co-simulation.} Our compact models allow for the simultaneous simulation of quantum and electrical variables within standard circuit simulators, (i.e., gate current and state probabilities in the top diagrams) granting the ability to carefully design quantum-classical interfaces for quantum computation (left), and hybrid analog circuits containing classical and quantum circuit elements (right), including, for example, frequency multipliers, mixers and parametric amplifiers.}
    \label{fig:uses}
\end{figure}

This work is structured in three sections. In section~\ref{sec:Theory}, we discuss a systematic way to express the quantum dynamics within a compact model, with particular emphasis on QD-based devices, which may include charge, spin and Majorana qubits~\cite{maman2020}. In section~\ref{sec:models}, we showcase two particular QD compact models, which represent the building blocks of any QD-based quantum device: (i) a QD exchanging single electrons with a reservoir and (ii) a double-QD (DQD) charge qubit. We perform simulations of the quantum devices alone, which demonstrate excellent agreement with the theoretical expectations.
Finally, in section~\ref{sec:circuits}, we present the co-simulations of circuits formed of both classical and quantum components. In particular, (i) we exploit the nonlinear voltage-dependent impedance of the QD-to-reservoir transition to design a frequency multiplier suitable for operation at cryogenic temperatures, and (ii) we simulate qubit readout considering a charge qubit coupled to a high-$Q$ microwave resonator, modeling the reflected signal in the adiabatic and resonant regimes.
All co-simulations presented in this work are performed using Cadence Spectre\registeredmark---an industry-standard circuit simulator---and, wherever possible, compared with theoretical expectations obtained with standard Crank-Nicholson-based Lindblad simulations \cite{Am-Shallem_Levy_Schaefer_Kosloff_2015,Riesch_Jirauschek_2019, Minganti_Huybrechts_2022, Oakes_Peri_2023}.

\section{Quantum Dynamics in Analog Compact Models}
\label{sec:Theory}

\begin{figure}[htb!]
    \centering
    \includegraphics[width=\linewidth]{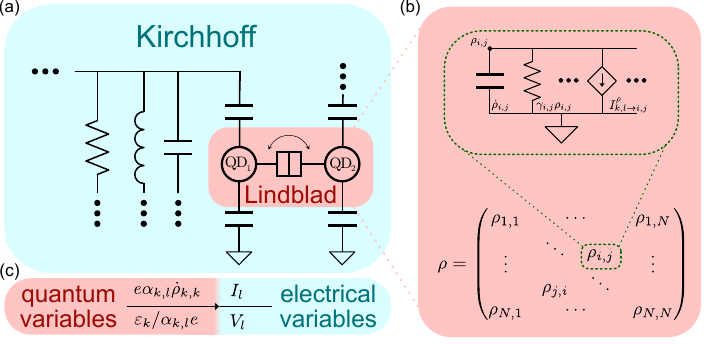}
    \caption{\textbf{Quantum-classical co-simulation.} (a) The differential problem solved by the circuit simulator is conceptually spit into two realms: classical (blue), following Kirchhoff's laws, and quantum (red), described by a Lindblad master equation. (b) The quantum evolution is recast into many equivalent subcircuits based on voltage-controlled current sources (VCCSs), while the model specifies how to convert between quantum and classical variables (panel c). $\alpha_{k,l}$ represents the respective lever arm of each gate. Summation over repeated indices is implied.}
    \label{fig:scheme}
\end{figure}

Conceptually, co-simulating quantum dynamics within an electric circuit is an exercise in simultaneously simulating two \textit{worlds} with different sets of variables, which evolve according to different laws, while providing a \textit{translation} layer to allow for the exchange of signals at the boundary between the two (\figr{fig:scheme}). 
Electronic components are generally described by a potential nature (voltage) and a flow nature (current), whose dynamics is described by Kirchhoff's circuit laws. 
Quantum devices, on the other hand, do not natively fit this picture (i.e., notice how a circuit diagram is not \textit{expressive} enough to indicate tunnel coupling between QDs in \figr{fig:scheme}a).
A better description may be found in terms of a density matrix $\rho(t)$, which fully characterizes the state of an (open) quantum system \cite{manzano_short_2020}. 
In particular, the diagonal elements of the matrix ($\rho_{i,i} = \tr{\rho\ket{i}\bra{i}}$) are known as \textit{populations}, and represent the probability of occupation of a (pure) state $\ket{i}$, while the off-diagonal elements of $\rho$, the \textit{coherences}, quantify the superposition between states \cite{Zoller_Gardiner_1997,Gardiner_Zoller_2004}. 

In this work, we model the evolution of the quantum dynamics via the Lindblad master equation (LME) \cite{manzano_short_2020,Albert_Bradlyn_Fraas_Jiang_2016,Albash_Boixo_Lidar_Zanardi_2012}, a Markovian approximation that allows for the modeling of the coherent (unitary) quantum behavior while also including decoherence caused by coupling of the quantum system with the environment \cite{Peri2024beyondadiabatic,Oakes_Peri_2023}.
The LME defines the evolution of the density matrix as 
\begin{equation}
    \dot{\rho}(t) = \mathcal{L} \rho(t) ,
    \label{eq:LME_def}
\end{equation}
\noindent
where $\dot{\rho}$ indicates the time derivative of $\rho$, and $\mathcal{L}$ is the \mbox{(super-)operator} known as the \textit{Liouvillian} of the system, reading 
\begin{align}
    & \mathcal{L} \rho = -{\rm i} \left[H/\hbar, \rho\right] + \sum_l \Gamma_l \mathcal{D}\left(L_l\right)\rho, \label{eq:LME} \\
    & \mathcal{D}\left(L_l\right) \rho = L_l\rho L_l^\dagger - \frac{1}{2}\left\{L_l^\dagger L_l , \rho \right\}.
\end{align}
Here we define $H$ as the Hamiltonian of the quantum system, while $L_l$ is the \textit{jump} operator describing a particular decoherence process (i.e., relaxation or dephasing), occurring at rate $\Gamma_l$.
It is interesting to point out how, for each element of the density matrix, the linearity of the LME allows \eqr{eq:LME} to be written as
\begin{equation}
    \dot{\rho}_{i,j} + \gamma_{i,j} \rho_{i,j} = \sum_{k,l \neq i,j} \mathcal{L}_{klij} \rho_{k,l} = \sum_{k,l \neq i,j} I^\rho_{k,l\rightarrow i,j}
\end{equation}
\noindent
which, in circuit terms, describes a capacitor and a resistor in parallel driven by voltage-controlled current sources (VCCSs) whose value (linearly) depends on all other elements of the density matrix ($\rho_{k,l \neq i,j}$), see \figr{fig:scheme}b.
This description of the LME is particularly amenable to co-simulation \cite{Acharya_2021,spine_2019,Gys_2021} as it allows one to leverage electronics simulators' native capabilities of solving coupled differential equations \cite{McAndrew_2015}, while, most importantly, allowing the quantum and classical dynamics to be expressed simultaneously (i.e., within one single Jacobian) to avoid any numerical issues regarding convergence \cite{Bashir_2019}.
As an implementation note, we recall for the benefit of the reader that the density matrix is generally a complex Hermitian matrix ($\rho^\dagger = \rho$). Thus, one may need to exploit this symmetry in order to separate the real and imaginary parts to comply with the need for \textit{real-valued} voltages and currents in the equivalent circuit. 

Figure~\ref{fig:scheme}b can be used to gain an intuitive understanding of the quantum behavior. In particular, unlike previous descriptions in the literature \cite{Gys_2021,Acharya_2021}, our LME-based approach shows a resistive element in parallel to the capacitor representing the time derivative. This promptly finds a physical interpretation by defining a time constant ($1/\gamma_{i,j}$) over which excitations in the relevant branch will be exponentially damped. In fact, with a little algebra, it is possible to show that the RC constant of diagonal elements ($i=j$) is equivalent to relaxation times ($T_1$), while for off-diagonal branches it corresponds to its dephasing time ($T_2^*$), providing a direct link between the quantum properties and the circuit representation of the system. 

Having described how the quantum and classical worlds may be implemented within a single differential problem, we are left with the task of describing how signals behave when crossing the boundary between the two worlds (\figr{fig:scheme}c). 
While this depends heavily on the physics of the devices being modeled, it is generally a matter of \textit{charge bookkeeping}. 
Charge is globally conserved. Thus, every charge entering the domain of quantum mechanics must be \textit{paid for} by classical currents, and, likewise, each charge exiting the quantum world must be \textit{sourced} as current by the terminals to then propagate within the electrical network. 
Charge conservation is a foundational requirement (and typical pain point) of compact modeling \cite{McAndrew_2015}.

When it comes to QD systems, it is particularly important to draw the world boundary \textit{just outside} the quantum device and include the screening charges arising from capacitive couplings between different elements of the nanostructure.
It is necessary, in fact, to keep track of the fact that charge redistribution events in a QD system will also necessarily appear as a redistribution of the screening charge accumulated on the gates to which the QDs are capacitively coupled. 
If we consider a QD array where the $l^{\rm th}$ terminal (gate) is coupled to the $k^{\rm th}$ level via a capacitance $C_{k,l}$, at equilibrium, the respective screening charge will read \cite{Vigneau_2023}
\begin{equation}
  \frac{Q_{k,l}}{C_{k,l}} = \frac{e \rho_{k,k}}{C_{\Sigma_k} } ,
\end{equation}
\noindent
where $C_{\Sigma_k}$ is the QD's total capacitance. 
Therefore, the current at each terminal arising from charge movements within the quantum system reads \cite{Peri_2023}
\begin{equation}
  I_{l}(t) = e\sum_k \alpha_{k,l} \dot{\rho}_{k,k}(t) ,
  \label{eq:rho_to_I}
\end{equation}
\noindent 
where we define the lever arm $\alpha_{k,l} = C_{k,l}/C_{\Sigma_k}$.
This is the physical description of gate currents \cite{Peri_2023}, which will be one of the key observables in our subsequent showcasing of QD compact models.

Equation~\eqref{eq:rho_to_I} links the flow at the model's terminals to the rate of change of quantum variables. However, we must also describe the effect of the terminal's potential on the quantum dynamics. Once again, this is highly dependent on the physics of the system being modeled. 
To first order, however, and within the constant interaction model, the main effect of applying a voltage to a terminal is to vary the electrochemical potential of each site $\varepsilon_k$.
This arises from the fact that the primary effect on a QD is the coupling of the electric field to the \textit{dipole} of the system, neglecting any effect that alters the spatial envelope of the wave function. If the latter effect may not be ignored (e.g., if the QD system presents barrier gates), further modeling is necessary. In the simplest case, however, the on-site electrostatic energy reads \cite{Oakes_Peri_2023,Mizuta2017,Esterli_Otxoa_Gonzalez-Zalba_2019}
\begin{equation}
    \varepsilon_k = - e \sum_l \alpha_{k,l} V_l .
\end{equation}
This quantity enters in the definition of the Liouvillian (\eqr{eq:LME_def}), making it possible for the classical signals to alter the quantum dynamics and thus inextricably coupling the differential equations governing the classical and quantum evolutions.

\section{Quantum-Dot Models}
\label{sec:models}

In this Section, we apply the co-simulation strategy discussed above and showcase the capability of compact models for two of the simplest basic blocks of QD circuits: (i) a QD exchanging single electrons with a reservoir, and (ii) a DQD charge qubit.

\subsection*{The Single-Electron Box}

\begin{figure}[htb!]
    \centering
    \includegraphics[width=0.9\linewidth]{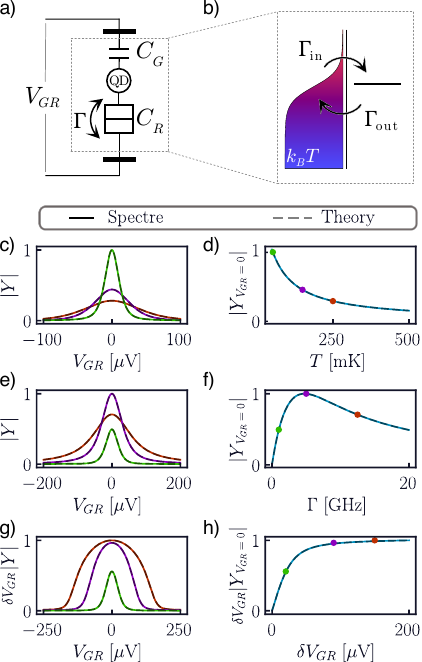}
    \caption{\textbf{Admittance of a single-electron box.} Model of a SEB from the circuit (a) and quantum (b) perspectives.
    (c-h) Lineshapes (left) and zero-detuning admittances (right) at $f=1$\,GHz in the thermal (c, d), lifetime (e, f), and power (g, h) broadening regimes. When not swept, the values of $T=100$\,mK, $\Gamma=0.5$\,GHz, and $\delta V_{\rm GR}=1$\,$\mu$V are assumed.}
    \label{fig:SEB_Simple}
\end{figure}

The first model we discuss is the single-electron box (SEB), where a QD cyclically exchanges single electrons with a reservoir at thermal equilibrium at temperature $T$ (\figr{fig:SEB_Simple}). 
The QD is capacitively coupled to a gate and to the reservoir ($C_{\rm G}$ and $C_{\rm R}$ in \figr{fig:SEB_Simple}a respectively), and charge tunneling events occur with a (total) tunnel rate $\Gamma$.

From a Lindblad perspective, an SEB is best modeled as a two-level system, with the two states representing the state in which an electron either occupies or does not occupy the QD \cite{Oakes_Peri_2023,Peri2024beyondadiabatic}. 
The Hamiltonian of this system reads, up to an arbitrary trace defining the zero of energy,
\begin{equation}
    H = \frac{1}{2} \begin{pmatrix}
        -\varepsilon_{\rm SEB} & 0\\
        0 & \varepsilon_{\rm SEB}\\
    \end{pmatrix}
\end{equation}
where 
\begin{equation}
    \varepsilon_{\rm SEB} /e = \alpha_{\rm G} V_{\rm G} - (1 - \alpha_{\rm R}) V_{\rm R},
\end{equation}
\noindent
is the QD energy detuning with respect to the Fermi level of the reservoir.
The parameters $\alpha_{{\rm G}(R)}$ indicate the QD-gate (reservoir) lever arm.
The tunneling process in and out of the QD can then be represented by the jump operators \cite{Oakes_Peri_2023}
\begin{equation}
    L_{\rm in} = \begin{pmatrix}
        0 & 1\\
        0 & 0\\
    \end{pmatrix} \hspace*{0.1\textwidth} L_{\rm out} = \begin{pmatrix}
        0 & 0\\
        1 & 0\\
    \end{pmatrix},
\end{equation}
and their respective tunnel rates then read \cite{Peri2024beyondadiabatic}
\begin{equation}
    \Gamma_{\rm in} = \Gamma \mathcal{F}(\varepsilon_{\rm SEB}) \hspace*{0.07\textwidth} 
    \Gamma_{\rm out} = \Gamma \left(1 - \mathcal{F}(\varepsilon_{\rm SEB})\right).
\end{equation}

The function
\begin{equation}
    \begin{aligned}
    \mathcal{F}(\varepsilon_{\rm SEB}) &= \mathcal{D}_{\rm QD}(\varepsilon_{\rm SEB};\Gamma) * f_{\rm R}(\varepsilon_{\rm SEB};k_{\rm B} T) \\
    &= \frac{1}{2} + \frac{1}{\pi} \psi_0 \left(\frac{1}{2} + \frac{\Gamma + {\rm i} \varepsilon_{\rm SEB}}{2 \pi k_{\rm B} T}\right)
    \end{aligned}
    \label{eq:LBFD}
\end{equation}
\noindent
represents the convolution of the (Lorentzian) effective density of state of the QD and the Fermi-Dirac distribution in the reservoir at thermal equilibrium\footnote{$\psi_0(z)$ represents Euler's digamma function \cite{Peri2024beyondadiabatic}.}. 

To begin with, we investigate the small-signal properties of the compact model. We apply a small sinusoidal voltage ($e\alpha_{\rm G} \delta V_{\rm GR} \ll k_{\rm B} T, h\Gamma$) at the gate terminal and monitor the resulting gate current to determine the equivalent admittance seen by the gate.
As shown in \figr{fig:SEB_Simple}c-f, we find excellent agreement of the model with the theoretical expectations. Generally, the admittance as a function of gate-reservoir potential difference takes the form of a zero-centered single peak (\figr{fig:SEB_Simple}c, e), whose width and height depend on the model's parameters \cite{Ahmed_2018,Peri2024beyondadiabatic,Vigneau_2023,Gonzalez_Zalba_2015}.
To showcase the model's capabilities, we first vary the simulation temperature. This results in a thermal broadening of the peak (\figr{fig:SEB_Simple}c), with the subsequent reduction in the maximum peak height (inversely proportional to the reservoir temperature \cite{Ahmed_2018}) shown in \figr{fig:SEB_Simple}d.

Our inclusion of the effective density of state in \eqr{eq:LBFD} allows the compact model to further include a phenomenon known as lifetime broadening \cite{Ahmed_2018,Peri2024beyondadiabatic}. 
This effect is shown in \figr{fig:SEB_Simple}e, f, where we sweep the tunnel rate for $T = 100$\,mK. When first increasing $\Gamma$, we see a simple increase in peak height (green and purple traces in \figr{fig:SEB_Simple}e). This arises from the fact that faster tunneling results in a larger number of tunneling events per ac cycle, thus increasing the gate current \cite{Peri2024beyondadiabatic}. 
As the tunnel rates become faster, however, we notice the peak start to broaden and the height drops (red trace).
This effect originates from the metastable nature of the electron level as the QD is coupled to the reservoir.
The finite lifetime of electrons in the QD causes a Heisenberg broadening of the energy level with width $h \Gamma$. When the tunnel rates become comparable to (or faster than) the reservoir temperature, the short electron lifetime leads to a further broadening and lowering of the admittance peak due to the smearing of the effective electron density of states \cite{Ahmed_2018,Oakes_Peri_2023,Peri2024beyondadiabatic,vonHorstig_floquet_2024}.

Finally, we explore the properties of the model in the large-signal regime (\figr{fig:SEB_Simple}g,h). To do so, we fix $\Gamma = 2$\,GHz and $k_{\rm B} T=100$\,mK and increase the amplitude of the ac stimuli. From the resulting gate current, we extract the average admittance, defined as the first harmonic component of the gate current divided by the voltage amplitude \cite{Peri2024beyondadiabatic}.
As shown in \figr{fig:SEB_Simple}g, when increasing the amplitude of the ac excitation, the gate current ($\delta V_{\rm GR} |Y|$) increases, resulting in a higher and broader peak---a phenomenon known as power broadening \cite{Oakes_Peri_2023}. 
Observing the maximum peak height (\figr{fig:SEB_Simple}h), we notice how the increase is at first linear with respect to the  voltage amplitude (small-signal regime), and saturates at large amplitudes. This is easily understood by noting that, once the voltage swing is comparable to the lifetime and thermal broadening, the QD completely empties and fills once per ac cycle \cite{Oakes_Peri_2023}. Therefore, increasing the voltage amplitude only affects the tails of the peak, but leads to progressively negligible increase of the gate current at zero detuning. 
This interpretation also implies nonlinearities in the gate current at large powers \cite{Oakes_Peri_2023}, which we shall investigate and exploit for the purpose of frequency multiplication later in this work.

Notably, we see how all the effects discussed in this section are captured by our SEB compact model and are in excellent agreement with the expectations from theory \cite{Peri2024beyondadiabatic,Peri_2023,Oakes_2023}.

\subsection*{Double Quantum Dot}

\begin{figure}[htb!]
    \centering
    \includegraphics[width=0.9\linewidth]{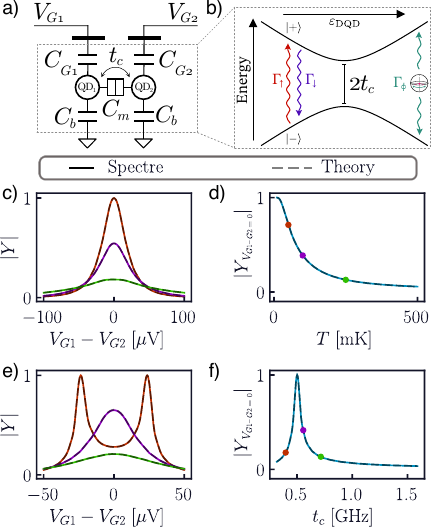}
    \caption{\textbf{Admittance of a double quantum-dot charge qubit.}
    Model of a DQD from ircuit (a) and quantum (b) perspectives. (c-f) Admittance lineshapes (left) and value at zero-detuning (right) at $f=1$\,GHz when varying temperature (c,d) and tunnel coupling (e,f). When not specified, the values $T=100$\,mK, $t_{\rm c}=8$\,GHz, $\Gamma_{\rm cr}=0.5$\,GHz, and $\Gamma_{\phi}=0$ are assumed.}
    \label{fig:DQD_Simple}
\end{figure}

As a second example of a QD device, we demonstrate a compact model for a DQD charge qubit (\figr{fig:DQD_Simple}). 
This is composed of two QDs coupled capacitively (with capacitance $C_{\rm m}$) and through tunneling (with overlap $t_{\rm c}$), each also capacitively coupled to its own gate. The model allows for cross-capacitances between QDs and gates (not shown), as well as a stray capacitance ($C_{\rm b}$) to ground. 
From a quantum perspective, the DQD is modelled as a two-level system, whose Hamiltonian reads \cite{Peri_2023}:
\begin{equation}
    H = \frac{1}{2} \begin{pmatrix}
        -\varepsilon_{\rm DQD} & 2 t_{\rm c}\\
        2 t_{\rm c} & \varepsilon_{\rm DQD}\\
    \end{pmatrix}  ,
\end{equation}
\noindent where
\begin{equation}
    \varepsilon_{\rm DQD} /e = \alpha_{{\rm G}_1} V_{{\rm G}_1} + \alpha_{{\rm G}_2} V_{{\rm G}_2}
\end{equation}
\noindent
is the DQD detuning.
The presence of a finite tunnel coupling between the two QDs gives rise to an avoided crossing at zero detuning between the two levels (\figr{fig:DQD_Simple}b). Therefore, the eigenstates of the system ($\ket{\pm}$) are now dependent on detuning (and hence on voltage). This is in stark contrast to the previous case of the SEB, and, as we shall demonstrate, allows for coherence phenomena in response to changes in voltage.

Our compact model includes the presence of decoherence processes, treated in a Born approximation known as the instantaneous eigenvalue approximation \cite{Peri2024beyondadiabatic,Yamaguchi_Yuge_Ogawa_2017}, which assumes stochastic processes to be well-described in the instantaneous eigenbasis of the (time-dependent) Hamiltonian \cite{Albash_Boixo_Lidar_Zanardi_2012,Cattaneo_2019,Mori_2023,Ikeda_Chinzei_Sato_2021}.
In particular, we follow a standard model in the literature for the DQD \cite{Mizuta2017,Esterli_Otxoa_Gonzalez-Zalba_2019,Peri_2023}, where relaxation processes are dominated by coupling with the phonon bath in the lattice, reading 
\begin{equation}
    L_{\uparrow}' = \begin{pmatrix}
        0 & 1\\
        0 & 0\\
    \end{pmatrix} \hspace*{0.1\textwidth} L_{\downarrow}' = \begin{pmatrix}
        0 & 0\\
        1 & 0\\
    \end{pmatrix},
\end{equation}
\noindent
where the primes indicate that these are defined in the instantaneous eigenbasis. Their respective relaxation rates read
\begin{equation}
    \Gamma_{\uparrow} = \Gamma_{\rm cr} n(\varepsilon_{\rm DQD}) \hspace*{0.07\textwidth} 
    \Gamma_{\downarrow} = \Gamma_{\rm cr} \left(n(\varepsilon_{\rm DQD}) + 1 \right),
    \label{eq:gamma_DQD}
\end{equation}
\noindent
where $\Gamma_{\rm cr}$ is the charge relaxation rate quantifying the charge-phonon coupling, and
\begin{equation}
    n(\varepsilon_{\rm DQD}) = \left(1 - \exp\left(\frac{\sqrt{\varepsilon_{\rm DQD}^2 + 4t_{\rm c}^2}}{k_{\rm B} T}\right)\right)^{-1}
\end{equation}
\noindent
is the Bose-Einstein statistics of the phonon bath at the (instantaneous) $\ket{\pm}$ energy separation.
We also include the possibility of pure dephasing processes, described by the jump operator
\begin{equation}
    L_{\phi}' = \frac{1}{\sqrt{2}} \begin{pmatrix}
        1 & 0\\
        0 & -1\\
    \end{pmatrix},
\end{equation}
\noindent
with (energy-independent) rate $\Gamma_\phi$.
The total decoherence rate reads $\gamma/2\pi = \Gamma_{\rm cr} \left(n(\varepsilon_{\rm DQD}) + 1/2 \right) + \Gamma_\phi$ \cite{Peri_2023}.

To begin with, we consider the small-signal (ac) admittance of the model seen by one of the gates at frequency $f=1$\,GHz. We consider slow relaxation ($\Gamma_{\rm cr} = 0.5$\,GHz) with no dephasing.
Firstly, we explore the system's response to an increase in temperature (\figr{fig:DQD_Simple}c, d) in the adiabatic regime $t_{\rm c} = 5$\,GHz~$\gg f$.
Similarly to the SEB, the admittance takes the form of a zero-centered peak, whose height decreases as the temperature increases, in accordance with the theoretical expectations. The physical origin of this phenomenon is that, in this regime, the admittance is dominated by the system's \textit{quantum capacitance}, deriving from the redistribution of electron occupation within the QDs due to the avoided crossing \cite{Mizuta2017}. This quantity is generally proportional to the second derivative of the eigenenergies \cite{Esterli_Otxoa_Gonzalez-Zalba_2019}, and thus is opposite in sign for ground and excited states. As temperature increases, the system depolarizes and there is (exponentially) more probability of occupation of the excited state. For temperatures larger than the energy splitting, the $\ket{\pm}$ quantum capacitances cancel out, leading to a vanishing admittance.
However, from \figr{fig:DQD_Simple}c we see that the peak not only shrinks, but also \textit{broadens}. This is due to the energy dependence of the relaxation rates in \eqr{eq:gamma_DQD}, which gives rise to a tunnelling capacitance and Sisyphus resistance \cite{Esterli_Otxoa_Gonzalez-Zalba_2019,Mizuta2017,Vigneau_2023}.
The faithful reproduction of this phenomenon in the Spectre\registeredmark~simulation demonstrates the ability of our model to include Sisyphus phenomena and the full capabilities of the presented approach to implement Liouvillians within the instantaneous-eigenvalue approximation.

\begin{figure}[htb!]
    \centering
    \includegraphics[width=0.9\linewidth]{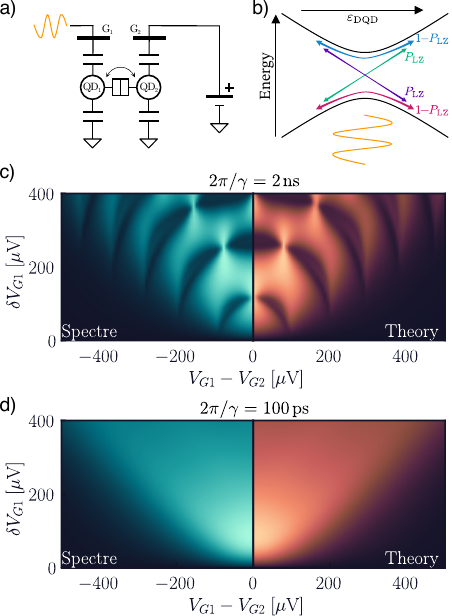}
    \caption{\textbf{LSZM interference in a charge qubit.} 
    Model of LSZM interference from the circuit (a) and quantum (b) perspectives: the gate of a charge qubit is strongly driven to cause diabatic transitions and accumulation of a dynamical phase.
    (c-d) First-harmonic gate current ($\delta V_{\rm G1} |Y|$) in the case where the coherence time $2\pi/\gamma$ is much longer (c) and much shorter (d) than the excitation frequency $f=1$\,GHz.}
    \label{fig:DQD_LZSM}
\end{figure}

Secondly, we fix the simulation temperature at $T=100$\,mK, and explore the response for varying tunnel coupling (\figr{fig:DQD_Simple}e, f). As $t_{\rm c}$ becomes smaller, the anticrossing becomes narrower and sharper, as does the DQD admittance. This trend, however, only holds while the system is in the adiabatic regime. As $2t_{\rm c}$ decreases below the excitation frequency, the excitation is able to coherently drive the charge qubit. This leads to the admittance splitting into two peaks (\figr{fig:DQD_Simple}e), with sharp features at the detuning, where the charge qubit is resonant with the ac stimuli \cite{Peri_2023}. This can be understood as a resonant driving of the charge qubit by the sinusoidal excitation \cite{kohler_dispersive_2017,kohler_dispersive_2017,Benito_Mi_2017,vonhorstig2024electrical,Mi_2018}. Decreasing $t_{\rm c}$ further decreases the dipole of the system, thus leading to a sharp drop in the zero-detuning admittance (\figr{fig:DQD_Simple}f) \cite{Peri_vonHorstig_2024}.
Figure~\ref{fig:DQD_Simple} is a demonstration of the capability of a compact model to reproduce resonant quantum behavior within a standard classical electronics simulator.

\begin{figure}[htb!]
    \centering
    \includegraphics[width=0.9\linewidth]{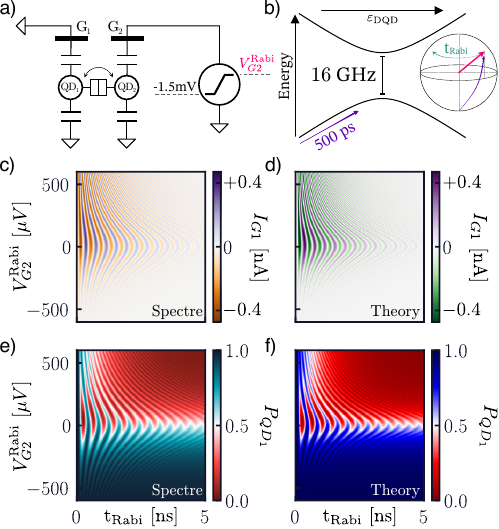}
    \caption{\textbf{Damped Rabi oscillations of a charge qubit.}
    Circuit (a) and quantum (b) model for generating Rabi oscillations: the gate of a charge qubit is pulsed towards the anticrossing (up to $V^{\rm Rabi}$) to generate a coherent superposition of ground and excited state. (c-f) Circuit simulations (left) and theory expectations (right) of the gate current (c, d) and QD occupation probability (e, f) after the pulse.
    }
    \label{fig:DQD_Rabi}
\end{figure}

We now explore the large-signal regime (\figr{fig:DQD_LZSM}), further discussing the description of coherent phenomena. 
From the quantum perspective, the biggest difference compared with the SEB is the presence of an anticrossing, which, when rapidly changing detuning in its proximity, may give rise to diabatic transitions into other states, known as Landau-Zener transitions, with probability $P_{\rm LZ}$ (\figr{fig:DQD_LZSM}b) \cite{Petta_Lu_Gossard_2010,Glasbrenner_Schleich_2023}.
If the voltages are changed periodically, even if the system is initialized in its ground state, periodic fast passages across the anticrossing (\figr{fig:DQD_LZSM}b) may generate a coherent superposition of the ground and excited states. This leads to the accumulation of a (dynamical) quantum phase in the system, which in turn leads to self-interference of the electron, in a process known as Landau-Zener-St\"uckelberg-Majorana (LZSM) interference \cite{Shevchenko_2010,Ivakhnenko_2023,GonzalezZalba_2016}.
In \figr{fig:DQD_LZSM}c, we explore this phenomenon in our compact model. In particular, we set $1/\Gamma_{\rm cr} = 2$\,ns and neglect dephasing, to allow the coherence time to be longer than the period of the sinusoidal excitation ($f=1$\,GHz). As shown, this results in the emergence of LZSM interference fringes in the equivalent DQD admittance, corresponding to constructive and destructive interference of the accumulated phase, in excellent agreement with the theory.
To explore the role of dephasing, we now increase $\Gamma_\phi$ such that the coherence time of the system is shorter than the ac period. As shown in \figr{fig:DQD_LZSM}d, this results in the disappearance of the interference fringes, which blur into one single broad peak, in accordance with the theory and experimental findings \cite{Shevchenko_2010,Ivakhnenko_2023, Oakes_Peri_2023,vonHorstig_floquet_2024}.
This further demonstrates the possibility of fully capturing coherent quantum phenomenon in a traditional mixed-signal electronic simulation.

Finally, we explore the coherent evolution of our model in the time domain (\figr{fig:DQD_Rabi}). To do so, we first initialize the system in its ground state far from the anticrossing, so that the charge is strongly localized in one of the QDs. We then pulse one of the gates within $500$\,ps to a (variable) voltage near the charge transition, where we allow the system to evolve freely (\figr{fig:DQD_Rabi}a, b) while recording the gate current of one of the gates and the occupation probability of the associated QD.
Figure~\ref{fig:DQD_Rabi} shows the characteristic pattern of the Rabi chevron \cite{Yoneda_2014} recreated in the gate current (\figr{fig:DQD_Rabi}c, d) while Rabi oscillations are clearly visible in the population of the QD (\figr{fig:DQD_Rabi}e, f). 
These are due to the fast pulse diabatically creating a superposition of ground and excited states, outside the $z$-axis of the (instantaneous) Bloch sphere at $V^{\rm Rabi}$. This state then naturally precesses around the $z$-axis, generating Rabi oscillations within the charge occupation, exponentially damped in the Lindblad formalism by the finite coherence rate.
As shown in \figr{fig:DQD_Rabi}, the simulation results are in excellent agreement with the theory, showcasing the ability to reproduce coherent effects in time-domain transients.

\begin{figure*}[htb!]
    \centering
    \includegraphics[width=0.9\textwidth]{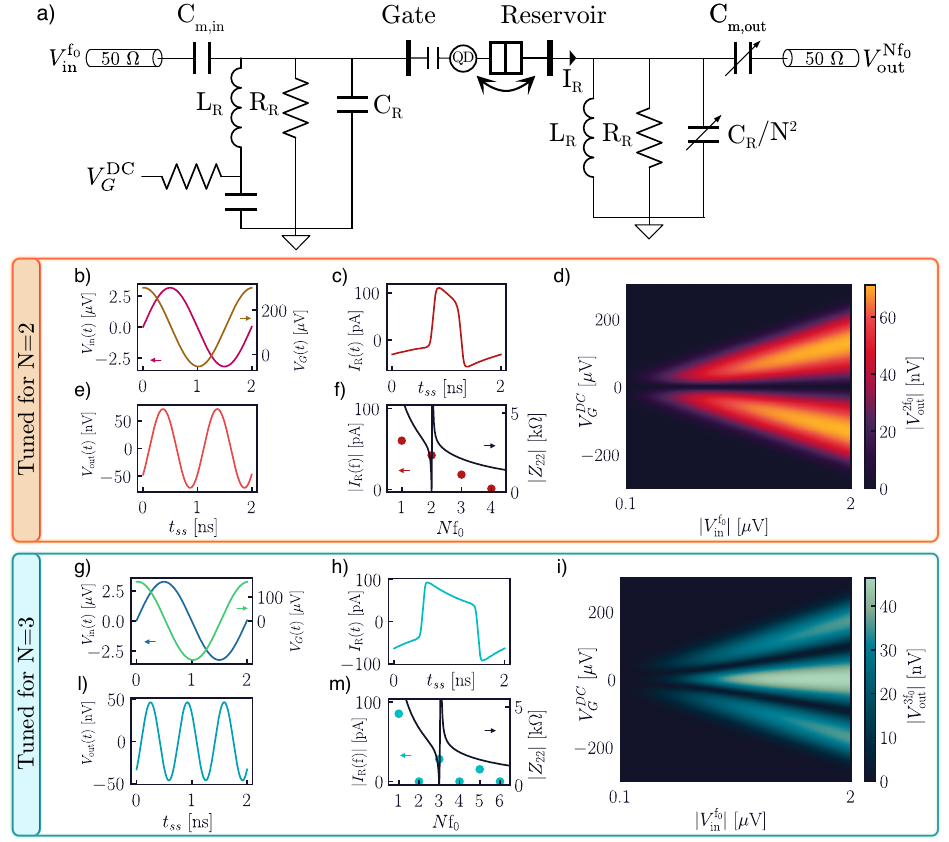}
    \caption{\textbf{Frequency multiplier circuit based on a single-electron box.} (a) Circuit schematic of the frequency multiplier. (b-f) The panels show the input, gate (b), and output voltages (e) in the time domain, and the time and frequency domains of the reservoir current (c, f). Panel~e shows the characteristic two-lobed fan of the output voltage with respect to SEB detuning and input amplitude \cite{Oakes_Peri_2023}. (g-m) The same quantities when the circuit is tuned for $N=3$.}
    \label{fig:Fmul}
\end{figure*}

\section{Co-Simulation of Quantum Dots and Analog circuits}
\label{sec:circuits}

Having investigated the behavior of the compact models, validating them against theory, we now showcase our co-simulation capabilities by implementing QD-based analog circuits composed of a QD compact model and linear circuit elements. 
In particular, we demonstrate (i) a SEB-based frequency multiplier, leveraging the nonlinear properties of the quantum system, and (ii) a charge qubit coupled to an RLC resonator, investigating the frequency response in the adiabatic and resonant regimes.

\subsection*{Single-Electron Box Frequency Multiplier}

The first hybrid quantum-classical device we discuss is a frequency multiplier based on the QD-to-reservoir transition of a SEB (\figr{fig:Fmul}).
Our implementation is based on traditional designs of analog circuits for frequency multiplication \cite{Suarez_2006,Bava_1979,Diamond_1963}, and it can be understood as essentially composed of two independent resonant current loops. 
On the input side (left \figr{fig:Fmul}a), we find an RLC resonator tuned for the input frequency (in our design $f_0\sim0.5$\,MHz), with the aim of providing a monochromatic driving for the SEB gate, while simultaneously increasing the amplitude of the oscillating gate voltage (\figr{fig:Fmul}b, g). A bias tee provides the ability to vary the dc detuning of the gate.
On the output side (right \figr{fig:Fmul}a), we find a similar design, where the SEB reservoir is connected to a parallel RLC resonator featuring a variable capacitor. If the same inductor is used in both loops, setting the output resonator capacitor to $C_{\rm R}/N^2$ causes the output loop to resonate at the $N^{\rm th}$ harmonic of the input frequency $f_0$, filtering out unwanted Fourier components. Both the input and the output are coupled to $50\,\Omega$ lines via coupling capacitors.

The frequency-multiplication capabilities of this circuit stem from the inherent nonlinearity of the SEB. In particular, when the amplitude of the gate-reservoir voltage swing is larger than any other broadening ($\delta \varepsilon \gg h \Gamma, k_{\rm  B} T$), the excitation will produce one tunneling event per cycle. Thus, the probability of occupation of the QD will resemble a square wave (or, more generally, a sawtooth, depending on the tunnel rates) \cite{Oakes_Peri_2023,Hogg_2023,Cochrane_2024}.
From \eqr{eq:rho_to_I}, the reservoir current $I_{\rm R}$ is proportional to the time derivative of the QD occupation, causing it to have a rich Fourier decomposition (\figr{fig:Fmul}c, f, h, m). Most importantly, from the above discussion it is clear how the \textit{duty cycle} of the QM occupation square wave depends on the dc detuning of the QD with respect to the reservoir, thus rendering the reservoir current electrically tunable.

In \figr{fig:Fmul}, we explore the frequency-multiplication performance of the circuit, analyzed via harmonic-balance simulations. 
In \figr{fig:Fmul}b-f, we tune the circuit for a multiplication factor of $N=2$. Panel b shows both the passive amplification of the input resonator and the addition of the dc voltage, generating the highly nonlinear reservoir current in panel c. Panel f shows the Fourier decomposition of $I_{\rm R}$ and the impedance $Z_{22}$ seen by the output line, which drops to $\sim50$\,$\Omega$ for $2f_0$. The output resonator thus efficiently acts as a band-pass filter, ensuring high harmonic purity of the output, shown in \figr{fig:Fmul}e.
In \figr{fig:Fmul}d, we sweep input amplitude and dc detuning of the system, showing the characteristic $N$-lobed fan experimentally observed in similar SEB-base systems \cite{Oakes_Peri_2023,Hogg_2023}.
We show similar results in \figr{fig:Fmul}g-m, in which we vary the values of the capacitors for the output branch to be resonant and $50\,\Omega$-matched at $N=3$, resulting in excellent harmonic purity of the third harmonic at the output (\figr{fig:Fmul}l).
Moreover, in \figr{fig:Fmul}i, we show the expected three-lobed fan in the output voltage when sweeping detuning and input amplitude, similarly to the $N=2$ case.

\subsection*{Circuit Quantum Electrodynamics With a Charge Qubit and a high-Q Resonator}

\begin{figure}[htb!]
    \centering
    \includegraphics[width=0.9\linewidth]{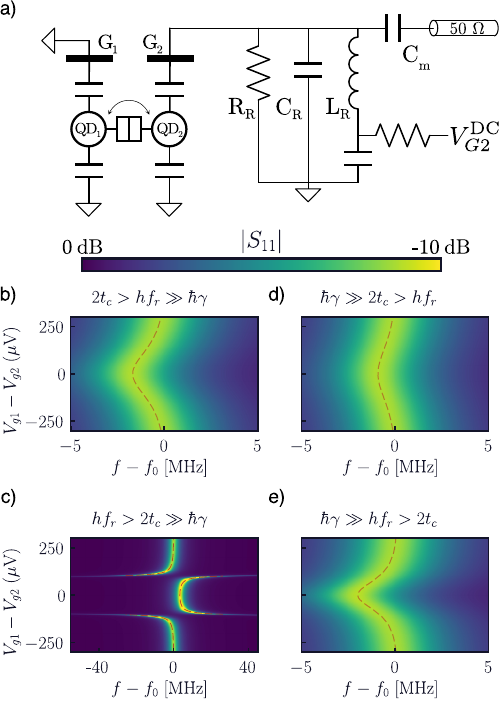}
    \caption{\textbf{Adiabatic and resonant dispersive sensing of a charge qubit.} (a) Circuit schematic of a DQD charge qubit coupled to a high-$Q$ ($Q=1000$) microwave resonator ($f_0=2$\,GHz). (b-e) $|S_{11}|$ parameter for the circuit as a function of frequency and detuning, showing the dispersive shift of the RLC resonator caused by the qubit in the adiabatic ($hf_0 \gg 2t_{\rm c}$, top) and resonant ($hf_0 < 2t_c$, bottom) regimes, in the presence of decoherence that is slow ($\hbar \gamma \ll 2t_{\rm c}$, left) or fast ($\hbar \gamma \gg 2t_{\rm c}$, right).}
    \label{fig:DQD_refl}
\end{figure}

As a last example, we discuss the case of a DQD charge qubit coupled to a microwave resonator ($f_0=2$\,GHz). 
Similarly to the previous circuit, we consider a parallel RLC resonator (\figr{fig:DQD_refl}a) with a dc bias tee to vary the DQD detuning. The system is coupled to a 50\,$\Omega$ lead via a coupling capacitor, mimicking a traditional radio-frequency reflectometry setup \cite{Vigneau_2023,Oakes_2023}, and we study the reflected signal ($|S_{11}|$) while varying DQD parameters, comparing the simulated resonant frequency shift with the expectations from circuit quantum electrodynamics (cQED) \cite{Ibberson_2021,Peri_2023,kohler_dispersive_2017}, shown as black dashed lines in \figr{fig:DQD_refl}.

Firstly, we consider the case in which the charge qubit has a long coherence time compared to the cavity ($2\pi/\gamma = 5$\,ns\,$\gg 1/f_0$). 
In the adiabatic regime, where $2t_{\rm c} = 10$\,GHz (\figr{fig:DQD_refl}b), we see the traditional result of a pull of the resonance towards lower frequencies, due to the quantum capacitance of the DQD in parallel to $C_{\rm R}$ \cite{Ibberson_2021}. 
If we lower $2t_{\rm c}$ below the resonator frequency, however, we see a starkly different response. In \figr{fig:DQD_refl}c, we observe a large shift in the resonant frequency where the qubit frequency is resonant with the cavity, characterized by a change in sign of quantum capacitance. This leads to a pull of the resonance towards higher frequencies in the region where the qubit frequency drops below $f_0$, in excellent agreement with cQED \cite{Ibberson_2021,Peri_2023}.
The behavior in \figr{fig:DQD_refl}c may be understood effectively as a vacuum Rabi splitting caused by the qubit--resonator coupling, which we show faithfully reproduced by the Spectre{\registeredmark} simulation.

Finally, we explore the influence of DQD dephasing on the reflected signal (\figr{fig:DQD_refl}d, e). To do so, we increase the dephasing rate $\Gamma_\phi$ to $2\pi/\gamma = 1$\,ps\,$\ll 1/f_0$. In the adiabatic regime (\figr{fig:DQD_refl}d), we merely see a reduction in the frequency pull, caused by the response of the qubit becoming increasingly resistive because of the increased losses \cite{Peri_2023,kohler_dispersive_2017}.
The effect is far more dramatic in the resonant case (\figr{fig:DQD_refl}e), where we observe the peak splitting and the region of negative capacitance disappear, in favor of a zero-centered peak of positive capacitance resembling the adiabatic case. The Spectre{\registeredmark} simulations correctly reproduce the behavior predicted by our unified linear-response theory of quantum systems \cite{Peri_2023}, demonstrating the full capabilities of our Lindblad-based compact models.

\section{Conclusions and Outlook}

We have shown that Verilog-A compact models of QD devices---a SEB and a DQD charge qubit---are able to reproduce coherent phenomena while remaining fully compatible with industry-standard analog circuit simulators. Furthermore, we have demonstrated the utility of these models for the design of two quantum-classical hybrid circuits, a SEB-based frequency multiplier and a charge qubit embedded in an LC resonator for quantum state readout, both circuits fully simulated in Cadence Spectre\registeredmark.
Overall, our work proposes a systematic method to produce a compact model for any arbitrary multilevel quantum system, as long as it can be described by a Lindblad master equation, which can then be co-simulated alongside classical circuit components in most available circuit simulators. 
This methodology bridges the existing gap in simulation capabilities between quantum and classical electronics, and enables quantum circuits to leverage the electronic design and automation tools that have enabled the very large-scale integration achieved by classical electronics, ushering in a new paradigm for the design optimization of the quantum-classical interfaces and unlocking the design of novel hybrid quantum-classical analog circuits.

\section*{Acknowledgements}
This research was supported by the European Union's Horizon 2020 research and innovation programme under grant agreement no.~951852 (QLSI), and by the UK's Engineering and Physical Sciences Research Council (EPSRC) via the Cambridge NanoDTC (EP/L015978/1). 
M.F.G.Z. acknowledges a UKRI Future Leaders Fellowship [MR/V023284/1]. 
L.P. acknowledges funding from the Winton Programme for the Physics of Sustainability.
A.G.-S. acknowledges an Industrial Fellowship from the Royal Commission for the Exhibition of 1851.

\section*{Author Contributions}
L.P. devised the quantum formalism, and L.P. and A.G.-S. implemented the models.
L.P. performed the simulations with input from A.G.-S. and M.F.G.Z. M.F.G.Z. and C.J.B.F. supervised the work.
All authors contributed to the discussion and interpretation of results.


\begin{thebibliography}{80}%
    \makeatletter
    \providecommand \@ifxundefined [1]{%
     \@ifx{#1\undefined}
    }%
    \providecommand \@ifnum [1]{%
     \ifnum #1\expandafter \@firstoftwo
     \else \expandafter \@secondoftwo
     \fi
    }%
    \providecommand \@ifx [1]{%
     \ifx #1\expandafter \@firstoftwo
     \else \expandafter \@secondoftwo
     \fi
    }%
    \providecommand \natexlab [1]{#1}%
    \providecommand \enquote  [1]{``#1''}%
    \providecommand \bibnamefont  [1]{#1}%
    \providecommand \bibfnamefont [1]{#1}%
    \providecommand \citenamefont [1]{#1}%
    \providecommand \href@noop [0]{\@secondoftwo}%
    \providecommand \href [0]{\begingroup \@sanitize@url \@href}%
    \providecommand \@href[1]{\@@startlink{#1}\@@href}%
    \providecommand \@@href[1]{\endgroup#1\@@endlink}%
    \providecommand \@sanitize@url [0]{\catcode `\\12\catcode `\$12\catcode
      `\&12\catcode `\#12\catcode `\^12\catcode `\_12\catcode `\%12\relax}%
    \providecommand \@@startlink[1]{}%
    \providecommand \@@endlink[0]{}%
    \providecommand \url  [0]{\begingroup\@sanitize@url \@url }%
    \providecommand \@url [1]{\endgroup\@href {#1}{\urlprefix }}%
    \providecommand \urlprefix  [0]{URL }%
    \providecommand \Eprint [0]{\href }%
    \providecommand \doibase [0]{https://doi.org/}%
    \providecommand \selectlanguage [0]{\@gobble}%
    \providecommand \bibinfo  [0]{\@secondoftwo}%
    \providecommand \bibfield  [0]{\@secondoftwo}%
    \providecommand \translation [1]{[#1]}%
    \providecommand \BibitemOpen [0]{}%
    \providecommand \bibitemStop [0]{}%
    \providecommand \bibitemNoStop [0]{.\EOS\space}%
    \providecommand \EOS [0]{\spacefactor3000\relax}%
    \providecommand \BibitemShut  [1]{\csname bibitem#1\endcsname}%
    \let\auto@bib@innerbib\@empty
    \bibitem [{\citenamefont {Arute}\ \emph {et~al.}(2019)\citenamefont {Arute},
      \citenamefont {Arya}, \citenamefont {Babbush}, \citenamefont {Bacon},
      \citenamefont {Bardin}, \citenamefont {Barends}, \citenamefont {Biswas},
      \citenamefont {Boixo}, \citenamefont {Brandao}, \citenamefont {Buell},
      \citenamefont {Burkett}, \citenamefont {Chen}, \citenamefont {Chen},
      \citenamefont {Chiaro}, \citenamefont {Collins}, \citenamefont {Courtney},
      \citenamefont {Dunsworth}, \citenamefont {Farhi}, \citenamefont {Foxen},
      \citenamefont {Fowler}, \citenamefont {Gidney}, \citenamefont {Giustina},
      \citenamefont {Graff}, \citenamefont {Guerin}, \citenamefont {Habegger},
      \citenamefont {Harrigan}, \citenamefont {Hartmann}, \citenamefont {Ho},
      \citenamefont {Hoffmann}, \citenamefont {Huang}, \citenamefont {Humble},
      \citenamefont {Isakov}, \citenamefont {Jeffrey}, \citenamefont {Jiang},
      \citenamefont {Kafri}, \citenamefont {Kechedzhi}, \citenamefont {Kelly},
      \citenamefont {Klimov}, \citenamefont {Knysh}, \citenamefont {Korotkov},
      \citenamefont {Kostritsa}, \citenamefont {Landhuis}, \citenamefont
      {Lindmark}, \citenamefont {Lucero}, \citenamefont {Lyakh}, \citenamefont
      {Mandrà}, \citenamefont {McClean}, \citenamefont {McEwen}, \citenamefont
      {Megrant}, \citenamefont {Mi}, \citenamefont {Michielsen}, \citenamefont
      {Mohseni}, \citenamefont {Mutus}, \citenamefont {Naaman}, \citenamefont
      {Neeley}, \citenamefont {Neill}, \citenamefont {Niu}, \citenamefont {Ostby},
      \citenamefont {Petukhov}, \citenamefont {Platt}, \citenamefont {Quintana},
      \citenamefont {Rieffel}, \citenamefont {Roushan}, \citenamefont {Rubin},
      \citenamefont {Sank}, \citenamefont {Satzinger}, \citenamefont {Smelyanskiy},
      \citenamefont {Sung}, \citenamefont {Trevithick}, \citenamefont
      {Vainsencher}, \citenamefont {Villalonga}, \citenamefont {White},
      \citenamefont {Yao}, \citenamefont {Yeh}, \citenamefont {Zalcman},
      \citenamefont {Neven},\ and\ \citenamefont {Martinis}}]{Arute_2019}%
      \BibitemOpen
      \bibfield  {author} {\bibinfo {author} {\bibfnamefont {F.}~\bibnamefont
      {Arute}}, \bibinfo {author} {\bibfnamefont {K.}~\bibnamefont {Arya}},
      \bibinfo {author} {\bibfnamefont {R.}~\bibnamefont {Babbush}}, \bibinfo
      {author} {\bibfnamefont {D.}~\bibnamefont {Bacon}}, \bibinfo {author}
      {\bibfnamefont {J.~C.}\ \bibnamefont {Bardin}}, \bibinfo {author}
      {\bibfnamefont {R.}~\bibnamefont {Barends}}, \bibinfo {author} {\bibfnamefont
      {R.}~\bibnamefont {Biswas}}, \bibinfo {author} {\bibfnamefont
      {S.}~\bibnamefont {Boixo}}, \bibinfo {author} {\bibfnamefont {F.~G. S.~L.}\
      \bibnamefont {Brandao}}, \bibinfo {author} {\bibfnamefont {D.~A.}\
      \bibnamefont {Buell}}, \bibinfo {author} {\bibfnamefont {B.}~\bibnamefont
      {Burkett}}, \bibinfo {author} {\bibfnamefont {Y.}~\bibnamefont {Chen}},
      \bibinfo {author} {\bibfnamefont {Z.}~\bibnamefont {Chen}}, \bibinfo {author}
      {\bibfnamefont {B.}~\bibnamefont {Chiaro}}, \bibinfo {author} {\bibfnamefont
      {R.}~\bibnamefont {Collins}}, \bibinfo {author} {\bibfnamefont
      {W.}~\bibnamefont {Courtney}}, \bibinfo {author} {\bibfnamefont
      {A.}~\bibnamefont {Dunsworth}}, \bibinfo {author} {\bibfnamefont
      {E.}~\bibnamefont {Farhi}}, \bibinfo {author} {\bibfnamefont
      {B.}~\bibnamefont {Foxen}}, \bibinfo {author} {\bibfnamefont
      {A.}~\bibnamefont {Fowler}}, \bibinfo {author} {\bibfnamefont
      {C.}~\bibnamefont {Gidney}}, \bibinfo {author} {\bibfnamefont
      {M.}~\bibnamefont {Giustina}}, \bibinfo {author} {\bibfnamefont
      {R.}~\bibnamefont {Graff}}, \bibinfo {author} {\bibfnamefont
      {K.}~\bibnamefont {Guerin}}, \bibinfo {author} {\bibfnamefont
      {S.}~\bibnamefont {Habegger}}, \bibinfo {author} {\bibfnamefont {M.~P.}\
      \bibnamefont {Harrigan}}, \bibinfo {author} {\bibfnamefont {M.~J.}\
      \bibnamefont {Hartmann}}, \bibinfo {author} {\bibfnamefont {A.}~\bibnamefont
      {Ho}}, \bibinfo {author} {\bibfnamefont {M.}~\bibnamefont {Hoffmann}},
      \bibinfo {author} {\bibfnamefont {T.}~\bibnamefont {Huang}}, \bibinfo
      {author} {\bibfnamefont {T.~S.}\ \bibnamefont {Humble}}, \bibinfo {author}
      {\bibfnamefont {S.~V.}\ \bibnamefont {Isakov}}, \bibinfo {author}
      {\bibfnamefont {E.}~\bibnamefont {Jeffrey}}, \bibinfo {author} {\bibfnamefont
      {Z.}~\bibnamefont {Jiang}}, \bibinfo {author} {\bibfnamefont
      {D.}~\bibnamefont {Kafri}}, \bibinfo {author} {\bibfnamefont
      {K.}~\bibnamefont {Kechedzhi}}, \bibinfo {author} {\bibfnamefont
      {J.}~\bibnamefont {Kelly}}, \bibinfo {author} {\bibfnamefont {P.~V.}\
      \bibnamefont {Klimov}}, \bibinfo {author} {\bibfnamefont {S.}~\bibnamefont
      {Knysh}}, \bibinfo {author} {\bibfnamefont {A.}~\bibnamefont {Korotkov}},
      \bibinfo {author} {\bibfnamefont {F.}~\bibnamefont {Kostritsa}}, \bibinfo
      {author} {\bibfnamefont {D.}~\bibnamefont {Landhuis}}, \bibinfo {author}
      {\bibfnamefont {M.}~\bibnamefont {Lindmark}}, \bibinfo {author}
      {\bibfnamefont {E.}~\bibnamefont {Lucero}}, \bibinfo {author} {\bibfnamefont
      {D.}~\bibnamefont {Lyakh}}, \bibinfo {author} {\bibfnamefont
      {S.}~\bibnamefont {Mandrà}}, \bibinfo {author} {\bibfnamefont {J.~R.}\
      \bibnamefont {McClean}}, \bibinfo {author} {\bibfnamefont {M.}~\bibnamefont
      {McEwen}}, \bibinfo {author} {\bibfnamefont {A.}~\bibnamefont {Megrant}},
      \bibinfo {author} {\bibfnamefont {X.}~\bibnamefont {Mi}}, \bibinfo {author}
      {\bibfnamefont {K.}~\bibnamefont {Michielsen}}, \bibinfo {author}
      {\bibfnamefont {M.}~\bibnamefont {Mohseni}}, \bibinfo {author} {\bibfnamefont
      {J.}~\bibnamefont {Mutus}}, \bibinfo {author} {\bibfnamefont
      {O.}~\bibnamefont {Naaman}}, \bibinfo {author} {\bibfnamefont
      {M.}~\bibnamefont {Neeley}}, \bibinfo {author} {\bibfnamefont
      {C.}~\bibnamefont {Neill}}, \bibinfo {author} {\bibfnamefont {M.~Y.}\
      \bibnamefont {Niu}}, \bibinfo {author} {\bibfnamefont {E.}~\bibnamefont
      {Ostby}}, \bibinfo {author} {\bibfnamefont {A.}~\bibnamefont {Petukhov}},
      \bibinfo {author} {\bibfnamefont {J.~C.}\ \bibnamefont {Platt}}, \bibinfo
      {author} {\bibfnamefont {C.}~\bibnamefont {Quintana}}, \bibinfo {author}
      {\bibfnamefont {E.~G.}\ \bibnamefont {Rieffel}}, \bibinfo {author}
      {\bibfnamefont {P.}~\bibnamefont {Roushan}}, \bibinfo {author} {\bibfnamefont
      {N.~C.}\ \bibnamefont {Rubin}}, \bibinfo {author} {\bibfnamefont
      {D.}~\bibnamefont {Sank}}, \bibinfo {author} {\bibfnamefont {K.~J.}\
      \bibnamefont {Satzinger}}, \bibinfo {author} {\bibfnamefont {V.}~\bibnamefont
      {Smelyanskiy}}, \bibinfo {author} {\bibfnamefont {K.~J.}\ \bibnamefont
      {Sung}}, \bibinfo {author} {\bibfnamefont {M.~D.}\ \bibnamefont
      {Trevithick}}, \bibinfo {author} {\bibfnamefont {A.}~\bibnamefont
      {Vainsencher}}, \bibinfo {author} {\bibfnamefont {B.}~\bibnamefont
      {Villalonga}}, \bibinfo {author} {\bibfnamefont {T.}~\bibnamefont {White}},
      \bibinfo {author} {\bibfnamefont {Z.~J.}\ \bibnamefont {Yao}}, \bibinfo
      {author} {\bibfnamefont {P.}~\bibnamefont {Yeh}}, \bibinfo {author}
      {\bibfnamefont {A.}~\bibnamefont {Zalcman}}, \bibinfo {author} {\bibfnamefont
      {H.}~\bibnamefont {Neven}},\ and\ \bibinfo {author} {\bibfnamefont {J.~M.}\
      \bibnamefont {Martinis}},\ }\href {https://doi.org/10.1038/s41586-019-1666-5}
      {\bibfield  {journal} {\bibinfo  {journal} {Nature}\ }\textbf {\bibinfo
      {volume} {574}},\ \bibinfo {pages} {505–510} (\bibinfo {year}
      {2019})}\BibitemShut {NoStop}%
    \bibitem [{\citenamefont {Preskill}(2023)}]{Preskill_2023}%
      \BibitemOpen
      \bibfield  {author} {\bibinfo {author} {\bibfnamefont {J.}~\bibnamefont
      {Preskill}},\ }\href {https://arxiv.org/abs/2106.10522} {\bibinfo {title}
      {Quantum computing 40 years later}} (\bibinfo {year} {2023}),\ \Eprint
      {https://arxiv.org/abs/2106.10522} {arXiv:2106.10522} \BibitemShut
      {NoStop}%
    \bibitem [{\citenamefont {Bluvstein}\ \emph {et~al.}(2024)\citenamefont
      {Bluvstein}, \citenamefont {Evered}, \citenamefont {Geim}, \citenamefont
      {Li}, \citenamefont {Zhou}, \citenamefont {Manovitz}, \citenamefont {Ebadi},
      \citenamefont {Cain}, \citenamefont {Kalinowski}, \citenamefont {Hangleiter},
      \citenamefont {Bonilla~Ataides}, \citenamefont {Maskara}, \citenamefont
      {Cong}, \citenamefont {Gao}, \citenamefont {Sales~Rodriguez}, \citenamefont
      {Karolyshyn}, \citenamefont {Semeghini}, \citenamefont {Gullans},
      \citenamefont {Greiner}, \citenamefont {Vuleti{\'{c}}},\ and\ \citenamefont
      {Lukin}}]{Bluvstein2024}%
      \BibitemOpen
      \bibfield  {author} {\bibinfo {author} {\bibfnamefont {D.}~\bibnamefont
      {Bluvstein}}, \bibinfo {author} {\bibfnamefont {S.~J.}\ \bibnamefont
      {Evered}}, \bibinfo {author} {\bibfnamefont {A.~A.}\ \bibnamefont {Geim}},
      \bibinfo {author} {\bibfnamefont {S.~H.}\ \bibnamefont {Li}}, \bibinfo
      {author} {\bibfnamefont {H.}~\bibnamefont {Zhou}}, \bibinfo {author}
      {\bibfnamefont {T.}~\bibnamefont {Manovitz}}, \bibinfo {author}
      {\bibfnamefont {S.}~\bibnamefont {Ebadi}}, \bibinfo {author} {\bibfnamefont
      {M.}~\bibnamefont {Cain}}, \bibinfo {author} {\bibfnamefont {M.}~\bibnamefont
      {Kalinowski}}, \bibinfo {author} {\bibfnamefont {D.}~\bibnamefont
      {Hangleiter}}, \bibinfo {author} {\bibfnamefont {J.~P.}\ \bibnamefont
      {Bonilla~Ataides}}, \bibinfo {author} {\bibfnamefont {N.}~\bibnamefont
      {Maskara}}, \bibinfo {author} {\bibfnamefont {I.}~\bibnamefont {Cong}},
      \bibinfo {author} {\bibfnamefont {X.}~\bibnamefont {Gao}}, \bibinfo {author}
      {\bibfnamefont {P.}~\bibnamefont {Sales~Rodriguez}}, \bibinfo {author}
      {\bibfnamefont {T.}~\bibnamefont {Karolyshyn}}, \bibinfo {author}
      {\bibfnamefont {G.}~\bibnamefont {Semeghini}}, \bibinfo {author}
      {\bibfnamefont {M.~J.}\ \bibnamefont {Gullans}}, \bibinfo {author}
      {\bibfnamefont {M.}~\bibnamefont {Greiner}}, \bibinfo {author} {\bibfnamefont
      {V.}~\bibnamefont {Vuleti{\'{c}}}},\ and\ \bibinfo {author} {\bibfnamefont
      {M.~D.}\ \bibnamefont {Lukin}},\ }\href
      {https://doi.org/10.1038/s41586-023-06927-3} {\bibfield  {journal} {\bibinfo
      {journal} {Nature}\ }\textbf {\bibinfo {volume} {626}},\ \bibinfo {pages}
      {58} (\bibinfo {year} {2024})}\BibitemShut {NoStop}%
    \bibitem [{\citenamefont {Bravyi}\ \emph {et~al.}(2024)\citenamefont {Bravyi},
      \citenamefont {Cross}, \citenamefont {Gambetta}, \citenamefont {Maslov},
      \citenamefont {Rall},\ and\ \citenamefont {Yoder}}]{Bravyi2024}%
      \BibitemOpen
      \bibfield  {author} {\bibinfo {author} {\bibfnamefont {S.}~\bibnamefont
      {Bravyi}}, \bibinfo {author} {\bibfnamefont {A.~W.}\ \bibnamefont {Cross}},
      \bibinfo {author} {\bibfnamefont {J.~M.}\ \bibnamefont {Gambetta}}, \bibinfo
      {author} {\bibfnamefont {D.}~\bibnamefont {Maslov}}, \bibinfo {author}
      {\bibfnamefont {P.}~\bibnamefont {Rall}},\ and\ \bibinfo {author}
      {\bibfnamefont {T.~J.}\ \bibnamefont {Yoder}},\ }\href
      {https://doi.org/10.1038/s41586-024-07107-7} {\bibfield  {journal} {\bibinfo
      {journal} {Nature}\ }\textbf {\bibinfo {volume} {627}},\ \bibinfo {pages}
      {778} (\bibinfo {year} {2024})}\BibitemShut {NoStop}%
    \bibitem [{\citenamefont {Acharya}\ \emph {et~al.}(2024)\citenamefont
      {Acharya}, \citenamefont {Abanin}, \citenamefont {Aghababaie-Beni},
      \citenamefont {Aleiner}, \citenamefont {Andersen}, \citenamefont {Ansmann},
      \citenamefont {Arute}, \citenamefont {Arya}, \citenamefont {Asfaw},
      \citenamefont {Astrakhantsev}, \citenamefont {Atalaya}, \citenamefont
      {Babbush}, \citenamefont {Bacon}, \citenamefont {Ballard}, \citenamefont
      {Bardin}, \citenamefont {Bausch}, \citenamefont {Bengtsson}, \citenamefont
      {Bilmes}, \citenamefont {Blackwell}, \citenamefont {Boixo}, \citenamefont
      {Bortoli}, \citenamefont {Bourassa}, \citenamefont {Bovaird}, \citenamefont
      {Brill}, \citenamefont {Broughton}, \citenamefont {Browne}, \citenamefont
      {Buchea}, \citenamefont {Buckley}, \citenamefont {Buell}, \citenamefont
      {Burger}, \citenamefont {Burkett}, \citenamefont {Bushnell}, \citenamefont
      {Cabrera}, \citenamefont {Campero}, \citenamefont {Chang}, \citenamefont
      {Chen}, \citenamefont {Chen}, \citenamefont {Chiaro}, \citenamefont {Chik},
      \citenamefont {Chou}, \citenamefont {Claes}, \citenamefont {Cleland},
      \citenamefont {Cogan}, \citenamefont {Collins}, \citenamefont {Conner},
      \citenamefont {Courtney}, \citenamefont {Crook}, \citenamefont {Curtin},
      \citenamefont {Das}, \citenamefont {Davies}, \citenamefont {De~Lorenzo},
      \citenamefont {Debroy}, \citenamefont {Demura}, \citenamefont {Devoret},
      \citenamefont {Di~Paolo}, \citenamefont {Donohoe}, \citenamefont {Drozdov},
      \citenamefont {Dunsworth}, \citenamefont {Earle}, \citenamefont {Edlich},
      \citenamefont {Eickbusch}, \citenamefont {Elbag}, \citenamefont {Elzouka},
      \citenamefont {Erickson}, \citenamefont {Faoro}, \citenamefont {Farhi},
      \citenamefont {Ferreira}, \citenamefont {Burgos}, \citenamefont {Forati},
      \citenamefont {Fowler}, \citenamefont {Foxen}, \citenamefont {Ganjam},
      \citenamefont {Garcia}, \citenamefont {Gasca}, \citenamefont {Genois},
      \citenamefont {Giang}, \citenamefont {Gidney}, \citenamefont {Gilboa},
      \citenamefont {Gosula}, \citenamefont {Dau}, \citenamefont {Graumann},
      \citenamefont {Greene}, \citenamefont {Gross}, \citenamefont {Habegger},
      \citenamefont {Hall}, \citenamefont {Hamilton}, \citenamefont {Hansen},
      \citenamefont {Harrigan}, \citenamefont {Harrington}, \citenamefont {Heras},
      \citenamefont {Heslin}, \citenamefont {Heu}, \citenamefont {Higgott},
      \citenamefont {Hill}, \citenamefont {Hilton}, \citenamefont {Holland},
      \citenamefont {Hong}, \citenamefont {Huang}, \citenamefont {Huff},
      \citenamefont {Huggins}, \citenamefont {Ioffe}, \citenamefont {Isakov},
      \citenamefont {Iveland}, \citenamefont {Jeffrey}, \citenamefont {Jiang},
      \citenamefont {Jones}, \citenamefont {Jordan}, \citenamefont {Joshi},
      \citenamefont {Juhas}, \citenamefont {Kafri}, \citenamefont {Kang},
      \citenamefont {Karamlou}, \citenamefont {Kechedzhi}, \citenamefont {Kelly},
      \citenamefont {Khaire}, \citenamefont {Khattar}, \citenamefont {Khezri},
      \citenamefont {Kim}, \citenamefont {Klimov}, \citenamefont {Klots},
      \citenamefont {Kobrin}, \citenamefont {Kohli}, \citenamefont {Korotkov},
      \citenamefont {Kostritsa}, \citenamefont {Kothari}, \citenamefont
      {Kozlovskii}, \citenamefont {Kreikebaum}, \citenamefont {Kurilovich},
      \citenamefont {Lacroix}, \citenamefont {Landhuis}, \citenamefont {Lange-Dei},
      \citenamefont {Langley}, \citenamefont {Laptev}, \citenamefont {Lau},
      \citenamefont {Le~Guevel}, \citenamefont {Ledford}, \citenamefont {Lee},
      \citenamefont {Lee}, \citenamefont {Lensky}, \citenamefont {Leon},
      \citenamefont {Lester}, \citenamefont {Li}, \citenamefont {Li}, \citenamefont
      {Lill}, \citenamefont {Liu}, \citenamefont {Livingston}, \citenamefont
      {Locharla}, \citenamefont {Lucero}, \citenamefont {Lundahl}, \citenamefont
      {Lunt}, \citenamefont {Madhuk}, \citenamefont {Malone}, \citenamefont
      {Maloney}, \citenamefont {Mandr{\`a}}, \citenamefont {Manyika}, \citenamefont
      {Martin}, \citenamefont {Martin}, \citenamefont {Martin}, \citenamefont
      {Maxfield}, \citenamefont {McClean}, \citenamefont {McEwen}, \citenamefont
      {Meeks}, \citenamefont {Megrant}, \citenamefont {Mi}, \citenamefont {Miao},
      \citenamefont {Mieszala}, \citenamefont {Molavi}, \citenamefont {Molina},
      \citenamefont {Montazeri}, \citenamefont {Morvan}, \citenamefont {Movassagh},
      \citenamefont {Mruczkiewicz}, \citenamefont {Naaman}, \citenamefont {Neeley},
      \citenamefont {Neill}, \citenamefont {Nersisyan}, \citenamefont {Neven},
      \citenamefont {Newman}, \citenamefont {Ng}, \citenamefont {Nguyen},
      \citenamefont {Nguyen}, \citenamefont {Ni}, \citenamefont {Niu},
      \citenamefont {O'Brien}, \citenamefont {Oliver}, \citenamefont {Opremcak},
      \citenamefont {Ottosson}, \citenamefont {Petukhov}, \citenamefont {Pizzuto},
      \citenamefont {Platt}, \citenamefont {Potter}, \citenamefont {Pritchard},
      \citenamefont {Pryadko}, \citenamefont {Quintana}, \citenamefont
      {Ramachandran}, \citenamefont {Reagor}, \citenamefont {Redding},
      \citenamefont {Rhodes}, \citenamefont {Roberts}, \citenamefont {Rosenberg},
      \citenamefont {Rosenfeld}, \citenamefont {Roushan}, \citenamefont {Rubin},
      \citenamefont {Saei}, \citenamefont {Sank}, \citenamefont {Sankaragomathi},
      \citenamefont {Satzinger}, \citenamefont {Schurkus}, \citenamefont
      {Schuster}, \citenamefont {Senior}, \citenamefont {Shearn}, \citenamefont
      {Shorter}, \citenamefont {Shutty}, \citenamefont {Shvarts}, \citenamefont
      {Singh}, \citenamefont {Sivak}, \citenamefont {Skruzny}, \citenamefont
      {Small}, \citenamefont {Smelyanskiy}, \citenamefont {Smith}, \citenamefont
      {Somma}, \citenamefont {Springer}, \citenamefont {Sterling}, \citenamefont
      {Strain}, \citenamefont {Suchard}, \citenamefont {Szasz}, \citenamefont
      {Sztein}, \citenamefont {Thor}, \citenamefont {Torres}, \citenamefont
      {Torunbalci}, \citenamefont {Vaishnav}, \citenamefont {Vargas}, \citenamefont
      {Vdovichev}, \citenamefont {Vidal}, \citenamefont {Villalonga}, \citenamefont
      {Heidweiller}, \citenamefont {Waltman}, \citenamefont {Wang}, \citenamefont
      {Ware}, \citenamefont {Weber}, \citenamefont {Weidel}, \citenamefont {White},
      \citenamefont {Wong}, \citenamefont {Woo}, \citenamefont {Xing},
      \citenamefont {Yao}, \citenamefont {Yeh}, \citenamefont {Ying}, \citenamefont
      {Yoo}, \citenamefont {Yosri}, \citenamefont {Young}, \citenamefont {Zalcman},
      \citenamefont {Zhang}, \citenamefont {Zhu}, \citenamefont {Zobrist},
      \citenamefont {AI},\ and\ \citenamefont {{Collaborators}}}]{Acharya2024}%
      \BibitemOpen
      \bibfield  {author} {\bibinfo {author} {\bibfnamefont {R.}~\bibnamefont
      {Acharya}}, \bibinfo {author} {\bibfnamefont {D.~A.}\ \bibnamefont {Abanin}},
      \bibinfo {author} {\bibfnamefont {L.}~\bibnamefont {Aghababaie-Beni}},
      \bibinfo {author} {\bibfnamefont {I.}~\bibnamefont {Aleiner}}, \bibinfo
      {author} {\bibfnamefont {T.~I.}\ \bibnamefont {Andersen}}, \bibinfo {author}
      {\bibfnamefont {M.}~\bibnamefont {Ansmann}}, \bibinfo {author} {\bibfnamefont
      {F.}~\bibnamefont {Arute}}, \bibinfo {author} {\bibfnamefont
      {K.}~\bibnamefont {Arya}}, \bibinfo {author} {\bibfnamefont {A.}~\bibnamefont
      {Asfaw}}, \bibinfo {author} {\bibfnamefont {N.}~\bibnamefont
      {Astrakhantsev}}, \bibinfo {author} {\bibfnamefont {J.}~\bibnamefont
      {Atalaya}}, \bibinfo {author} {\bibfnamefont {R.}~\bibnamefont {Babbush}},
      \bibinfo {author} {\bibfnamefont {D.}~\bibnamefont {Bacon}}, \bibinfo
      {author} {\bibfnamefont {B.}~\bibnamefont {Ballard}}, \bibinfo {author}
      {\bibfnamefont {J.~C.}\ \bibnamefont {Bardin}}, \bibinfo {author}
      {\bibfnamefont {J.}~\bibnamefont {Bausch}}, \bibinfo {author} {\bibfnamefont
      {A.}~\bibnamefont {Bengtsson}}, \bibinfo {author} {\bibfnamefont
      {A.}~\bibnamefont {Bilmes}}, \bibinfo {author} {\bibfnamefont
      {S.}~\bibnamefont {Blackwell}}, \bibinfo {author} {\bibfnamefont
      {S.}~\bibnamefont {Boixo}}, \bibinfo {author} {\bibfnamefont
      {G.}~\bibnamefont {Bortoli}}, \bibinfo {author} {\bibfnamefont
      {A.}~\bibnamefont {Bourassa}}, \bibinfo {author} {\bibfnamefont
      {J.}~\bibnamefont {Bovaird}}, \bibinfo {author} {\bibfnamefont
      {L.}~\bibnamefont {Brill}}, \bibinfo {author} {\bibfnamefont
      {M.}~\bibnamefont {Broughton}}, \bibinfo {author} {\bibfnamefont {D.~A.}\
      \bibnamefont {Browne}}, \bibinfo {author} {\bibfnamefont {B.}~\bibnamefont
      {Buchea}}, \bibinfo {author} {\bibfnamefont {B.~B.}\ \bibnamefont {Buckley}},
      \bibinfo {author} {\bibfnamefont {D.~A.}\ \bibnamefont {Buell}}, \bibinfo
      {author} {\bibfnamefont {T.}~\bibnamefont {Burger}}, \bibinfo {author}
      {\bibfnamefont {B.}~\bibnamefont {Burkett}}, \bibinfo {author} {\bibfnamefont
      {N.}~\bibnamefont {Bushnell}}, \bibinfo {author} {\bibfnamefont
      {A.}~\bibnamefont {Cabrera}}, \bibinfo {author} {\bibfnamefont
      {J.}~\bibnamefont {Campero}}, \bibinfo {author} {\bibfnamefont {H.-S.}\
      \bibnamefont {Chang}}, \bibinfo {author} {\bibfnamefont {Y.}~\bibnamefont
      {Chen}}, \bibinfo {author} {\bibfnamefont {Z.}~\bibnamefont {Chen}}, \bibinfo
      {author} {\bibfnamefont {B.}~\bibnamefont {Chiaro}}, \bibinfo {author}
      {\bibfnamefont {D.}~\bibnamefont {Chik}}, \bibinfo {author} {\bibfnamefont
      {C.}~\bibnamefont {Chou}}, \bibinfo {author} {\bibfnamefont {J.}~\bibnamefont
      {Claes}}, \bibinfo {author} {\bibfnamefont {A.~Y.}\ \bibnamefont {Cleland}},
      \bibinfo {author} {\bibfnamefont {J.}~\bibnamefont {Cogan}}, \bibinfo
      {author} {\bibfnamefont {R.}~\bibnamefont {Collins}}, \bibinfo {author}
      {\bibfnamefont {P.}~\bibnamefont {Conner}}, \bibinfo {author} {\bibfnamefont
      {W.}~\bibnamefont {Courtney}}, \bibinfo {author} {\bibfnamefont {A.~L.}\
      \bibnamefont {Crook}}, \bibinfo {author} {\bibfnamefont {B.}~\bibnamefont
      {Curtin}}, \bibinfo {author} {\bibfnamefont {S.}~\bibnamefont {Das}},
      \bibinfo {author} {\bibfnamefont {A.}~\bibnamefont {Davies}}, \bibinfo
      {author} {\bibfnamefont {L.}~\bibnamefont {De~Lorenzo}}, \bibinfo {author}
      {\bibfnamefont {D.~M.}\ \bibnamefont {Debroy}}, \bibinfo {author}
      {\bibfnamefont {S.}~\bibnamefont {Demura}}, \bibinfo {author} {\bibfnamefont
      {M.}~\bibnamefont {Devoret}}, \bibinfo {author} {\bibfnamefont
      {A.}~\bibnamefont {Di~Paolo}}, \bibinfo {author} {\bibfnamefont
      {P.}~\bibnamefont {Donohoe}}, \bibinfo {author} {\bibfnamefont
      {I.}~\bibnamefont {Drozdov}}, \bibinfo {author} {\bibfnamefont
      {A.}~\bibnamefont {Dunsworth}}, \bibinfo {author} {\bibfnamefont
      {C.}~\bibnamefont {Earle}}, \bibinfo {author} {\bibfnamefont
      {T.}~\bibnamefont {Edlich}}, \bibinfo {author} {\bibfnamefont
      {A.}~\bibnamefont {Eickbusch}}, \bibinfo {author} {\bibfnamefont {A.~M.}\
      \bibnamefont {Elbag}}, \bibinfo {author} {\bibfnamefont {M.}~\bibnamefont
      {Elzouka}}, \bibinfo {author} {\bibfnamefont {C.}~\bibnamefont {Erickson}},
      \bibinfo {author} {\bibfnamefont {L.}~\bibnamefont {Faoro}}, \bibinfo
      {author} {\bibfnamefont {E.}~\bibnamefont {Farhi}}, \bibinfo {author}
      {\bibfnamefont {V.~S.}\ \bibnamefont {Ferreira}}, \bibinfo {author}
      {\bibfnamefont {L.~F.}\ \bibnamefont {Burgos}}, \bibinfo {author}
      {\bibfnamefont {E.}~\bibnamefont {Forati}}, \bibinfo {author} {\bibfnamefont
      {A.~G.}\ \bibnamefont {Fowler}}, \bibinfo {author} {\bibfnamefont
      {B.}~\bibnamefont {Foxen}}, \bibinfo {author} {\bibfnamefont
      {S.}~\bibnamefont {Ganjam}}, \bibinfo {author} {\bibfnamefont
      {G.}~\bibnamefont {Garcia}}, \bibinfo {author} {\bibfnamefont
      {R.}~\bibnamefont {Gasca}}, \bibinfo {author} {\bibfnamefont
      {{\'E}.}~\bibnamefont {Genois}}, \bibinfo {author} {\bibfnamefont
      {W.}~\bibnamefont {Giang}}, \bibinfo {author} {\bibfnamefont
      {C.}~\bibnamefont {Gidney}}, \bibinfo {author} {\bibfnamefont
      {D.}~\bibnamefont {Gilboa}}, \bibinfo {author} {\bibfnamefont
      {R.}~\bibnamefont {Gosula}}, \bibinfo {author} {\bibfnamefont {A.~G.}\
      \bibnamefont {Dau}}, \bibinfo {author} {\bibfnamefont {D.}~\bibnamefont
      {Graumann}}, \bibinfo {author} {\bibfnamefont {A.}~\bibnamefont {Greene}},
      \bibinfo {author} {\bibfnamefont {J.~A.}\ \bibnamefont {Gross}}, \bibinfo
      {author} {\bibfnamefont {S.}~\bibnamefont {Habegger}}, \bibinfo {author}
      {\bibfnamefont {J.}~\bibnamefont {Hall}}, \bibinfo {author} {\bibfnamefont
      {M.~C.}\ \bibnamefont {Hamilton}}, \bibinfo {author} {\bibfnamefont
      {M.}~\bibnamefont {Hansen}}, \bibinfo {author} {\bibfnamefont {M.~P.}\
      \bibnamefont {Harrigan}}, \bibinfo {author} {\bibfnamefont {S.~D.}\
      \bibnamefont {Harrington}}, \bibinfo {author} {\bibfnamefont {F.~J.~H.}\
      \bibnamefont {Heras}}, \bibinfo {author} {\bibfnamefont {S.}~\bibnamefont
      {Heslin}}, \bibinfo {author} {\bibfnamefont {P.}~\bibnamefont {Heu}},
      \bibinfo {author} {\bibfnamefont {O.}~\bibnamefont {Higgott}}, \bibinfo
      {author} {\bibfnamefont {G.}~\bibnamefont {Hill}}, \bibinfo {author}
      {\bibfnamefont {J.}~\bibnamefont {Hilton}}, \bibinfo {author} {\bibfnamefont
      {G.}~\bibnamefont {Holland}}, \bibinfo {author} {\bibfnamefont
      {S.}~\bibnamefont {Hong}}, \bibinfo {author} {\bibfnamefont {H.-Y.}\
      \bibnamefont {Huang}}, \bibinfo {author} {\bibfnamefont {A.}~\bibnamefont
      {Huff}}, \bibinfo {author} {\bibfnamefont {W.~J.}\ \bibnamefont {Huggins}},
      \bibinfo {author} {\bibfnamefont {L.~B.}\ \bibnamefont {Ioffe}}, \bibinfo
      {author} {\bibfnamefont {S.~V.}\ \bibnamefont {Isakov}}, \bibinfo {author}
      {\bibfnamefont {J.}~\bibnamefont {Iveland}}, \bibinfo {author} {\bibfnamefont
      {E.}~\bibnamefont {Jeffrey}}, \bibinfo {author} {\bibfnamefont
      {Z.}~\bibnamefont {Jiang}}, \bibinfo {author} {\bibfnamefont
      {C.}~\bibnamefont {Jones}}, \bibinfo {author} {\bibfnamefont
      {S.}~\bibnamefont {Jordan}}, \bibinfo {author} {\bibfnamefont
      {C.}~\bibnamefont {Joshi}}, \bibinfo {author} {\bibfnamefont
      {P.}~\bibnamefont {Juhas}}, \bibinfo {author} {\bibfnamefont
      {D.}~\bibnamefont {Kafri}}, \bibinfo {author} {\bibfnamefont
      {H.}~\bibnamefont {Kang}}, \bibinfo {author} {\bibfnamefont {A.~H.}\
      \bibnamefont {Karamlou}}, \bibinfo {author} {\bibfnamefont {K.}~\bibnamefont
      {Kechedzhi}}, \bibinfo {author} {\bibfnamefont {J.}~\bibnamefont {Kelly}},
      \bibinfo {author} {\bibfnamefont {T.}~\bibnamefont {Khaire}}, \bibinfo
      {author} {\bibfnamefont {T.}~\bibnamefont {Khattar}}, \bibinfo {author}
      {\bibfnamefont {M.}~\bibnamefont {Khezri}}, \bibinfo {author} {\bibfnamefont
      {S.}~\bibnamefont {Kim}}, \bibinfo {author} {\bibfnamefont {P.~V.}\
      \bibnamefont {Klimov}}, \bibinfo {author} {\bibfnamefont {A.~R.}\
      \bibnamefont {Klots}}, \bibinfo {author} {\bibfnamefont {B.}~\bibnamefont
      {Kobrin}}, \bibinfo {author} {\bibfnamefont {P.}~\bibnamefont {Kohli}},
      \bibinfo {author} {\bibfnamefont {A.~N.}\ \bibnamefont {Korotkov}}, \bibinfo
      {author} {\bibfnamefont {F.}~\bibnamefont {Kostritsa}}, \bibinfo {author}
      {\bibfnamefont {R.}~\bibnamefont {Kothari}}, \bibinfo {author} {\bibfnamefont
      {B.}~\bibnamefont {Kozlovskii}}, \bibinfo {author} {\bibfnamefont {J.~M.}\
      \bibnamefont {Kreikebaum}}, \bibinfo {author} {\bibfnamefont {V.~D.}\
      \bibnamefont {Kurilovich}}, \bibinfo {author} {\bibfnamefont
      {N.}~\bibnamefont {Lacroix}}, \bibinfo {author} {\bibfnamefont
      {D.}~\bibnamefont {Landhuis}}, \bibinfo {author} {\bibfnamefont
      {T.}~\bibnamefont {Lange-Dei}}, \bibinfo {author} {\bibfnamefont {B.~W.}\
      \bibnamefont {Langley}}, \bibinfo {author} {\bibfnamefont {P.}~\bibnamefont
      {Laptev}}, \bibinfo {author} {\bibfnamefont {K.-M.}\ \bibnamefont {Lau}},
      \bibinfo {author} {\bibfnamefont {L.}~\bibnamefont {Le~Guevel}}, \bibinfo
      {author} {\bibfnamefont {J.}~\bibnamefont {Ledford}}, \bibinfo {author}
      {\bibfnamefont {J.}~\bibnamefont {Lee}}, \bibinfo {author} {\bibfnamefont
      {K.}~\bibnamefont {Lee}}, \bibinfo {author} {\bibfnamefont {Y.~D.}\
      \bibnamefont {Lensky}}, \bibinfo {author} {\bibfnamefont {S.}~\bibnamefont
      {Leon}}, \bibinfo {author} {\bibfnamefont {B.~J.}\ \bibnamefont {Lester}},
      \bibinfo {author} {\bibfnamefont {W.~Y.}\ \bibnamefont {Li}}, \bibinfo
      {author} {\bibfnamefont {Y.}~\bibnamefont {Li}}, \bibinfo {author}
      {\bibfnamefont {A.~T.}\ \bibnamefont {Lill}}, \bibinfo {author}
      {\bibfnamefont {W.}~\bibnamefont {Liu}}, \bibinfo {author} {\bibfnamefont
      {W.~P.}\ \bibnamefont {Livingston}}, \bibinfo {author} {\bibfnamefont
      {A.}~\bibnamefont {Locharla}}, \bibinfo {author} {\bibfnamefont
      {E.}~\bibnamefont {Lucero}}, \bibinfo {author} {\bibfnamefont
      {D.}~\bibnamefont {Lundahl}}, \bibinfo {author} {\bibfnamefont
      {A.}~\bibnamefont {Lunt}}, \bibinfo {author} {\bibfnamefont {S.}~\bibnamefont
      {Madhuk}}, \bibinfo {author} {\bibfnamefont {F.~D.}\ \bibnamefont {Malone}},
      \bibinfo {author} {\bibfnamefont {A.}~\bibnamefont {Maloney}}, \bibinfo
      {author} {\bibfnamefont {S.}~\bibnamefont {Mandr{\`a}}}, \bibinfo {author}
      {\bibfnamefont {J.}~\bibnamefont {Manyika}}, \bibinfo {author} {\bibfnamefont
      {L.~S.}\ \bibnamefont {Martin}}, \bibinfo {author} {\bibfnamefont
      {O.}~\bibnamefont {Martin}}, \bibinfo {author} {\bibfnamefont
      {S.}~\bibnamefont {Martin}}, \bibinfo {author} {\bibfnamefont
      {C.}~\bibnamefont {Maxfield}}, \bibinfo {author} {\bibfnamefont {J.~R.}\
      \bibnamefont {McClean}}, \bibinfo {author} {\bibfnamefont {M.}~\bibnamefont
      {McEwen}}, \bibinfo {author} {\bibfnamefont {S.}~\bibnamefont {Meeks}},
      \bibinfo {author} {\bibfnamefont {A.}~\bibnamefont {Megrant}}, \bibinfo
      {author} {\bibfnamefont {X.}~\bibnamefont {Mi}}, \bibinfo {author}
      {\bibfnamefont {K.~C.}\ \bibnamefont {Miao}}, \bibinfo {author}
      {\bibfnamefont {A.}~\bibnamefont {Mieszala}}, \bibinfo {author}
      {\bibfnamefont {R.}~\bibnamefont {Molavi}}, \bibinfo {author} {\bibfnamefont
      {S.}~\bibnamefont {Molina}}, \bibinfo {author} {\bibfnamefont
      {S.}~\bibnamefont {Montazeri}}, \bibinfo {author} {\bibfnamefont
      {A.}~\bibnamefont {Morvan}}, \bibinfo {author} {\bibfnamefont
      {R.}~\bibnamefont {Movassagh}}, \bibinfo {author} {\bibfnamefont
      {W.}~\bibnamefont {Mruczkiewicz}}, \bibinfo {author} {\bibfnamefont
      {O.}~\bibnamefont {Naaman}}, \bibinfo {author} {\bibfnamefont
      {M.}~\bibnamefont {Neeley}}, \bibinfo {author} {\bibfnamefont
      {C.}~\bibnamefont {Neill}}, \bibinfo {author} {\bibfnamefont
      {A.}~\bibnamefont {Nersisyan}}, \bibinfo {author} {\bibfnamefont
      {H.}~\bibnamefont {Neven}}, \bibinfo {author} {\bibfnamefont
      {M.}~\bibnamefont {Newman}}, \bibinfo {author} {\bibfnamefont {J.~H.}\
      \bibnamefont {Ng}}, \bibinfo {author} {\bibfnamefont {A.}~\bibnamefont
      {Nguyen}}, \bibinfo {author} {\bibfnamefont {M.}~\bibnamefont {Nguyen}},
      \bibinfo {author} {\bibfnamefont {C.-H.}\ \bibnamefont {Ni}}, \bibinfo
      {author} {\bibfnamefont {M.~Y.}\ \bibnamefont {Niu}}, \bibinfo {author}
      {\bibfnamefont {T.~E.}\ \bibnamefont {O'Brien}}, \bibinfo {author}
      {\bibfnamefont {W.~D.}\ \bibnamefont {Oliver}}, \bibinfo {author}
      {\bibfnamefont {A.}~\bibnamefont {Opremcak}}, \bibinfo {author}
      {\bibfnamefont {K.}~\bibnamefont {Ottosson}}, \bibinfo {author}
      {\bibfnamefont {A.}~\bibnamefont {Petukhov}}, \bibinfo {author}
      {\bibfnamefont {A.}~\bibnamefont {Pizzuto}}, \bibinfo {author} {\bibfnamefont
      {J.}~\bibnamefont {Platt}}, \bibinfo {author} {\bibfnamefont
      {R.}~\bibnamefont {Potter}}, \bibinfo {author} {\bibfnamefont
      {O.}~\bibnamefont {Pritchard}}, \bibinfo {author} {\bibfnamefont {L.~P.}\
      \bibnamefont {Pryadko}}, \bibinfo {author} {\bibfnamefont {C.}~\bibnamefont
      {Quintana}}, \bibinfo {author} {\bibfnamefont {G.}~\bibnamefont
      {Ramachandran}}, \bibinfo {author} {\bibfnamefont {M.~J.}\ \bibnamefont
      {Reagor}}, \bibinfo {author} {\bibfnamefont {J.}~\bibnamefont {Redding}},
      \bibinfo {author} {\bibfnamefont {D.~M.}\ \bibnamefont {Rhodes}}, \bibinfo
      {author} {\bibfnamefont {G.}~\bibnamefont {Roberts}}, \bibinfo {author}
      {\bibfnamefont {E.}~\bibnamefont {Rosenberg}}, \bibinfo {author}
      {\bibfnamefont {E.}~\bibnamefont {Rosenfeld}}, \bibinfo {author}
      {\bibfnamefont {P.}~\bibnamefont {Roushan}}, \bibinfo {author} {\bibfnamefont
      {N.~C.}\ \bibnamefont {Rubin}}, \bibinfo {author} {\bibfnamefont
      {N.}~\bibnamefont {Saei}}, \bibinfo {author} {\bibfnamefont {D.}~\bibnamefont
      {Sank}}, \bibinfo {author} {\bibfnamefont {K.}~\bibnamefont
      {Sankaragomathi}}, \bibinfo {author} {\bibfnamefont {K.~J.}\ \bibnamefont
      {Satzinger}}, \bibinfo {author} {\bibfnamefont {H.~F.}\ \bibnamefont
      {Schurkus}}, \bibinfo {author} {\bibfnamefont {C.}~\bibnamefont {Schuster}},
      \bibinfo {author} {\bibfnamefont {A.~W.}\ \bibnamefont {Senior}}, \bibinfo
      {author} {\bibfnamefont {M.~J.}\ \bibnamefont {Shearn}}, \bibinfo {author}
      {\bibfnamefont {A.}~\bibnamefont {Shorter}}, \bibinfo {author} {\bibfnamefont
      {N.}~\bibnamefont {Shutty}}, \bibinfo {author} {\bibfnamefont
      {V.}~\bibnamefont {Shvarts}}, \bibinfo {author} {\bibfnamefont
      {S.}~\bibnamefont {Singh}}, \bibinfo {author} {\bibfnamefont
      {V.}~\bibnamefont {Sivak}}, \bibinfo {author} {\bibfnamefont
      {J.}~\bibnamefont {Skruzny}}, \bibinfo {author} {\bibfnamefont
      {S.}~\bibnamefont {Small}}, \bibinfo {author} {\bibfnamefont
      {V.}~\bibnamefont {Smelyanskiy}}, \bibinfo {author} {\bibfnamefont {W.~C.}\
      \bibnamefont {Smith}}, \bibinfo {author} {\bibfnamefont {R.~D.}\ \bibnamefont
      {Somma}}, \bibinfo {author} {\bibfnamefont {S.}~\bibnamefont {Springer}},
      \bibinfo {author} {\bibfnamefont {G.}~\bibnamefont {Sterling}}, \bibinfo
      {author} {\bibfnamefont {D.}~\bibnamefont {Strain}}, \bibinfo {author}
      {\bibfnamefont {J.}~\bibnamefont {Suchard}}, \bibinfo {author} {\bibfnamefont
      {A.}~\bibnamefont {Szasz}}, \bibinfo {author} {\bibfnamefont
      {A.}~\bibnamefont {Sztein}}, \bibinfo {author} {\bibfnamefont
      {D.}~\bibnamefont {Thor}}, \bibinfo {author} {\bibfnamefont {A.}~\bibnamefont
      {Torres}}, \bibinfo {author} {\bibfnamefont {M.~M.}\ \bibnamefont
      {Torunbalci}}, \bibinfo {author} {\bibfnamefont {A.}~\bibnamefont
      {Vaishnav}}, \bibinfo {author} {\bibfnamefont {J.}~\bibnamefont {Vargas}},
      \bibinfo {author} {\bibfnamefont {S.}~\bibnamefont {Vdovichev}}, \bibinfo
      {author} {\bibfnamefont {G.}~\bibnamefont {Vidal}}, \bibinfo {author}
      {\bibfnamefont {B.}~\bibnamefont {Villalonga}}, \bibinfo {author}
      {\bibfnamefont {C.~V.}\ \bibnamefont {Heidweiller}}, \bibinfo {author}
      {\bibfnamefont {S.}~\bibnamefont {Waltman}}, \bibinfo {author} {\bibfnamefont
      {S.~X.}\ \bibnamefont {Wang}}, \bibinfo {author} {\bibfnamefont
      {B.}~\bibnamefont {Ware}}, \bibinfo {author} {\bibfnamefont {K.}~\bibnamefont
      {Weber}}, \bibinfo {author} {\bibfnamefont {T.}~\bibnamefont {Weidel}},
      \bibinfo {author} {\bibfnamefont {T.}~\bibnamefont {White}}, \bibinfo
      {author} {\bibfnamefont {K.}~\bibnamefont {Wong}}, \bibinfo {author}
      {\bibfnamefont {B.~W.~K.}\ \bibnamefont {Woo}}, \bibinfo {author}
      {\bibfnamefont {C.}~\bibnamefont {Xing}}, \bibinfo {author} {\bibfnamefont
      {Z.~J.}\ \bibnamefont {Yao}}, \bibinfo {author} {\bibfnamefont
      {P.}~\bibnamefont {Yeh}}, \bibinfo {author} {\bibfnamefont {B.}~\bibnamefont
      {Ying}}, \bibinfo {author} {\bibfnamefont {J.}~\bibnamefont {Yoo}}, \bibinfo
      {author} {\bibfnamefont {N.}~\bibnamefont {Yosri}}, \bibinfo {author}
      {\bibfnamefont {G.}~\bibnamefont {Young}}, \bibinfo {author} {\bibfnamefont
      {A.}~\bibnamefont {Zalcman}}, \bibinfo {author} {\bibfnamefont
      {Y.}~\bibnamefont {Zhang}}, \bibinfo {author} {\bibfnamefont
      {N.}~\bibnamefont {Zhu}}, \bibinfo {author} {\bibfnamefont {N.}~\bibnamefont
      {Zobrist}}, \bibinfo {author} {\bibfnamefont {G.~Q.}\ \bibnamefont {AI}},\
      and\ \bibinfo {author} {\bibnamefont {{Collaborators}}},\ }\bibfield
      {journal} {\bibinfo  {journal} {Nature}\ }\href
      {https://doi.org/10.1038/s41586-024-08449-y} {10.1038/s41586-024-08449-y}
      (\bibinfo {year} {2024})\BibitemShut {NoStop}%
    \bibitem [{\citenamefont {Paetznick}\ \emph {et~al.}(2024)\citenamefont
      {Paetznick}, \citenamefont {da~Silva}, \citenamefont {Ryan-Anderson},
      \citenamefont {Bello-Rivas}, \citenamefont {III}, \citenamefont
      {Chernoguzov}, \citenamefont {Dreiling}, \citenamefont {Foltz}, \citenamefont
      {Frachon}, \citenamefont {Gaebler}, \citenamefont {Gatterman}, \citenamefont
      {Grans-Samuelsson}, \citenamefont {Gresh}, \citenamefont {Hayes},
      \citenamefont {Hewitt}, \citenamefont {Holliman}, \citenamefont {Horst},
      \citenamefont {Johansen}, \citenamefont {Lucchetti}, \citenamefont
      {Matsuoka}, \citenamefont {Mills}, \citenamefont {Moses}, \citenamefont
      {Neyenhuis}, \citenamefont {Paz}, \citenamefont {Pino}, \citenamefont
      {Siegfried}, \citenamefont {Sundaram}, \citenamefont {Tom}, \citenamefont
      {Wernli}, \citenamefont {Zanner}, \citenamefont {Stutz},\ and\ \citenamefont
      {Svore}}]{paetznick2024}%
      \BibitemOpen
      \bibfield  {author} {\bibinfo {author} {\bibfnamefont {A.}~\bibnamefont
      {Paetznick}}, \bibinfo {author} {\bibfnamefont {M.~P.}\ \bibnamefont
      {da~Silva}}, \bibinfo {author} {\bibfnamefont {C.}~\bibnamefont
      {Ryan-Anderson}}, \bibinfo {author} {\bibfnamefont {J.~M.}\ \bibnamefont
      {Bello-Rivas}}, \bibinfo {author} {\bibfnamefont {J.~P.~C.}\ \bibnamefont
      {III}}, \bibinfo {author} {\bibfnamefont {A.}~\bibnamefont {Chernoguzov}},
      \bibinfo {author} {\bibfnamefont {J.~M.}\ \bibnamefont {Dreiling}}, \bibinfo
      {author} {\bibfnamefont {C.}~\bibnamefont {Foltz}}, \bibinfo {author}
      {\bibfnamefont {F.}~\bibnamefont {Frachon}}, \bibinfo {author} {\bibfnamefont
      {J.~P.}\ \bibnamefont {Gaebler}}, \bibinfo {author} {\bibfnamefont {T.~M.}\
      \bibnamefont {Gatterman}}, \bibinfo {author} {\bibfnamefont {L.}~\bibnamefont
      {Grans-Samuelsson}}, \bibinfo {author} {\bibfnamefont {D.}~\bibnamefont
      {Gresh}}, \bibinfo {author} {\bibfnamefont {D.}~\bibnamefont {Hayes}},
      \bibinfo {author} {\bibfnamefont {N.}~\bibnamefont {Hewitt}}, \bibinfo
      {author} {\bibfnamefont {C.}~\bibnamefont {Holliman}}, \bibinfo {author}
      {\bibfnamefont {C.~V.}\ \bibnamefont {Horst}}, \bibinfo {author}
      {\bibfnamefont {J.}~\bibnamefont {Johansen}}, \bibinfo {author}
      {\bibfnamefont {D.}~\bibnamefont {Lucchetti}}, \bibinfo {author}
      {\bibfnamefont {Y.}~\bibnamefont {Matsuoka}}, \bibinfo {author}
      {\bibfnamefont {M.}~\bibnamefont {Mills}}, \bibinfo {author} {\bibfnamefont
      {S.~A.}\ \bibnamefont {Moses}}, \bibinfo {author} {\bibfnamefont
      {B.}~\bibnamefont {Neyenhuis}}, \bibinfo {author} {\bibfnamefont
      {A.}~\bibnamefont {Paz}}, \bibinfo {author} {\bibfnamefont {J.}~\bibnamefont
      {Pino}}, \bibinfo {author} {\bibfnamefont {P.}~\bibnamefont {Siegfried}},
      \bibinfo {author} {\bibfnamefont {A.}~\bibnamefont {Sundaram}}, \bibinfo
      {author} {\bibfnamefont {D.}~\bibnamefont {Tom}}, \bibinfo {author}
      {\bibfnamefont {S.~J.}\ \bibnamefont {Wernli}}, \bibinfo {author}
      {\bibfnamefont {M.}~\bibnamefont {Zanner}}, \bibinfo {author} {\bibfnamefont
      {R.~P.}\ \bibnamefont {Stutz}},\ and\ \bibinfo {author} {\bibfnamefont
      {K.~M.}\ \bibnamefont {Svore}},\ }\href {https://arxiv.org/abs/2404.02280}
      {\bibinfo {title} {Demonstration of logical qubits and repeated error
      correction with better-than-physical error rates}} (\bibinfo {year} {2024}),\
      \Eprint {https://arxiv.org/abs/2404.02280} {arXiv:2404.02280}
      \BibitemShut {NoStop}%
    \bibitem [{\citenamefont {Putterman}\ \emph {et~al.}(2024)\citenamefont
      {Putterman}, \citenamefont {Noh}, \citenamefont {Hann}, \citenamefont
      {MacCabe}, \citenamefont {Aghaeimeibodi}, \citenamefont {Patel},
      \citenamefont {Lee}, \citenamefont {Jones}, \citenamefont {Moradinejad},
      \citenamefont {Rodriguez}, \citenamefont {Mahuli}, \citenamefont {Rose},
      \citenamefont {Owens}, \citenamefont {Levine}, \citenamefont {Rosenfeld},
      \citenamefont {Reinhold}, \citenamefont {Moncelsi}, \citenamefont {Alcid},
      \citenamefont {Alidoust}, \citenamefont {Arrangoiz-Arriola}, \citenamefont
      {Barnett}, \citenamefont {Bienias}, \citenamefont {Carson}, \citenamefont
      {Chen}, \citenamefont {Chen}, \citenamefont {Chinkezian}, \citenamefont
      {Chisholm}, \citenamefont {Chou}, \citenamefont {Clerk}, \citenamefont
      {Clifford}, \citenamefont {Cosmic}, \citenamefont {Curiel}, \citenamefont
      {Davis}, \citenamefont {DeLorenzo}, \citenamefont {D'Ewart}, \citenamefont
      {Diky}, \citenamefont {D'Souza}, \citenamefont {Dumitrescu}, \citenamefont
      {Eisenmann}, \citenamefont {Elkhouly}, \citenamefont {Evenbly}, \citenamefont
      {Fang}, \citenamefont {Fang}, \citenamefont {Fling}, \citenamefont {Fon},
      \citenamefont {Garcia}, \citenamefont {Gorshkov}, \citenamefont {Grant},
      \citenamefont {Gray}, \citenamefont {Grimberg}, \citenamefont {Grimsmo},
      \citenamefont {Haim}, \citenamefont {Hand}, \citenamefont {He}, \citenamefont
      {Hernandez}, \citenamefont {Hover}, \citenamefont {Hung}, \citenamefont
      {Hunt}, \citenamefont {Iverson}, \citenamefont {Jarrige}, \citenamefont
      {Jaskula}, \citenamefont {Jiang}, \citenamefont {Kalaee}, \citenamefont
      {Karabalin}, \citenamefont {Karalekas}, \citenamefont {Keller}, \citenamefont
      {Khalajhedayati}, \citenamefont {Kubica}, \citenamefont {Lee}, \citenamefont
      {Leroux}, \citenamefont {Lieu}, \citenamefont {Ly}, \citenamefont {Madrigal},
      \citenamefont {Marcaud}, \citenamefont {McCabe}, \citenamefont {Miles},
      \citenamefont {Milsted}, \citenamefont {Minguzzi}, \citenamefont {Mishra},
      \citenamefont {Mukherjee}, \citenamefont {Naghiloo}, \citenamefont
      {Oblepias}, \citenamefont {Ortuno}, \citenamefont {Pagdilao}, \citenamefont
      {Pancotti}, \citenamefont {Panduro}, \citenamefont {Paquette}, \citenamefont
      {Park}, \citenamefont {Peairs}, \citenamefont {Perello}, \citenamefont
      {Peterson}, \citenamefont {Ponte}, \citenamefont {Preskill}, \citenamefont
      {Qiao}, \citenamefont {Refael}, \citenamefont {Resnick}, \citenamefont
      {Retzker}, \citenamefont {Reyna}, \citenamefont {Runyan}, \citenamefont
      {Ryan}, \citenamefont {Sahmoud}, \citenamefont {Sanchez}, \citenamefont
      {Sanil}, \citenamefont {Sankar}, \citenamefont {Sato}, \citenamefont
      {Scaffidi}, \citenamefont {Siavoshi}, \citenamefont {Sivarajah},
      \citenamefont {Skogland}, \citenamefont {Su}, \citenamefont {Swenson},
      \citenamefont {Teo}, \citenamefont {Tomada}, \citenamefont {Torlai},
      \citenamefont {Wollack}, \citenamefont {Ye}, \citenamefont {Zerrudo},
      \citenamefont {Zhang}, \citenamefont {Brandão}, \citenamefont {Matheny},\
      and\ \citenamefont {Painter}}]{putterman2024}%
      \BibitemOpen
      \bibfield  {author} {\bibinfo {author} {\bibfnamefont {H.}~\bibnamefont
      {Putterman}}, \bibinfo {author} {\bibfnamefont {K.}~\bibnamefont {Noh}},
      \bibinfo {author} {\bibfnamefont {C.~T.}\ \bibnamefont {Hann}}, \bibinfo
      {author} {\bibfnamefont {G.~S.}\ \bibnamefont {MacCabe}}, \bibinfo {author}
      {\bibfnamefont {S.}~\bibnamefont {Aghaeimeibodi}}, \bibinfo {author}
      {\bibfnamefont {R.~N.}\ \bibnamefont {Patel}}, \bibinfo {author}
      {\bibfnamefont {M.}~\bibnamefont {Lee}}, \bibinfo {author} {\bibfnamefont
      {W.~M.}\ \bibnamefont {Jones}}, \bibinfo {author} {\bibfnamefont
      {H.}~\bibnamefont {Moradinejad}}, \bibinfo {author} {\bibfnamefont
      {R.}~\bibnamefont {Rodriguez}}, \bibinfo {author} {\bibfnamefont
      {N.}~\bibnamefont {Mahuli}}, \bibinfo {author} {\bibfnamefont
      {J.}~\bibnamefont {Rose}}, \bibinfo {author} {\bibfnamefont {J.~C.}\
      \bibnamefont {Owens}}, \bibinfo {author} {\bibfnamefont {H.}~\bibnamefont
      {Levine}}, \bibinfo {author} {\bibfnamefont {E.}~\bibnamefont {Rosenfeld}},
      \bibinfo {author} {\bibfnamefont {P.}~\bibnamefont {Reinhold}}, \bibinfo
      {author} {\bibfnamefont {L.}~\bibnamefont {Moncelsi}}, \bibinfo {author}
      {\bibfnamefont {J.~A.}\ \bibnamefont {Alcid}}, \bibinfo {author}
      {\bibfnamefont {N.}~\bibnamefont {Alidoust}}, \bibinfo {author}
      {\bibfnamefont {P.}~\bibnamefont {Arrangoiz-Arriola}}, \bibinfo {author}
      {\bibfnamefont {J.}~\bibnamefont {Barnett}}, \bibinfo {author} {\bibfnamefont
      {P.}~\bibnamefont {Bienias}}, \bibinfo {author} {\bibfnamefont {H.~A.}\
      \bibnamefont {Carson}}, \bibinfo {author} {\bibfnamefont {C.}~\bibnamefont
      {Chen}}, \bibinfo {author} {\bibfnamefont {L.}~\bibnamefont {Chen}}, \bibinfo
      {author} {\bibfnamefont {H.}~\bibnamefont {Chinkezian}}, \bibinfo {author}
      {\bibfnamefont {E.~M.}\ \bibnamefont {Chisholm}}, \bibinfo {author}
      {\bibfnamefont {M.-H.}\ \bibnamefont {Chou}}, \bibinfo {author}
      {\bibfnamefont {A.}~\bibnamefont {Clerk}}, \bibinfo {author} {\bibfnamefont
      {A.}~\bibnamefont {Clifford}}, \bibinfo {author} {\bibfnamefont
      {R.}~\bibnamefont {Cosmic}}, \bibinfo {author} {\bibfnamefont {A.~V.}\
      \bibnamefont {Curiel}}, \bibinfo {author} {\bibfnamefont {E.}~\bibnamefont
      {Davis}}, \bibinfo {author} {\bibfnamefont {L.}~\bibnamefont {DeLorenzo}},
      \bibinfo {author} {\bibfnamefont {J.~M.}\ \bibnamefont {D'Ewart}}, \bibinfo
      {author} {\bibfnamefont {A.}~\bibnamefont {Diky}}, \bibinfo {author}
      {\bibfnamefont {N.}~\bibnamefont {D'Souza}}, \bibinfo {author} {\bibfnamefont
      {P.~T.}\ \bibnamefont {Dumitrescu}}, \bibinfo {author} {\bibfnamefont
      {S.}~\bibnamefont {Eisenmann}}, \bibinfo {author} {\bibfnamefont
      {E.}~\bibnamefont {Elkhouly}}, \bibinfo {author} {\bibfnamefont
      {G.}~\bibnamefont {Evenbly}}, \bibinfo {author} {\bibfnamefont {M.~T.}\
      \bibnamefont {Fang}}, \bibinfo {author} {\bibfnamefont {Y.}~\bibnamefont
      {Fang}}, \bibinfo {author} {\bibfnamefont {M.~J.}\ \bibnamefont {Fling}},
      \bibinfo {author} {\bibfnamefont {W.}~\bibnamefont {Fon}}, \bibinfo {author}
      {\bibfnamefont {G.}~\bibnamefont {Garcia}}, \bibinfo {author} {\bibfnamefont
      {A.~V.}\ \bibnamefont {Gorshkov}}, \bibinfo {author} {\bibfnamefont {J.~A.}\
      \bibnamefont {Grant}}, \bibinfo {author} {\bibfnamefont {M.~J.}\ \bibnamefont
      {Gray}}, \bibinfo {author} {\bibfnamefont {S.}~\bibnamefont {Grimberg}},
      \bibinfo {author} {\bibfnamefont {A.~L.}\ \bibnamefont {Grimsmo}}, \bibinfo
      {author} {\bibfnamefont {A.}~\bibnamefont {Haim}}, \bibinfo {author}
      {\bibfnamefont {J.}~\bibnamefont {Hand}}, \bibinfo {author} {\bibfnamefont
      {Y.}~\bibnamefont {He}}, \bibinfo {author} {\bibfnamefont {M.}~\bibnamefont
      {Hernandez}}, \bibinfo {author} {\bibfnamefont {D.}~\bibnamefont {Hover}},
      \bibinfo {author} {\bibfnamefont {J.~S.~C.}\ \bibnamefont {Hung}}, \bibinfo
      {author} {\bibfnamefont {M.}~\bibnamefont {Hunt}}, \bibinfo {author}
      {\bibfnamefont {J.}~\bibnamefont {Iverson}}, \bibinfo {author} {\bibfnamefont
      {I.}~\bibnamefont {Jarrige}}, \bibinfo {author} {\bibfnamefont {J.-C.}\
      \bibnamefont {Jaskula}}, \bibinfo {author} {\bibfnamefont {L.}~\bibnamefont
      {Jiang}}, \bibinfo {author} {\bibfnamefont {M.}~\bibnamefont {Kalaee}},
      \bibinfo {author} {\bibfnamefont {R.}~\bibnamefont {Karabalin}}, \bibinfo
      {author} {\bibfnamefont {P.~J.}\ \bibnamefont {Karalekas}}, \bibinfo {author}
      {\bibfnamefont {A.~J.}\ \bibnamefont {Keller}}, \bibinfo {author}
      {\bibfnamefont {A.}~\bibnamefont {Khalajhedayati}}, \bibinfo {author}
      {\bibfnamefont {A.}~\bibnamefont {Kubica}}, \bibinfo {author} {\bibfnamefont
      {H.}~\bibnamefont {Lee}}, \bibinfo {author} {\bibfnamefont {C.}~\bibnamefont
      {Leroux}}, \bibinfo {author} {\bibfnamefont {S.}~\bibnamefont {Lieu}},
      \bibinfo {author} {\bibfnamefont {V.}~\bibnamefont {Ly}}, \bibinfo {author}
      {\bibfnamefont {K.~V.}\ \bibnamefont {Madrigal}}, \bibinfo {author}
      {\bibfnamefont {G.}~\bibnamefont {Marcaud}}, \bibinfo {author} {\bibfnamefont
      {G.}~\bibnamefont {McCabe}}, \bibinfo {author} {\bibfnamefont
      {C.}~\bibnamefont {Miles}}, \bibinfo {author} {\bibfnamefont
      {A.}~\bibnamefont {Milsted}}, \bibinfo {author} {\bibfnamefont
      {J.}~\bibnamefont {Minguzzi}}, \bibinfo {author} {\bibfnamefont
      {A.}~\bibnamefont {Mishra}}, \bibinfo {author} {\bibfnamefont
      {B.}~\bibnamefont {Mukherjee}}, \bibinfo {author} {\bibfnamefont
      {M.}~\bibnamefont {Naghiloo}}, \bibinfo {author} {\bibfnamefont
      {E.}~\bibnamefont {Oblepias}}, \bibinfo {author} {\bibfnamefont
      {G.}~\bibnamefont {Ortuno}}, \bibinfo {author} {\bibfnamefont
      {J.}~\bibnamefont {Pagdilao}}, \bibinfo {author} {\bibfnamefont
      {N.}~\bibnamefont {Pancotti}}, \bibinfo {author} {\bibfnamefont
      {A.}~\bibnamefont {Panduro}}, \bibinfo {author} {\bibfnamefont
      {J.}~\bibnamefont {Paquette}}, \bibinfo {author} {\bibfnamefont
      {M.}~\bibnamefont {Park}}, \bibinfo {author} {\bibfnamefont {G.~A.}\
      \bibnamefont {Peairs}}, \bibinfo {author} {\bibfnamefont {D.}~\bibnamefont
      {Perello}}, \bibinfo {author} {\bibfnamefont {E.~C.}\ \bibnamefont
      {Peterson}}, \bibinfo {author} {\bibfnamefont {S.}~\bibnamefont {Ponte}},
      \bibinfo {author} {\bibfnamefont {J.}~\bibnamefont {Preskill}}, \bibinfo
      {author} {\bibfnamefont {J.}~\bibnamefont {Qiao}}, \bibinfo {author}
      {\bibfnamefont {G.}~\bibnamefont {Refael}}, \bibinfo {author} {\bibfnamefont
      {R.}~\bibnamefont {Resnick}}, \bibinfo {author} {\bibfnamefont
      {A.}~\bibnamefont {Retzker}}, \bibinfo {author} {\bibfnamefont {O.~A.}\
      \bibnamefont {Reyna}}, \bibinfo {author} {\bibfnamefont {M.}~\bibnamefont
      {Runyan}}, \bibinfo {author} {\bibfnamefont {C.~A.}\ \bibnamefont {Ryan}},
      \bibinfo {author} {\bibfnamefont {A.}~\bibnamefont {Sahmoud}}, \bibinfo
      {author} {\bibfnamefont {E.}~\bibnamefont {Sanchez}}, \bibinfo {author}
      {\bibfnamefont {R.}~\bibnamefont {Sanil}}, \bibinfo {author} {\bibfnamefont
      {K.}~\bibnamefont {Sankar}}, \bibinfo {author} {\bibfnamefont
      {Y.}~\bibnamefont {Sato}}, \bibinfo {author} {\bibfnamefont {T.}~\bibnamefont
      {Scaffidi}}, \bibinfo {author} {\bibfnamefont {S.}~\bibnamefont {Siavoshi}},
      \bibinfo {author} {\bibfnamefont {P.}~\bibnamefont {Sivarajah}}, \bibinfo
      {author} {\bibfnamefont {T.}~\bibnamefont {Skogland}}, \bibinfo {author}
      {\bibfnamefont {C.-J.}\ \bibnamefont {Su}}, \bibinfo {author} {\bibfnamefont
      {L.~J.}\ \bibnamefont {Swenson}}, \bibinfo {author} {\bibfnamefont {S.~M.}\
      \bibnamefont {Teo}}, \bibinfo {author} {\bibfnamefont {A.}~\bibnamefont
      {Tomada}}, \bibinfo {author} {\bibfnamefont {G.}~\bibnamefont {Torlai}},
      \bibinfo {author} {\bibfnamefont {E.~A.}\ \bibnamefont {Wollack}}, \bibinfo
      {author} {\bibfnamefont {Y.}~\bibnamefont {Ye}}, \bibinfo {author}
      {\bibfnamefont {J.~A.}\ \bibnamefont {Zerrudo}}, \bibinfo {author}
      {\bibfnamefont {K.}~\bibnamefont {Zhang}}, \bibinfo {author} {\bibfnamefont
      {F.~G. S.~L.}\ \bibnamefont {Brandão}}, \bibinfo {author} {\bibfnamefont
      {M.~H.}\ \bibnamefont {Matheny}},\ and\ \bibinfo {author} {\bibfnamefont
      {O.}~\bibnamefont {Painter}},\ }\href {https://arxiv.org/abs/2409.13025}
      {\bibinfo {title} {Hardware-efficient quantum error correction using
      concatenated bosonic qubits}} (\bibinfo {year} {2024}),\ \Eprint
      {https://arxiv.org/abs/2409.13025} {arXiv:2409.13025} \BibitemShut
      {NoStop}%
    \bibitem [{\citenamefont {Feynman}(1960)}]{Feynman_1960}%
      \BibitemOpen
      \bibfield  {author} {\bibinfo {author} {\bibfnamefont {R.~P.}\ \bibnamefont
      {Feynman}},\ }\href@noop {} {\bibfield  {journal} {\bibinfo  {journal}
      {Engineering and Science}\ }\textbf {\bibinfo {volume} {23}},\ \bibinfo
      {pages} {22–36} (\bibinfo {year} {1960})}\BibitemShut {NoStop}%
    \bibitem [{\citenamefont {Fowler}\ \emph {et~al.}(2012)\citenamefont {Fowler},
      \citenamefont {Mariantoni}, \citenamefont {Martinis},\ and\ \citenamefont
      {Cleland}}]{Fowler_2012}%
      \BibitemOpen
      \bibfield  {author} {\bibinfo {author} {\bibfnamefont {A.~G.}\ \bibnamefont
      {Fowler}}, \bibinfo {author} {\bibfnamefont {M.}~\bibnamefont {Mariantoni}},
      \bibinfo {author} {\bibfnamefont {J.~M.}\ \bibnamefont {Martinis}},\ and\
      \bibinfo {author} {\bibfnamefont {A.~N.}\ \bibnamefont {Cleland}},\ }\href
      {https://doi.org/10.1103/PhysRevA.86.032324} {\bibfield  {journal} {\bibinfo
      {journal} {Physical Review A}\ }\textbf {\bibinfo {volume} {86}},\ \bibinfo
      {pages} {032324} (\bibinfo {year} {2012})}\BibitemShut {NoStop}%
    \bibitem [{\citenamefont {Campbell}\ \emph {et~al.}(2017)\citenamefont
      {Campbell}, \citenamefont {Terhal},\ and\ \citenamefont
      {Vuillot}}]{Campbell_Terhal_Vuillot_2017}%
      \BibitemOpen
      \bibfield  {author} {\bibinfo {author} {\bibfnamefont {E.~T.}\ \bibnamefont
      {Campbell}}, \bibinfo {author} {\bibfnamefont {B.~M.}\ \bibnamefont
      {Terhal}},\ and\ \bibinfo {author} {\bibfnamefont {C.}~\bibnamefont
      {Vuillot}},\ }\href {https://doi.org/10.1038/nature23460} {\bibfield
      {journal} {\bibinfo  {journal} {Nature}\ }\textbf {\bibinfo {volume} {549}},\
      \bibinfo {pages} {172–179} (\bibinfo {year} {2017})}\BibitemShut {NoStop}%
    \bibitem [{\citenamefont {Chan}\ \emph {et~al.}(2023)\citenamefont {Chan},
      \citenamefont {Meister}, \citenamefont {Goh},\ and\ \citenamefont
      {Koczor}}]{chan2023algorithmic}%
      \BibitemOpen
      \bibfield  {author} {\bibinfo {author} {\bibfnamefont {H.~H.~S.}\
      \bibnamefont {Chan}}, \bibinfo {author} {\bibfnamefont {R.}~\bibnamefont
      {Meister}}, \bibinfo {author} {\bibfnamefont {M.~L.}\ \bibnamefont {Goh}},\
      and\ \bibinfo {author} {\bibfnamefont {B.}~\bibnamefont {Koczor}},\
      }\href@noop {} {\bibinfo {title} {Algorithmic shadow spectroscopy}} (\bibinfo
      {year} {2023}),\ \Eprint {https://arxiv.org/abs/2212.11036} {arXiv:2212.11036} \BibitemShut {NoStop}%
    \bibitem [{\citenamefont {Huang}\ \emph {et~al.}(2020)\citenamefont {Huang},
      \citenamefont {Kueng},\ and\ \citenamefont
      {Preskill}}]{Huang_Kueng_Preskill_2020}%
      \BibitemOpen
      \bibfield  {author} {\bibinfo {author} {\bibfnamefont {H.-Y.}\ \bibnamefont
      {Huang}}, \bibinfo {author} {\bibfnamefont {R.}~\bibnamefont {Kueng}},\ and\
      \bibinfo {author} {\bibfnamefont {J.}~\bibnamefont {Preskill}},\ }\href
      {https://doi.org/10.1038/s41567-020-0932-7} {\bibfield  {journal} {\bibinfo
      {journal} {Nature Physics}\ }\textbf {\bibinfo {volume} {16}},\ \bibinfo
      {pages} {1050–1057} (\bibinfo {year} {2020})}\BibitemShut {NoStop}%
    \bibitem [{\citenamefont {Reilly}(2015)}]{Reilly2015}%
      \BibitemOpen
      \bibfield  {author} {\bibinfo {author} {\bibfnamefont {D.~J.}\ \bibnamefont
      {Reilly}},\ }\href {https://doi.org/10.1038/npjqi.2015.11} {\bibfield
      {journal} {\bibinfo  {journal} {npj Quantum Information}\ }\textbf {\bibinfo
      {volume} {1}},\ \bibinfo {pages} {15011} (\bibinfo {year}
      {2015})}\BibitemShut {NoStop}%
    \bibitem [{\citenamefont {Reilly}(2019)}]{Reilly2019}%
      \BibitemOpen
      \bibfield  {author} {\bibinfo {author} {\bibfnamefont {D.~J.}\ \bibnamefont
      {Reilly}},\ }in\ \href {https://doi.org/10.1109/IEDM19573.2019.8993497}
      {\emph {\bibinfo {booktitle} {2019 IEEE International Electron Devices
      Meeting (IEDM)}}}\ (\bibinfo {year} {2019})\ pp.\ \bibinfo {pages}
      {31.7.1--31.7.6}\BibitemShut {NoStop}%
    \bibitem [{\citenamefont {Gonzalez-Zalba}\ \emph {et~al.}(2021)\citenamefont
      {Gonzalez-Zalba}, \citenamefont {De~Franceschi}, \citenamefont {Charbon},
      \citenamefont {Meunier}, \citenamefont {Vinet},\ and\ \citenamefont
      {Dzurak}}]{Gonzalez-Zalba_2021}%
      \BibitemOpen
      \bibfield  {author} {\bibinfo {author} {\bibfnamefont {M.~F.}\ \bibnamefont
      {Gonzalez-Zalba}}, \bibinfo {author} {\bibfnamefont {S.}~\bibnamefont
      {De~Franceschi}}, \bibinfo {author} {\bibfnamefont {E.}~\bibnamefont
      {Charbon}}, \bibinfo {author} {\bibfnamefont {T.}~\bibnamefont {Meunier}},
      \bibinfo {author} {\bibfnamefont {M.}~\bibnamefont {Vinet}},\ and\ \bibinfo
      {author} {\bibfnamefont {A.~S.}\ \bibnamefont {Dzurak}},\ }\href
      {https://doi.org/10.1038/s41928-021-00681-y} {\bibfield  {journal} {\bibinfo
      {journal} {Nature Electronics}\ }\textbf {\bibinfo {volume} {4}},\ \bibinfo
      {pages} {872–884} (\bibinfo {year} {2021})}\BibitemShut {NoStop}%
    \bibitem [{\citenamefont {Pauka}\ \emph {et~al.}(2021)\citenamefont {Pauka},
      \citenamefont {Das}, \citenamefont {Kalra}, \citenamefont {Moini},
      \citenamefont {Yang}, \citenamefont {Trainer}, \citenamefont {Bousquet},
      \citenamefont {Cantaloube}, \citenamefont {Dick}, \citenamefont {Gardner},
      \citenamefont {Manfra},\ and\ \citenamefont {Reilly}}]{Pauka_2021}%
      \BibitemOpen
      \bibfield  {author} {\bibinfo {author} {\bibfnamefont {S.~J.}\ \bibnamefont
      {Pauka}}, \bibinfo {author} {\bibfnamefont {K.}~\bibnamefont {Das}}, \bibinfo
      {author} {\bibfnamefont {R.}~\bibnamefont {Kalra}}, \bibinfo {author}
      {\bibfnamefont {A.}~\bibnamefont {Moini}}, \bibinfo {author} {\bibfnamefont
      {Y.}~\bibnamefont {Yang}}, \bibinfo {author} {\bibfnamefont {M.}~\bibnamefont
      {Trainer}}, \bibinfo {author} {\bibfnamefont {A.}~\bibnamefont {Bousquet}},
      \bibinfo {author} {\bibfnamefont {C.}~\bibnamefont {Cantaloube}}, \bibinfo
      {author} {\bibfnamefont {N.}~\bibnamefont {Dick}}, \bibinfo {author}
      {\bibfnamefont {G.~C.}\ \bibnamefont {Gardner}}, \bibinfo {author}
      {\bibfnamefont {M.~J.}\ \bibnamefont {Manfra}},\ and\ \bibinfo {author}
      {\bibfnamefont {D.~J.}\ \bibnamefont {Reilly}},\ }\href
      {https://doi.org/10.1038/s41928-020-00528-y} {\bibfield  {journal} {\bibinfo
      {journal} {Nature Electronics}\ }\textbf {\bibinfo {volume} {4}},\ \bibinfo
      {pages} {64–70} (\bibinfo {year} {2021})}\BibitemShut {NoStop}%
    \bibitem [{\citenamefont {Hornibrook}\ \emph {et~al.}(2015)\citenamefont
      {Hornibrook}, \citenamefont {Colless}, \citenamefont {Conway~Lamb},
      \citenamefont {Pauka}, \citenamefont {Lu}, \citenamefont {Gossard},
      \citenamefont {Watson}, \citenamefont {Gardner}, \citenamefont {Fallahi},
      \citenamefont {Manfra},\ and\ \citenamefont {Reilly}}]{Hornibrook_2015}%
      \BibitemOpen
      \bibfield  {author} {\bibinfo {author} {\bibfnamefont {J.}~\bibnamefont
      {Hornibrook}}, \bibinfo {author} {\bibfnamefont {J.}~\bibnamefont {Colless}},
      \bibinfo {author} {\bibfnamefont {I.}~\bibnamefont {Conway~Lamb}}, \bibinfo
      {author} {\bibfnamefont {S.}~\bibnamefont {Pauka}}, \bibinfo {author}
      {\bibfnamefont {H.}~\bibnamefont {Lu}}, \bibinfo {author} {\bibfnamefont
      {A.}~\bibnamefont {Gossard}}, \bibinfo {author} {\bibfnamefont
      {J.}~\bibnamefont {Watson}}, \bibinfo {author} {\bibfnamefont
      {G.}~\bibnamefont {Gardner}}, \bibinfo {author} {\bibfnamefont
      {S.}~\bibnamefont {Fallahi}}, \bibinfo {author} {\bibfnamefont
      {M.}~\bibnamefont {Manfra}},\ and\ \bibinfo {author} {\bibfnamefont
      {D.}~\bibnamefont {Reilly}},\ }\href
      {https://doi.org/10.1103/PhysRevApplied.3.024010} {\bibfield  {journal}
      {\bibinfo  {journal} {Physical Review Applied}\ }\textbf {\bibinfo {volume}
      {3}},\ \bibinfo {pages} {024010} (\bibinfo {year} {2015})}\BibitemShut
      {NoStop}%
    \bibitem [{\citenamefont {Charbon}\ \emph {et~al.}(2016)\citenamefont
      {Charbon}, \citenamefont {Sebastiano}, \citenamefont {Vladimirescu},
      \citenamefont {Homulle}, \citenamefont {Visser}, \citenamefont {Song},\ and\
      \citenamefont {Incandela}}]{Charbon_2016}%
      \BibitemOpen
      \bibfield  {author} {\bibinfo {author} {\bibfnamefont {E.}~\bibnamefont
      {Charbon}}, \bibinfo {author} {\bibfnamefont {F.}~\bibnamefont {Sebastiano}},
      \bibinfo {author} {\bibfnamefont {A.}~\bibnamefont {Vladimirescu}}, \bibinfo
      {author} {\bibfnamefont {H.}~\bibnamefont {Homulle}}, \bibinfo {author}
      {\bibfnamefont {S.}~\bibnamefont {Visser}}, \bibinfo {author} {\bibfnamefont
      {L.}~\bibnamefont {Song}},\ and\ \bibinfo {author} {\bibfnamefont {R.~M.}\
      \bibnamefont {Incandela}},\ }in\ \href
      {https://doi.org/10.1109/IEDM.2016.7838410} {\emph {\bibinfo {booktitle}
      {2016 IEEE International Electron Devices Meeting (IEDM)}}}\ (\bibinfo {year}
      {2016})\ pp.\ \bibinfo {pages} {13.5.1--13.5.4}\BibitemShut {NoStop}%
    \bibitem [{\citenamefont {Sebastiano}\ \emph {et~al.}(2017)\citenamefont
      {Sebastiano}, \citenamefont {Homulle}, \citenamefont {Patra}, \citenamefont
      {Incandela}, \citenamefont {van Dijk}, \citenamefont {Song}, \citenamefont
      {Babaie}, \citenamefont {Vladimirescu},\ and\ \citenamefont
      {Charbon}}]{Sebastiano_2017}%
      \BibitemOpen
      \bibfield  {author} {\bibinfo {author} {\bibfnamefont {F.}~\bibnamefont
      {Sebastiano}}, \bibinfo {author} {\bibfnamefont {H.}~\bibnamefont {Homulle}},
      \bibinfo {author} {\bibfnamefont {B.}~\bibnamefont {Patra}}, \bibinfo
      {author} {\bibfnamefont {R.}~\bibnamefont {Incandela}}, \bibinfo {author}
      {\bibfnamefont {J.}~\bibnamefont {van Dijk}}, \bibinfo {author}
      {\bibfnamefont {L.}~\bibnamefont {Song}}, \bibinfo {author} {\bibfnamefont
      {M.}~\bibnamefont {Babaie}}, \bibinfo {author} {\bibfnamefont
      {A.}~\bibnamefont {Vladimirescu}},\ and\ \bibinfo {author} {\bibfnamefont
      {E.}~\bibnamefont {Charbon}},\ }in\ \href
      {https://doi.org/10.1145/3061639.3072948} {\emph {\bibinfo {booktitle}
      {Proceedings of the 54th Annual Design Automation Conference 2017}}},\
      \bibinfo {series and number} {DAC '17}\ (\bibinfo  {publisher} {Association
      for Computing Machinery},\ \bibinfo {address} {New York, NY, USA},\ \bibinfo
      {year} {2017})\ p.\ \bibinfo {pages} {1–6}\BibitemShut {NoStop}%
    \bibitem [{\citenamefont {Park}\ \emph {et~al.}(2021)\citenamefont {Park},
      \citenamefont {Subramanian}, \citenamefont {Lampert}, \citenamefont
      {Mladenov}, \citenamefont {Klotchkov}, \citenamefont {Kurian}, \citenamefont
      {Juarez-Hernandez}, \citenamefont {Esparza}, \citenamefont {Kale},
      \citenamefont {K.~T.}, \citenamefont {Premaratne}, \citenamefont {Watson},
      \citenamefont {Suzuki}, \citenamefont {Rahman}, \citenamefont {Timbadiya},
      \citenamefont {Soni},\ and\ \citenamefont
      {Pellerano}}]{Park_Subramanian_2021}%
      \BibitemOpen
      \bibfield  {author} {\bibinfo {author} {\bibfnamefont {J.}~\bibnamefont
      {Park}}, \bibinfo {author} {\bibfnamefont {S.}~\bibnamefont {Subramanian}},
      \bibinfo {author} {\bibfnamefont {L.}~\bibnamefont {Lampert}}, \bibinfo
      {author} {\bibfnamefont {T.}~\bibnamefont {Mladenov}}, \bibinfo {author}
      {\bibfnamefont {I.}~\bibnamefont {Klotchkov}}, \bibinfo {author}
      {\bibfnamefont {D.~J.}\ \bibnamefont {Kurian}}, \bibinfo {author}
      {\bibfnamefont {E.}~\bibnamefont {Juarez-Hernandez}}, \bibinfo {author}
      {\bibfnamefont {B.~P.}\ \bibnamefont {Esparza}}, \bibinfo {author}
      {\bibfnamefont {S.~R.}\ \bibnamefont {Kale}}, \bibinfo {author}
      {\bibfnamefont {A.~B.}\ \bibnamefont {K.~T.}}, \bibinfo {author}
      {\bibfnamefont {S.~P.}\ \bibnamefont {Premaratne}}, \bibinfo {author}
      {\bibfnamefont {T.~F.}\ \bibnamefont {Watson}}, \bibinfo {author}
      {\bibfnamefont {S.}~\bibnamefont {Suzuki}}, \bibinfo {author} {\bibfnamefont
      {M.}~\bibnamefont {Rahman}}, \bibinfo {author} {\bibfnamefont {J.~B.}\
      \bibnamefont {Timbadiya}}, \bibinfo {author} {\bibfnamefont {S.}~\bibnamefont
      {Soni}},\ and\ \bibinfo {author} {\bibfnamefont {S.}~\bibnamefont
      {Pellerano}},\ }\href {https://doi.org/10.1109/JSSC.2021.3115988} {\bibfield
      {journal} {\bibinfo  {journal} {IEEE Journal of Solid-State Circuits}\
      }\textbf {\bibinfo {volume} {56}},\ \bibinfo {pages} {3289–3306} (\bibinfo
      {year} {2021})}\BibitemShut {NoStop}%
    \bibitem [{\citenamefont {van Dijk}\ \emph
      {et~al.}(2019{\natexlab{a}})\citenamefont {van Dijk}, \citenamefont
      {Kawakami}, \citenamefont {Schouten}, \citenamefont {Veldhorst},
      \citenamefont {Vandersypen}, \citenamefont {Babaie}, \citenamefont
      {Charbon},\ and\ \citenamefont {Sebastiano}}]{vanDijk_2019}%
      \BibitemOpen
      \bibfield  {author} {\bibinfo {author} {\bibfnamefont {J.}~\bibnamefont {van
      Dijk}}, \bibinfo {author} {\bibfnamefont {E.}~\bibnamefont {Kawakami}},
      \bibinfo {author} {\bibfnamefont {R.}~\bibnamefont {Schouten}}, \bibinfo
      {author} {\bibfnamefont {M.}~\bibnamefont {Veldhorst}}, \bibinfo {author}
      {\bibfnamefont {L.}~\bibnamefont {Vandersypen}}, \bibinfo {author}
      {\bibfnamefont {M.}~\bibnamefont {Babaie}}, \bibinfo {author} {\bibfnamefont
      {E.}~\bibnamefont {Charbon}},\ and\ \bibinfo {author} {\bibfnamefont
      {F.}~\bibnamefont {Sebastiano}},\ }\href
      {https://doi.org/10.1103/PhysRevApplied.12.044054} {\bibfield  {journal}
      {\bibinfo  {journal} {Physical Review Applied}\ }\textbf {\bibinfo {volume}
      {12}},\ \bibinfo {pages} {044054} (\bibinfo {year}
      {2019}{\natexlab{a}})}\BibitemShut {NoStop}%
    \bibitem [{\citenamefont {Eggli}\ \emph {et~al.}(2024)\citenamefont {Eggli},
      \citenamefont {Patlatiuk}, \citenamefont {Kelly}, \citenamefont {Orekhov},
      \citenamefont {Salis}, \citenamefont {Warburton}, \citenamefont {Zumbühl},\
      and\ \citenamefont {Kuhlmann}}]{Eggli_2024}%
      \BibitemOpen
      \bibfield  {author} {\bibinfo {author} {\bibfnamefont {R.~S.}\ \bibnamefont
      {Eggli}}, \bibinfo {author} {\bibfnamefont {T.}~\bibnamefont {Patlatiuk}},
      \bibinfo {author} {\bibfnamefont {E.~G.}\ \bibnamefont {Kelly}}, \bibinfo
      {author} {\bibfnamefont {A.}~\bibnamefont {Orekhov}}, \bibinfo {author}
      {\bibfnamefont {G.}~\bibnamefont {Salis}}, \bibinfo {author} {\bibfnamefont
      {R.~J.}\ \bibnamefont {Warburton}}, \bibinfo {author} {\bibfnamefont {D.~M.}\
      \bibnamefont {Zumbühl}},\ and\ \bibinfo {author} {\bibfnamefont {A.~V.}\
      \bibnamefont {Kuhlmann}},\ }\href {https://doi.org/10.48550/arXiv.2407.21484}
      {\bibinfo {title} {All-electrical operation of a spin qubit coupled to a
      high-q resonator}} (\bibinfo {year} {2024}),\ \bibinfo {note}
      {arXiv:2407.21484}\BibitemShut {NoStop}%
    \bibitem [{\citenamefont {Sliwa}\ \emph {et~al.}(2015)\citenamefont {Sliwa},
      \citenamefont {Hatridge}, \citenamefont {Narla}, \citenamefont {Shankar},
      \citenamefont {Frunzio}, \citenamefont {Schoelkopf},\ and\ \citenamefont
      {Devoret}}]{Sliwa2015}%
      \BibitemOpen
      \bibfield  {author} {\bibinfo {author} {\bibfnamefont {K.~M.}\ \bibnamefont
      {Sliwa}}, \bibinfo {author} {\bibfnamefont {M.}~\bibnamefont {Hatridge}},
      \bibinfo {author} {\bibfnamefont {A.}~\bibnamefont {Narla}}, \bibinfo
      {author} {\bibfnamefont {S.}~\bibnamefont {Shankar}}, \bibinfo {author}
      {\bibfnamefont {L.}~\bibnamefont {Frunzio}}, \bibinfo {author} {\bibfnamefont
      {R.~J.}\ \bibnamefont {Schoelkopf}},\ and\ \bibinfo {author} {\bibfnamefont
      {M.~H.}\ \bibnamefont {Devoret}},\ }\href
      {https://doi.org/10.1103/PhysRevX.5.041020} {\bibfield  {journal} {\bibinfo
      {journal} {Phys. Rev. X}\ }\textbf {\bibinfo {volume} {5}},\ \bibinfo {pages}
      {041020} (\bibinfo {year} {2015})}\BibitemShut {NoStop}%
    \bibitem [{\citenamefont {Ahmed}\ \emph {et~al.}(2018)\citenamefont {Ahmed},
      \citenamefont {Chatterjee}, \citenamefont {Barraud}, \citenamefont {Morton},
      \citenamefont {Haigh},\ and\ \citenamefont {Gonzalez-Zalba}}]{Ahmed_2018}%
      \BibitemOpen
      \bibfield  {author} {\bibinfo {author} {\bibfnamefont {I.}~\bibnamefont
      {Ahmed}}, \bibinfo {author} {\bibfnamefont {A.}~\bibnamefont {Chatterjee}},
      \bibinfo {author} {\bibfnamefont {S.}~\bibnamefont {Barraud}}, \bibinfo
      {author} {\bibfnamefont {J.~J.~L.}\ \bibnamefont {Morton}}, \bibinfo {author}
      {\bibfnamefont {J.~A.}\ \bibnamefont {Haigh}},\ and\ \bibinfo {author}
      {\bibfnamefont {M.~F.}\ \bibnamefont {Gonzalez-Zalba}},\ }\href
      {https://doi.org/10.1038/s42005-018-0066-8} {\bibfield  {journal} {\bibinfo
      {journal} {Communications Physics}\ }\textbf {\bibinfo {volume} {1}},\
      \bibinfo {pages} {1–7} (\bibinfo {year} {2018})}\BibitemShut {NoStop}%
    \bibitem [{\citenamefont {Aumentado}(2020)}]{Aumentado2020}%
      \BibitemOpen
      \bibfield  {author} {\bibinfo {author} {\bibfnamefont {J.}~\bibnamefont
      {Aumentado}},\ }\href {https://doi.org/10.1109/MMM.2020.2993476} {\bibfield
      {journal} {\bibinfo  {journal} {IEEE Microwave Magazine}\ }\textbf {\bibinfo
      {volume} {21}},\ \bibinfo {pages} {45} (\bibinfo {year} {2020})}\BibitemShut
      {NoStop}%
    \bibitem [{\citenamefont {Cochrane}\ \emph {et~al.}(2022)\citenamefont
      {Cochrane}, \citenamefont {Lundberg}, \citenamefont {Ibberson}, \citenamefont
      {Ibberson}, \citenamefont {Hutin}, \citenamefont {Bertrand}, \citenamefont
      {Stelmashenko}, \citenamefont {Robinson}, \citenamefont {Vinet},
      \citenamefont {Seshia},\ and\ \citenamefont
      {Gonzalez-Zalba}}]{Cochrane_2022}%
      \BibitemOpen
      \bibfield  {author} {\bibinfo {author} {\bibfnamefont {L.}~\bibnamefont
      {Cochrane}}, \bibinfo {author} {\bibfnamefont {T.}~\bibnamefont {Lundberg}},
      \bibinfo {author} {\bibfnamefont {D.~J.}\ \bibnamefont {Ibberson}}, \bibinfo
      {author} {\bibfnamefont {L.~A.}\ \bibnamefont {Ibberson}}, \bibinfo {author}
      {\bibfnamefont {L.}~\bibnamefont {Hutin}}, \bibinfo {author} {\bibfnamefont
      {B.}~\bibnamefont {Bertrand}}, \bibinfo {author} {\bibfnamefont
      {N.}~\bibnamefont {Stelmashenko}}, \bibinfo {author} {\bibfnamefont {J.~W.}\
      \bibnamefont {Robinson}}, \bibinfo {author} {\bibfnamefont {M.}~\bibnamefont
      {Vinet}}, \bibinfo {author} {\bibfnamefont {A.~A.}\ \bibnamefont {Seshia}},\
      and\ \bibinfo {author} {\bibfnamefont {M.~F.}\ \bibnamefont
      {Gonzalez-Zalba}},\ }\href {https://doi.org/10.1103/PhysRevLett.128.197701}
      {\bibfield  {journal} {\bibinfo  {journal} {Physical Review Letters}\
      }\textbf {\bibinfo {volume} {128}},\ \bibinfo {pages} {197701} (\bibinfo
      {year} {2022})}\BibitemShut {NoStop}%
    \bibitem [{\citenamefont {Oakes}\ \emph
      {et~al.}(2023{\natexlab{a}})\citenamefont {Oakes}, \citenamefont {Peri},
      \citenamefont {Cochrane}, \citenamefont {Martins}, \citenamefont {Hutin},
      \citenamefont {Bertrand}, \citenamefont {Vinet}, \citenamefont {Gomez~Saiz},
      \citenamefont {Ford}, \citenamefont {Smith},\ and\ \citenamefont
      {Gonzalez-Zalba}}]{Oakes_Peri_2023}%
      \BibitemOpen
      \bibfield  {author} {\bibinfo {author} {\bibfnamefont {G.}~\bibnamefont
      {Oakes}}, \bibinfo {author} {\bibfnamefont {L.}~\bibnamefont {Peri}},
      \bibinfo {author} {\bibfnamefont {L.}~\bibnamefont {Cochrane}}, \bibinfo
      {author} {\bibfnamefont {F.}~\bibnamefont {Martins}}, \bibinfo {author}
      {\bibfnamefont {L.}~\bibnamefont {Hutin}}, \bibinfo {author} {\bibfnamefont
      {B.}~\bibnamefont {Bertrand}}, \bibinfo {author} {\bibfnamefont
      {M.}~\bibnamefont {Vinet}}, \bibinfo {author} {\bibfnamefont
      {A.}~\bibnamefont {Gomez~Saiz}}, \bibinfo {author} {\bibfnamefont
      {C.}~\bibnamefont {Ford}}, \bibinfo {author} {\bibfnamefont {C.}~\bibnamefont
      {Smith}},\ and\ \bibinfo {author} {\bibfnamefont {M.}~\bibnamefont
      {Gonzalez-Zalba}},\ }\href {https://doi.org/10.1103/PRXQuantum.4.020346}
      {\bibfield  {journal} {\bibinfo  {journal} {PRX Quantum}\ }\textbf {\bibinfo
      {volume} {4}},\ \bibinfo {pages} {020346} (\bibinfo {year}
      {2023}{\natexlab{a}})}\BibitemShut {NoStop}%
    \bibitem [{\citenamefont {Hogg}\ \emph {et~al.}(2023)\citenamefont {Hogg},
      \citenamefont {House}, \citenamefont {Pakkiam},\ and\ \citenamefont
      {Simmons}}]{Hogg_2023}%
      \BibitemOpen
      \bibfield  {author} {\bibinfo {author} {\bibfnamefont {M.~R.}\ \bibnamefont
      {Hogg}}, \bibinfo {author} {\bibfnamefont {M.}~\bibnamefont {House}},
      \bibinfo {author} {\bibfnamefont {P.}~\bibnamefont {Pakkiam}},\ and\ \bibinfo
      {author} {\bibfnamefont {M.}~\bibnamefont {Simmons}},\ }\href
      {https://doi.org/10.1103/PhysRevApplied.20.034066} {\bibfield  {journal}
      {\bibinfo  {journal} {Physical Review Applied}\ }\textbf {\bibinfo {volume}
      {20}},\ \bibinfo {pages} {034066} (\bibinfo {year} {2023})}\BibitemShut
      {NoStop}%
    \bibitem [{\citenamefont {Phan}\ \emph {et~al.}(2023)\citenamefont {Phan},
      \citenamefont {Falthansl-Scheinecker}, \citenamefont {Mishra}, \citenamefont
      {Strickland}, \citenamefont {Langone}, \citenamefont {Shabani},\ and\
      \citenamefont {Higginbotham}}]{Phan_2023}%
      \BibitemOpen
      \bibfield  {author} {\bibinfo {author} {\bibfnamefont {D.}~\bibnamefont
      {Phan}}, \bibinfo {author} {\bibfnamefont {P.}~\bibnamefont
      {Falthansl-Scheinecker}}, \bibinfo {author} {\bibfnamefont {U.}~\bibnamefont
      {Mishra}}, \bibinfo {author} {\bibfnamefont {W.}~\bibnamefont {Strickland}},
      \bibinfo {author} {\bibfnamefont {D.}~\bibnamefont {Langone}}, \bibinfo
      {author} {\bibfnamefont {J.}~\bibnamefont {Shabani}},\ and\ \bibinfo {author}
      {\bibfnamefont {A.}~\bibnamefont {Higginbotham}},\ }\href
      {https://doi.org/10.1103/PhysRevApplied.19.064032} {\bibfield  {journal}
      {\bibinfo  {journal} {Physical Review Applied}\ }\textbf {\bibinfo {volume}
      {19}},\ \bibinfo {pages} {064032} (\bibinfo {year} {2023})}\BibitemShut
      {NoStop}%
    \bibitem [{\citenamefont {Navarathna}\ \emph {et~al.}(2023)\citenamefont
      {Navarathna}, \citenamefont {Le}, \citenamefont {Hamann}, \citenamefont
      {Nguyen}, \citenamefont {Stace},\ and\ \citenamefont
      {Fedorov}}]{Navarathna2023}%
      \BibitemOpen
      \bibfield  {author} {\bibinfo {author} {\bibfnamefont {R.}~\bibnamefont
      {Navarathna}}, \bibinfo {author} {\bibfnamefont {D.~T.}\ \bibnamefont {Le}},
      \bibinfo {author} {\bibfnamefont {A.~R.}\ \bibnamefont {Hamann}}, \bibinfo
      {author} {\bibfnamefont {H.~D.}\ \bibnamefont {Nguyen}}, \bibinfo {author}
      {\bibfnamefont {T.~M.}\ \bibnamefont {Stace}},\ and\ \bibinfo {author}
      {\bibfnamefont {A.}~\bibnamefont {Fedorov}},\ }\href
      {https://doi.org/10.1103/PhysRevLett.130.037001} {\bibfield  {journal}
      {\bibinfo  {journal} {Phys. Rev. Lett.}\ }\textbf {\bibinfo {volume} {130}},\
      \bibinfo {pages} {037001} (\bibinfo {year} {2023})}\BibitemShut {NoStop}%
    \bibitem [{\citenamefont {Csurgay}\ and\ \citenamefont
      {Porod}(2001)}]{Csurgay_Porod_2001}%
      \BibitemOpen
      \bibfield  {author} {\bibinfo {author} {\bibfnamefont {A.}~\bibnamefont
      {Csurgay}}\ and\ \bibinfo {author} {\bibfnamefont {W.}~\bibnamefont
      {Porod}},\ }\href
      {https://doi.org/10.1002/1097-007X(200101/02)29:1<3::AID-CTA130>3.0.CO;2-Y}
      {\bibfield  {journal} {\bibinfo  {journal} {International Journal of Circuit
      Theory and Applications}\ }\textbf {\bibinfo {volume} {29}},\ \bibinfo
      {pages} {3–35} (\bibinfo {year} {2001})}\BibitemShut {NoStop}%
    \bibitem [{\citenamefont {Csurgay}(2007)}]{Csurgay_2007}%
      \BibitemOpen
      \bibfield  {author} {\bibinfo {author} {\bibfnamefont {A.~I.}\ \bibnamefont
      {Csurgay}},\ }\href {https://doi.org/10.1002/cta.444} {\bibfield  {journal}
      {\bibinfo  {journal} {International Journal of Circuit Theory and
      Applications}\ }\textbf {\bibinfo {volume} {35}},\ \bibinfo {pages}
      {471–484} (\bibinfo {year} {2007})}\BibitemShut {NoStop}%
    \bibitem [{\citenamefont {van Dijk}\ \emph
      {et~al.}(2019{\natexlab{b}})\citenamefont {van Dijk}, \citenamefont
      {Vladimirescu}, \citenamefont {Babaie}, \citenamefont {Charbon},\ and\
      \citenamefont {Sebastiano}}]{spine_2019}%
      \BibitemOpen
      \bibfield  {author} {\bibinfo {author} {\bibfnamefont {J.}~\bibnamefont {van
      Dijk}}, \bibinfo {author} {\bibfnamefont {A.}~\bibnamefont {Vladimirescu}},
      \bibinfo {author} {\bibfnamefont {M.}~\bibnamefont {Babaie}}, \bibinfo
      {author} {\bibfnamefont {E.}~\bibnamefont {Charbon}},\ and\ \bibinfo {author}
      {\bibfnamefont {F.}~\bibnamefont {Sebastiano}},\ }in\ \href
      {https://doi.org/10.1109/IWASI.2019.8791334} {\emph {\bibinfo {booktitle}
      {2019 IEEE 8th International Workshop on Advances in Sensors and Interfaces
      (IWASI)}}}\ (\bibinfo {year} {2019})\ p.\ \bibinfo {pages}
      {23–28}\BibitemShut {NoStop}%
    \bibitem [{\citenamefont {Acharya}\ \emph {et~al.}(2021)\citenamefont
      {Acharya}, \citenamefont {Mohiyaddin}, \citenamefont {{Potočnik}},
      \citenamefont {De~Greve}, \citenamefont {Govoreanu}, \citenamefont {Radu},
      \citenamefont {Gielen},\ and\ \citenamefont {Catthoor}}]{Acharya_2021}%
      \BibitemOpen
      \bibfield  {author} {\bibinfo {author} {\bibfnamefont {R.}~\bibnamefont
      {Acharya}}, \bibinfo {author} {\bibfnamefont {F.~A.}\ \bibnamefont
      {Mohiyaddin}}, \bibinfo {author} {\bibfnamefont {A.}~\bibnamefont
      {{Potočnik}}}, \bibinfo {author} {\bibfnamefont {K.}~\bibnamefont
      {De~Greve}}, \bibinfo {author} {\bibfnamefont {B.}~\bibnamefont {Govoreanu}},
      \bibinfo {author} {\bibfnamefont {I.~P.}\ \bibnamefont {Radu}}, \bibinfo
      {author} {\bibfnamefont {G.}~\bibnamefont {Gielen}},\ and\ \bibinfo {author}
      {\bibfnamefont {F.}~\bibnamefont {Catthoor}},\ }in\ \href
      {https://doi.org/10.23919/DATE51398.2021.9474086} {\emph {\bibinfo
      {booktitle} {2021 Design, Automation and Test in Europe Conference and
      Exhibition (DATE)}}}\ (\bibinfo {year} {2021})\ p.\ \bibinfo {pages}
      {968–973}\BibitemShut {NoStop}%
    \bibitem [{\citenamefont {Gys}\ \emph {et~al.}(2021)\citenamefont {Gys},
      \citenamefont {Mohiyaddin}, \citenamefont {Acharya}, \citenamefont {Li},
      \citenamefont {De~Greve}, \citenamefont {Gielen}, \citenamefont {Govoreanu},
      \citenamefont {Radu},\ and\ \citenamefont {Catthoor}}]{Gys_2021}%
      \BibitemOpen
      \bibfield  {author} {\bibinfo {author} {\bibfnamefont {B.}~\bibnamefont
      {Gys}}, \bibinfo {author} {\bibfnamefont {F.~A.}\ \bibnamefont {Mohiyaddin}},
      \bibinfo {author} {\bibfnamefont {R.}~\bibnamefont {Acharya}}, \bibinfo
      {author} {\bibfnamefont {R.}~\bibnamefont {Li}}, \bibinfo {author}
      {\bibfnamefont {K.}~\bibnamefont {De~Greve}}, \bibinfo {author}
      {\bibfnamefont {G.}~\bibnamefont {Gielen}}, \bibinfo {author} {\bibfnamefont
      {B.}~\bibnamefont {Govoreanu}}, \bibinfo {author} {\bibfnamefont {I.~P.}\
      \bibnamefont {Radu}},\ and\ \bibinfo {author} {\bibfnamefont
      {F.}~\bibnamefont {Catthoor}},\ }in\ \href
      {https://doi.org/10.1109/ESSDERC53440.2021.9631776} {\emph {\bibinfo
      {booktitle} {ESSDERC 2021 - IEEE 51st European Solid-State Device Research
      Conference (ESSDERC)}}}\ (\bibinfo {year} {2021})\ p.\ \bibinfo {pages}
      {63–66}\BibitemShut {NoStop}%
    \bibitem [{\citenamefont {{Pešić}}\ \emph {et~al.}(2024)\citenamefont
      {{Pešić}}, \citenamefont {Wright},\ and\ \citenamefont
      {Charbon}}]{Pesic_2024}%
      \BibitemOpen
      \bibfield  {author} {\bibinfo {author} {\bibfnamefont {V.}~\bibnamefont
      {{Pešić}}}, \bibinfo {author} {\bibfnamefont {A.}~\bibnamefont {Wright}},\
      and\ \bibinfo {author} {\bibfnamefont {E.}~\bibnamefont {Charbon}},\ }in\
      \href {https://doi.org/10.23919/DATE58400.2024.10546593} {\emph {\bibinfo
      {booktitle} {2024 Design, Automation and Test in Europe Conference and
      Exhibition (DATE)}}}\ (\bibinfo {year} {2024})\ p.\ \bibinfo {pages}
      {1–6}\BibitemShut {NoStop}%
    \bibitem [{\citenamefont {McAndrew}\ \emph {et~al.}(2015)\citenamefont
      {McAndrew}, \citenamefont {Coram}, \citenamefont {Gullapalli}, \citenamefont
      {Jones}, \citenamefont {Nagel}, \citenamefont {Roy}, \citenamefont
      {Roychowdhury}, \citenamefont {Scholten}, \citenamefont {Smit}, \citenamefont
      {Wang},\ and\ \citenamefont {Yoshitomi}}]{McAndrew_2015}%
      \BibitemOpen
      \bibfield  {author} {\bibinfo {author} {\bibfnamefont {C.~C.}\ \bibnamefont
      {McAndrew}}, \bibinfo {author} {\bibfnamefont {G.~J.}\ \bibnamefont {Coram}},
      \bibinfo {author} {\bibfnamefont {K.~K.}\ \bibnamefont {Gullapalli}},
      \bibinfo {author} {\bibfnamefont {J.~R.}\ \bibnamefont {Jones}}, \bibinfo
      {author} {\bibfnamefont {L.~W.}\ \bibnamefont {Nagel}}, \bibinfo {author}
      {\bibfnamefont {A.~S.}\ \bibnamefont {Roy}}, \bibinfo {author} {\bibfnamefont
      {J.}~\bibnamefont {Roychowdhury}}, \bibinfo {author} {\bibfnamefont {A.~J.}\
      \bibnamefont {Scholten}}, \bibinfo {author} {\bibfnamefont {G.~D.~J.}\
      \bibnamefont {Smit}}, \bibinfo {author} {\bibfnamefont {X.}~\bibnamefont
      {Wang}},\ and\ \bibinfo {author} {\bibfnamefont {S.}~\bibnamefont
      {Yoshitomi}},\ }\href {https://doi.org/10.1109/JEDS.2015.2455342} {\bibfield
      {journal} {\bibinfo  {journal} {IEEE Journal of the Electron Devices
      Society}\ }\textbf {\bibinfo {volume} {3}},\ \bibinfo {pages} {383–396}
      (\bibinfo {year} {2015})}\BibitemShut {NoStop}%
    \bibitem [{\citenamefont {Persson}\ \emph {et~al.}(2010)\citenamefont
      {Persson}, \citenamefont {Wilson}, \citenamefont {Sandberg}, \citenamefont
      {Johansson},\ and\ \citenamefont {Delsing}}]{Persson_2010}%
      \BibitemOpen
      \bibfield  {author} {\bibinfo {author} {\bibfnamefont {F.}~\bibnamefont
      {Persson}}, \bibinfo {author} {\bibfnamefont {C.~M.}\ \bibnamefont {Wilson}},
      \bibinfo {author} {\bibfnamefont {M.}~\bibnamefont {Sandberg}}, \bibinfo
      {author} {\bibfnamefont {G.}~\bibnamefont {Johansson}},\ and\ \bibinfo
      {author} {\bibfnamefont {P.}~\bibnamefont {Delsing}},\ }\href
      {https://doi.org/10.1021/nl903887x} {\bibfield  {journal} {\bibinfo
      {journal} {Nano Letters}\ }\textbf {\bibinfo {volume} {10}},\ \bibinfo
      {pages} {953–957} (\bibinfo {year} {2010})},\ \bibinfo {note}
      {arXiv:0902.4316}\BibitemShut {NoStop}%
    \bibitem [{\citenamefont {Xue}\ \emph {et~al.}(2022)\citenamefont {Xue},
      \citenamefont {Russ}, \citenamefont {Samkharadze}, \citenamefont {Undseth},
      \citenamefont {Sammak}, \citenamefont {Scappucci},\ and\ \citenamefont
      {Vandersypen}}]{Xue_2022}%
      \BibitemOpen
      \bibfield  {author} {\bibinfo {author} {\bibfnamefont {X.}~\bibnamefont
      {Xue}}, \bibinfo {author} {\bibfnamefont {M.}~\bibnamefont {Russ}}, \bibinfo
      {author} {\bibfnamefont {N.}~\bibnamefont {Samkharadze}}, \bibinfo {author}
      {\bibfnamefont {B.}~\bibnamefont {Undseth}}, \bibinfo {author} {\bibfnamefont
      {A.}~\bibnamefont {Sammak}}, \bibinfo {author} {\bibfnamefont
      {G.}~\bibnamefont {Scappucci}},\ and\ \bibinfo {author} {\bibfnamefont
      {L.~M.~K.}\ \bibnamefont {Vandersypen}},\ }\href
      {https://doi.org/10.1038/s41586-021-04273-w} {\bibfield  {journal} {\bibinfo
      {journal} {Nature}\ }\textbf {\bibinfo {volume} {601}},\ \bibinfo {pages}
      {343–347} (\bibinfo {year} {2022})}\BibitemShut {NoStop}%
    \bibitem [{\citenamefont {Noiri}\ \emph {et~al.}(2022)\citenamefont {Noiri},
      \citenamefont {Takeda}, \citenamefont {Nakajima}, \citenamefont {Kobayashi},
      \citenamefont {Sammak}, \citenamefont {Scappucci},\ and\ \citenamefont
      {Tarucha}}]{Noiri_2022}%
      \BibitemOpen
      \bibfield  {author} {\bibinfo {author} {\bibfnamefont {A.}~\bibnamefont
      {Noiri}}, \bibinfo {author} {\bibfnamefont {K.}~\bibnamefont {Takeda}},
      \bibinfo {author} {\bibfnamefont {T.}~\bibnamefont {Nakajima}}, \bibinfo
      {author} {\bibfnamefont {T.}~\bibnamefont {Kobayashi}}, \bibinfo {author}
      {\bibfnamefont {A.}~\bibnamefont {Sammak}}, \bibinfo {author} {\bibfnamefont
      {G.}~\bibnamefont {Scappucci}},\ and\ \bibinfo {author} {\bibfnamefont
      {S.}~\bibnamefont {Tarucha}},\ }\href
      {https://doi.org/10.1038/s41586-021-04182-y} {\bibfield  {journal} {\bibinfo
      {journal} {Nature}\ }\textbf {\bibinfo {volume} {601}},\ \bibinfo {pages}
      {338} (\bibinfo {year} {2022})}\BibitemShut {NoStop}%
    \bibitem [{\citenamefont {Burkard}\ \emph {et~al.}(2023)\citenamefont
      {Burkard}, \citenamefont {Ladd}, \citenamefont {Pan}, \citenamefont
      {Nichol},\ and\ \citenamefont {Petta}}]{Burkard_2023}%
      \BibitemOpen
      \bibfield  {author} {\bibinfo {author} {\bibfnamefont {G.}~\bibnamefont
      {Burkard}}, \bibinfo {author} {\bibfnamefont {T.~D.}\ \bibnamefont {Ladd}},
      \bibinfo {author} {\bibfnamefont {A.}~\bibnamefont {Pan}}, \bibinfo {author}
      {\bibfnamefont {J.~M.}\ \bibnamefont {Nichol}},\ and\ \bibinfo {author}
      {\bibfnamefont {J.~R.}\ \bibnamefont {Petta}},\ }\href
      {https://doi.org/10.1103/RevModPhys.95.025003} {\bibfield  {journal}
      {\bibinfo  {journal} {Reviews of Modern Physics}\ }\textbf {\bibinfo {volume}
      {95}},\ \bibinfo {pages} {025003} (\bibinfo {year} {2023})}\BibitemShut
      {NoStop}%
    \bibitem [{\citenamefont {Aghaee}\ \emph {et~al.}(2023)\citenamefont {Aghaee},
      \citenamefont {Akkala}, \citenamefont {Alam}, \citenamefont {Ali},
      \citenamefont {Alcaraz~Ramirez}, \citenamefont {Andrzejczuk}, \citenamefont
      {Antipov}, \citenamefont {Aseev}, \citenamefont {Astafev}, \citenamefont
      {Bauer}, \citenamefont {Becker}, \citenamefont {Boddapati}, \citenamefont
      {Boekhout}, \citenamefont {Bommer}, \citenamefont {Bosma}, \citenamefont
      {Bourdet}, \citenamefont {Boutin}, \citenamefont {Caroff}, \citenamefont
      {Casparis}, \citenamefont {Cassidy}, \citenamefont {Chatoor}, \citenamefont
      {Christensen}, \citenamefont {Clay}, \citenamefont {Cole}, \citenamefont
      {Corsetti}, \citenamefont {Cui}, \citenamefont {Dalampiras}, \citenamefont
      {Dokania}, \citenamefont {de~Lange}, \citenamefont {de~Moor}, \citenamefont
      {Estrada Salda\~na}, \citenamefont {Fallahi}, \citenamefont {Fathabad},
      \citenamefont {Gamble}, \citenamefont {Gardner}, \citenamefont {Govender},
      \citenamefont {Griggio}, \citenamefont {Grigoryan}, \citenamefont {Gronin},
      \citenamefont {Gukelberger}, \citenamefont {Hansen}, \citenamefont {Heedt},
      \citenamefont {Herranz~Zamorano}, \citenamefont {Ho}, \citenamefont
      {Holgaard}, \citenamefont {Ingerslev}, \citenamefont {Johansson},
      \citenamefont {Jones}, \citenamefont {Kallaher}, \citenamefont {Karimi},
      \citenamefont {Karzig}, \citenamefont {King}, \citenamefont {Kloster},
      \citenamefont {Knapp}, \citenamefont {Kocon}, \citenamefont {Koski},
      \citenamefont {Kostamo}, \citenamefont {Krogstrup}, \citenamefont {Kumar},
      \citenamefont {Laeven}, \citenamefont {Larsen}, \citenamefont {Li},
      \citenamefont {Lindemann}, \citenamefont {Love}, \citenamefont {Lutchyn},
      \citenamefont {Madsen}, \citenamefont {Manfra}, \citenamefont {Markussen},
      \citenamefont {Martinez}, \citenamefont {McNeil}, \citenamefont {Memisevic},
      \citenamefont {Morgan}, \citenamefont {Mullally}, \citenamefont {Nayak},
      \citenamefont {Nielsen}, \citenamefont {Nielsen}, \citenamefont {Nijholt},
      \citenamefont {Nurmohamed}, \citenamefont {O'Farrell}, \citenamefont {Otani},
      \citenamefont {Pauka}, \citenamefont {Petersson}, \citenamefont {Petit},
      \citenamefont {Pikulin}, \citenamefont {Preiss}, \citenamefont
      {Quintero-Perez}, \citenamefont {Rajpalke}, \citenamefont {Rasmussen},
      \citenamefont {Razmadze}, \citenamefont {Reentila}, \citenamefont {Reilly},
      \citenamefont {Rouse}, \citenamefont {Sadovskyy}, \citenamefont {Sainiemi},
      \citenamefont {Schreppler}, \citenamefont {Sidorkin}, \citenamefont {Singh},
      \citenamefont {Singh}, \citenamefont {Sinha}, \citenamefont {Sohr},
      \citenamefont {Stankevi\ifmmode~\check{c}\else \v{c}\fi{}}, \citenamefont
      {Stek}, \citenamefont {Suominen}, \citenamefont {Suter}, \citenamefont
      {Svidenko}, \citenamefont {Teicher}, \citenamefont {Temuerhan}, \citenamefont
      {Thiyagarajah}, \citenamefont {Tholapi}, \citenamefont {Thomas},
      \citenamefont {Toomey}, \citenamefont {Upadhyay}, \citenamefont {Urban},
      \citenamefont {Vaitiek\ifmmode~\dot{e}\else \.{e}\fi{}nas}, \citenamefont
      {Van~Hoogdalem}, \citenamefont {Van~Woerkom}, \citenamefont {Viazmitinov},
      \citenamefont {Vogel}, \citenamefont {Waddy}, \citenamefont {Watson},
      \citenamefont {Weston}, \citenamefont {Winkler}, \citenamefont {Yang},
      \citenamefont {Yau}, \citenamefont {Yi}, \citenamefont {Yucelen},
      \citenamefont {Webster}, \citenamefont {Zeisel},\ and\ \citenamefont
      {Zhao}}]{Aghaee2024}%
      \BibitemOpen
      \bibfield  {author} {\bibinfo {author} {\bibfnamefont {M.}~\bibnamefont
      {Aghaee}}, \bibinfo {author} {\bibfnamefont {A.}~\bibnamefont {Akkala}},
      \bibinfo {author} {\bibfnamefont {Z.}~\bibnamefont {Alam}}, \bibinfo {author}
      {\bibfnamefont {R.}~\bibnamefont {Ali}}, \bibinfo {author} {\bibfnamefont
      {A.}~\bibnamefont {Alcaraz~Ramirez}}, \bibinfo {author} {\bibfnamefont
      {M.}~\bibnamefont {Andrzejczuk}}, \bibinfo {author} {\bibfnamefont {A.~E.}\
      \bibnamefont {Antipov}}, \bibinfo {author} {\bibfnamefont {P.}~\bibnamefont
      {Aseev}}, \bibinfo {author} {\bibfnamefont {M.}~\bibnamefont {Astafev}},
      \bibinfo {author} {\bibfnamefont {B.}~\bibnamefont {Bauer}}, \bibinfo
      {author} {\bibfnamefont {J.}~\bibnamefont {Becker}}, \bibinfo {author}
      {\bibfnamefont {S.}~\bibnamefont {Boddapati}}, \bibinfo {author}
      {\bibfnamefont {F.}~\bibnamefont {Boekhout}}, \bibinfo {author}
      {\bibfnamefont {J.}~\bibnamefont {Bommer}}, \bibinfo {author} {\bibfnamefont
      {T.}~\bibnamefont {Bosma}}, \bibinfo {author} {\bibfnamefont
      {L.}~\bibnamefont {Bourdet}}, \bibinfo {author} {\bibfnamefont
      {S.}~\bibnamefont {Boutin}}, \bibinfo {author} {\bibfnamefont
      {P.}~\bibnamefont {Caroff}}, \bibinfo {author} {\bibfnamefont
      {L.}~\bibnamefont {Casparis}}, \bibinfo {author} {\bibfnamefont
      {M.}~\bibnamefont {Cassidy}}, \bibinfo {author} {\bibfnamefont
      {S.}~\bibnamefont {Chatoor}}, \bibinfo {author} {\bibfnamefont {A.~W.}\
      \bibnamefont {Christensen}}, \bibinfo {author} {\bibfnamefont
      {N.}~\bibnamefont {Clay}}, \bibinfo {author} {\bibfnamefont {W.~S.}\
      \bibnamefont {Cole}}, \bibinfo {author} {\bibfnamefont {F.}~\bibnamefont
      {Corsetti}}, \bibinfo {author} {\bibfnamefont {A.}~\bibnamefont {Cui}},
      \bibinfo {author} {\bibfnamefont {P.}~\bibnamefont {Dalampiras}}, \bibinfo
      {author} {\bibfnamefont {A.}~\bibnamefont {Dokania}}, \bibinfo {author}
      {\bibfnamefont {G.}~\bibnamefont {de~Lange}}, \bibinfo {author}
      {\bibfnamefont {M.}~\bibnamefont {de~Moor}}, \bibinfo {author} {\bibfnamefont
      {J.~C.}\ \bibnamefont {Estrada Salda\~na}}, \bibinfo {author} {\bibfnamefont
      {S.}~\bibnamefont {Fallahi}}, \bibinfo {author} {\bibfnamefont {Z.~H.}\
      \bibnamefont {Fathabad}}, \bibinfo {author} {\bibfnamefont {J.}~\bibnamefont
      {Gamble}}, \bibinfo {author} {\bibfnamefont {G.}~\bibnamefont {Gardner}},
      \bibinfo {author} {\bibfnamefont {D.}~\bibnamefont {Govender}}, \bibinfo
      {author} {\bibfnamefont {F.}~\bibnamefont {Griggio}}, \bibinfo {author}
      {\bibfnamefont {R.}~\bibnamefont {Grigoryan}}, \bibinfo {author}
      {\bibfnamefont {S.}~\bibnamefont {Gronin}}, \bibinfo {author} {\bibfnamefont
      {J.}~\bibnamefont {Gukelberger}}, \bibinfo {author} {\bibfnamefont {E.~B.}\
      \bibnamefont {Hansen}}, \bibinfo {author} {\bibfnamefont {S.}~\bibnamefont
      {Heedt}}, \bibinfo {author} {\bibfnamefont {J.}~\bibnamefont
      {Herranz~Zamorano}}, \bibinfo {author} {\bibfnamefont {S.}~\bibnamefont
      {Ho}}, \bibinfo {author} {\bibfnamefont {U.~L.}\ \bibnamefont {Holgaard}},
      \bibinfo {author} {\bibfnamefont {H.}~\bibnamefont {Ingerslev}}, \bibinfo
      {author} {\bibfnamefont {L.}~\bibnamefont {Johansson}}, \bibinfo {author}
      {\bibfnamefont {J.}~\bibnamefont {Jones}}, \bibinfo {author} {\bibfnamefont
      {R.}~\bibnamefont {Kallaher}}, \bibinfo {author} {\bibfnamefont
      {F.}~\bibnamefont {Karimi}}, \bibinfo {author} {\bibfnamefont
      {T.}~\bibnamefont {Karzig}}, \bibinfo {author} {\bibfnamefont
      {E.}~\bibnamefont {King}}, \bibinfo {author} {\bibfnamefont {M.~E.}\
      \bibnamefont {Kloster}}, \bibinfo {author} {\bibfnamefont {C.}~\bibnamefont
      {Knapp}}, \bibinfo {author} {\bibfnamefont {D.}~\bibnamefont {Kocon}},
      \bibinfo {author} {\bibfnamefont {J.}~\bibnamefont {Koski}}, \bibinfo
      {author} {\bibfnamefont {P.}~\bibnamefont {Kostamo}}, \bibinfo {author}
      {\bibfnamefont {P.}~\bibnamefont {Krogstrup}}, \bibinfo {author}
      {\bibfnamefont {M.}~\bibnamefont {Kumar}}, \bibinfo {author} {\bibfnamefont
      {T.}~\bibnamefont {Laeven}}, \bibinfo {author} {\bibfnamefont
      {T.}~\bibnamefont {Larsen}}, \bibinfo {author} {\bibfnamefont
      {K.}~\bibnamefont {Li}}, \bibinfo {author} {\bibfnamefont {T.}~\bibnamefont
      {Lindemann}}, \bibinfo {author} {\bibfnamefont {J.}~\bibnamefont {Love}},
      \bibinfo {author} {\bibfnamefont {R.}~\bibnamefont {Lutchyn}}, \bibinfo
      {author} {\bibfnamefont {M.~H.}\ \bibnamefont {Madsen}}, \bibinfo {author}
      {\bibfnamefont {M.}~\bibnamefont {Manfra}}, \bibinfo {author} {\bibfnamefont
      {S.}~\bibnamefont {Markussen}}, \bibinfo {author} {\bibfnamefont
      {E.}~\bibnamefont {Martinez}}, \bibinfo {author} {\bibfnamefont
      {R.}~\bibnamefont {McNeil}}, \bibinfo {author} {\bibfnamefont
      {E.}~\bibnamefont {Memisevic}}, \bibinfo {author} {\bibfnamefont
      {T.}~\bibnamefont {Morgan}}, \bibinfo {author} {\bibfnamefont
      {A.}~\bibnamefont {Mullally}}, \bibinfo {author} {\bibfnamefont
      {C.}~\bibnamefont {Nayak}}, \bibinfo {author} {\bibfnamefont
      {J.}~\bibnamefont {Nielsen}}, \bibinfo {author} {\bibfnamefont {W.~H.~P.}\
      \bibnamefont {Nielsen}}, \bibinfo {author} {\bibfnamefont {B.}~\bibnamefont
      {Nijholt}}, \bibinfo {author} {\bibfnamefont {A.}~\bibnamefont {Nurmohamed}},
      \bibinfo {author} {\bibfnamefont {E.}~\bibnamefont {O'Farrell}}, \bibinfo
      {author} {\bibfnamefont {K.}~\bibnamefont {Otani}}, \bibinfo {author}
      {\bibfnamefont {S.}~\bibnamefont {Pauka}}, \bibinfo {author} {\bibfnamefont
      {K.}~\bibnamefont {Petersson}}, \bibinfo {author} {\bibfnamefont
      {L.}~\bibnamefont {Petit}}, \bibinfo {author} {\bibfnamefont {D.~I.}\
      \bibnamefont {Pikulin}}, \bibinfo {author} {\bibfnamefont {F.}~\bibnamefont
      {Preiss}}, \bibinfo {author} {\bibfnamefont {M.}~\bibnamefont
      {Quintero-Perez}}, \bibinfo {author} {\bibfnamefont {M.}~\bibnamefont
      {Rajpalke}}, \bibinfo {author} {\bibfnamefont {K.}~\bibnamefont {Rasmussen}},
      \bibinfo {author} {\bibfnamefont {D.}~\bibnamefont {Razmadze}}, \bibinfo
      {author} {\bibfnamefont {O.}~\bibnamefont {Reentila}}, \bibinfo {author}
      {\bibfnamefont {D.}~\bibnamefont {Reilly}}, \bibinfo {author} {\bibfnamefont
      {R.}~\bibnamefont {Rouse}}, \bibinfo {author} {\bibfnamefont
      {I.}~\bibnamefont {Sadovskyy}}, \bibinfo {author} {\bibfnamefont
      {L.}~\bibnamefont {Sainiemi}}, \bibinfo {author} {\bibfnamefont
      {S.}~\bibnamefont {Schreppler}}, \bibinfo {author} {\bibfnamefont
      {V.}~\bibnamefont {Sidorkin}}, \bibinfo {author} {\bibfnamefont
      {A.}~\bibnamefont {Singh}}, \bibinfo {author} {\bibfnamefont
      {S.}~\bibnamefont {Singh}}, \bibinfo {author} {\bibfnamefont
      {S.}~\bibnamefont {Sinha}}, \bibinfo {author} {\bibfnamefont
      {P.}~\bibnamefont {Sohr}}, \bibinfo {author} {\bibfnamefont {T.~c.~v.}\
      \bibnamefont {Stankevi\ifmmode~\check{c}\else \v{c}\fi{}}}, \bibinfo {author}
      {\bibfnamefont {L.}~\bibnamefont {Stek}}, \bibinfo {author} {\bibfnamefont
      {H.}~\bibnamefont {Suominen}}, \bibinfo {author} {\bibfnamefont
      {J.}~\bibnamefont {Suter}}, \bibinfo {author} {\bibfnamefont
      {V.}~\bibnamefont {Svidenko}}, \bibinfo {author} {\bibfnamefont
      {S.}~\bibnamefont {Teicher}}, \bibinfo {author} {\bibfnamefont
      {M.}~\bibnamefont {Temuerhan}}, \bibinfo {author} {\bibfnamefont
      {N.}~\bibnamefont {Thiyagarajah}}, \bibinfo {author} {\bibfnamefont
      {R.}~\bibnamefont {Tholapi}}, \bibinfo {author} {\bibfnamefont
      {M.}~\bibnamefont {Thomas}}, \bibinfo {author} {\bibfnamefont
      {E.}~\bibnamefont {Toomey}}, \bibinfo {author} {\bibfnamefont
      {S.}~\bibnamefont {Upadhyay}}, \bibinfo {author} {\bibfnamefont
      {I.}~\bibnamefont {Urban}}, \bibinfo {author} {\bibfnamefont
      {S.}~\bibnamefont {Vaitiek\ifmmode~\dot{e}\else \.{e}\fi{}nas}}, \bibinfo
      {author} {\bibfnamefont {K.}~\bibnamefont {Van~Hoogdalem}}, \bibinfo {author}
      {\bibfnamefont {D.}~\bibnamefont {Van~Woerkom}}, \bibinfo {author}
      {\bibfnamefont {D.~V.}\ \bibnamefont {Viazmitinov}}, \bibinfo {author}
      {\bibfnamefont {D.}~\bibnamefont {Vogel}}, \bibinfo {author} {\bibfnamefont
      {S.}~\bibnamefont {Waddy}}, \bibinfo {author} {\bibfnamefont
      {J.}~\bibnamefont {Watson}}, \bibinfo {author} {\bibfnamefont
      {J.}~\bibnamefont {Weston}}, \bibinfo {author} {\bibfnamefont {G.~W.}\
      \bibnamefont {Winkler}}, \bibinfo {author} {\bibfnamefont {C.~K.}\
      \bibnamefont {Yang}}, \bibinfo {author} {\bibfnamefont {S.}~\bibnamefont
      {Yau}}, \bibinfo {author} {\bibfnamefont {D.}~\bibnamefont {Yi}}, \bibinfo
      {author} {\bibfnamefont {E.}~\bibnamefont {Yucelen}}, \bibinfo {author}
      {\bibfnamefont {A.}~\bibnamefont {Webster}}, \bibinfo {author} {\bibfnamefont
      {R.}~\bibnamefont {Zeisel}},\ and\ \bibinfo {author} {\bibfnamefont
      {R.}~\bibnamefont {Zhao}} (\bibinfo {collaboration} {Microsoft Quantum}),\
      }\href {https://doi.org/10.1103/PhysRevB.107.245423} {\bibfield  {journal}
      {\bibinfo  {journal} {Phys. Rev. B}\ }\textbf {\bibinfo {volume} {107}},\
      \bibinfo {pages} {245423} (\bibinfo {year} {2023})}\BibitemShut {NoStop}%
    \bibitem [{\citenamefont {Maman}\ \emph {et~al.}(2020)\citenamefont {Maman},
      \citenamefont {Gonzalez-Zalba},\ and\ \citenamefont {Pályi}}]{maman2020}%
      \BibitemOpen
      \bibfield  {author} {\bibinfo {author} {\bibfnamefont {V.~D.}\ \bibnamefont
      {Maman}}, \bibinfo {author} {\bibfnamefont {M.~F.}\ \bibnamefont
      {Gonzalez-Zalba}},\ and\ \bibinfo {author} {\bibfnamefont {A.}~\bibnamefont
      {Pályi}},\ }\href@noop {} {\bibfield  {journal} {\bibinfo  {journal} {Phys.
      Rev. Applied}\ }\textbf {\bibinfo {volume} {14}},\ \bibinfo {pages} {64024}
      (\bibinfo {year} {2020})}\BibitemShut {NoStop}%
    \bibitem [{\citenamefont {Am-Shallem}\ \emph {et~al.}(2015)\citenamefont
      {Am-Shallem}, \citenamefont {Levy}, \citenamefont {Schaefer},\ and\
      \citenamefont {Kosloff}}]{Am-Shallem_Levy_Schaefer_Kosloff_2015}%
      \BibitemOpen
      \bibfield  {author} {\bibinfo {author} {\bibfnamefont {M.}~\bibnamefont
      {Am-Shallem}}, \bibinfo {author} {\bibfnamefont {A.}~\bibnamefont {Levy}},
      \bibinfo {author} {\bibfnamefont {I.}~\bibnamefont {Schaefer}},\ and\
      \bibinfo {author} {\bibfnamefont {R.}~\bibnamefont {Kosloff}},\ }\href
      {http://arxiv.org/abs/1510.08634} {\bibinfo {title} {Three approaches for
      representing lindblad dynamics by a matrix-vector notation}} (\bibinfo {year}
      {2015}),\ \bibinfo {note} {arXiv:1510.08634}\BibitemShut {NoStop}%
    \bibitem [{\citenamefont {Riesch}\ and\ \citenamefont
      {Jirauschek}(2019)}]{Riesch_Jirauschek_2019}%
      \BibitemOpen
      \bibfield  {author} {\bibinfo {author} {\bibfnamefont {M.}~\bibnamefont
      {Riesch}}\ and\ \bibinfo {author} {\bibfnamefont {C.}~\bibnamefont
      {Jirauschek}},\ }\href {https://doi.org/10.1016/j.jcp.2019.04.006} {\bibfield
       {journal} {\bibinfo  {journal} {Journal of Computational Physics}\ }\textbf
      {\bibinfo {volume} {390}},\ \bibinfo {pages} {290–296} (\bibinfo {year}
      {2019})},\ \bibinfo {note} {arXiv:1808.00416}\BibitemShut {NoStop}%
    \bibitem [{\citenamefont {Minganti}\ and\ \citenamefont
      {Huybrechts}(2022)}]{Minganti_Huybrechts_2022}%
      \BibitemOpen
      \bibfield  {author} {\bibinfo {author} {\bibfnamefont {F.}~\bibnamefont
      {Minganti}}\ and\ \bibinfo {author} {\bibfnamefont {D.}~\bibnamefont
      {Huybrechts}},\ }\href {https://doi.org/10.22331/q-2022-02-10-649} {\bibinfo
      {title} {Arnoldi-lindblad time evolution: Faster-than-the-clock algorithm for
      the spectrum of time-independent and floquet open quantum systems}} (\bibinfo
      {year} {2022}),\ \bibinfo {note} {arXiv:2109.01648}\BibitemShut
      {NoStop}%
    \bibitem [{\citenamefont {Manzano}(2020)}]{manzano_short_2020}%
      \BibitemOpen
      \bibfield  {author} {\bibinfo {author} {\bibfnamefont {D.}~\bibnamefont
      {Manzano}},\ }\href {https://doi.org/10.1063/1.5115323} {\bibfield  {journal}
      {\bibinfo  {journal} {AIP Advances}\ }\textbf {\bibinfo {volume} {10}},\
      \bibinfo {pages} {025106} (\bibinfo {year} {2020})},\ \Eprint
      {https://arxiv.org/abs/https://doi.org/10.1063/1.5115323}
      {https://doi.org/10.1063/1.5115323} \BibitemShut {NoStop}%
    \bibitem [{\citenamefont {Zoller}\ and\ \citenamefont
      {Gardiner}(1997)}]{Zoller_Gardiner_1997}%
      \BibitemOpen
      \bibfield  {author} {\bibinfo {author} {\bibfnamefont {P.}~\bibnamefont
      {Zoller}}\ and\ \bibinfo {author} {\bibfnamefont {C.~W.}\ \bibnamefont
      {Gardiner}},\ }\href {http://arxiv.org/abs/quant-ph/9702030} {\bibinfo
      {title} {Quantum noise in quantum optics: the {Stochastic} {Schr\"odinger}
      equation}} (\bibinfo {year} {1997}),\ \bibinfo {note}
      {arXiv:quant-ph/9702030}\BibitemShut {NoStop}%
    \bibitem [{\citenamefont {Gardiner}\ and\ \citenamefont
      {Zoller}(2004)}]{Gardiner_Zoller_2004}%
      \BibitemOpen
      \bibfield  {author} {\bibinfo {author} {\bibfnamefont {C.~W.}\ \bibnamefont
      {Gardiner}}\ and\ \bibinfo {author} {\bibfnamefont {P.}~\bibnamefont
      {Zoller}},\ }\href@noop {} {\emph {\bibinfo {title} {Quantum noise: a
      handbook of Markovian and non-Markovian quantum stochastic methods with
      applications to quantum optics}}},\ \bibinfo {edition} {3rd}\ ed.,\ Springer
      series in synergetics\ (\bibinfo  {publisher} {Springer},\ \bibinfo {year}
      {2004})\BibitemShut {NoStop}%
    \bibitem [{\citenamefont {Albert}\ \emph {et~al.}(2016)\citenamefont {Albert},
      \citenamefont {Bradlyn}, \citenamefont {Fraas},\ and\ \citenamefont
      {Jiang}}]{Albert_Bradlyn_Fraas_Jiang_2016}%
      \BibitemOpen
      \bibfield  {author} {\bibinfo {author} {\bibfnamefont {V.~V.}\ \bibnamefont
      {Albert}}, \bibinfo {author} {\bibfnamefont {B.}~\bibnamefont {Bradlyn}},
      \bibinfo {author} {\bibfnamefont {M.}~\bibnamefont {Fraas}},\ and\ \bibinfo
      {author} {\bibfnamefont {L.}~\bibnamefont {Jiang}},\ }\href
      {https://doi.org/10.1103/PhysRevX.6.041031} {\bibfield  {journal} {\bibinfo
      {journal} {Physical Review X}\ }\textbf {\bibinfo {volume} {6}},\ \bibinfo
      {pages} {041031} (\bibinfo {year} {2016})},\ \bibinfo {note}
      {arXiv:1512.08079}\BibitemShut
      {NoStop}%
    \bibitem [{\citenamefont {Albash}\ \emph {et~al.}(2012)\citenamefont {Albash},
      \citenamefont {Boixo}, \citenamefont {Lidar},\ and\ \citenamefont
      {Zanardi}}]{Albash_Boixo_Lidar_Zanardi_2012}%
      \BibitemOpen
      \bibfield  {author} {\bibinfo {author} {\bibfnamefont {T.}~\bibnamefont
      {Albash}}, \bibinfo {author} {\bibfnamefont {S.}~\bibnamefont {Boixo}},
      \bibinfo {author} {\bibfnamefont {D.~A.}\ \bibnamefont {Lidar}},\ and\
      \bibinfo {author} {\bibfnamefont {P.}~\bibnamefont {Zanardi}},\ }\href
      {https://doi.org/10.1088/1367-2630/14/12/123016} {\bibfield  {journal}
      {\bibinfo  {journal} {New Journal of Physics}\ }\textbf {\bibinfo {volume}
      {14}},\ \bibinfo {pages} {123016} (\bibinfo {year} {2012})}\BibitemShut
      {NoStop}%
    \bibitem [{\citenamefont {Peri}\ \emph
      {et~al.}(2024{\natexlab{a}})\citenamefont {Peri}, \citenamefont {Oakes},
      \citenamefont {Cochrane}, \citenamefont {Ford},\ and\ \citenamefont
      {Gonzalez-Zalba}}]{Peri2024beyondadiabatic}%
      \BibitemOpen
      \bibfield  {author} {\bibinfo {author} {\bibfnamefont {L.}~\bibnamefont
      {Peri}}, \bibinfo {author} {\bibfnamefont {G.~A.}\ \bibnamefont {Oakes}},
      \bibinfo {author} {\bibfnamefont {L.}~\bibnamefont {Cochrane}}, \bibinfo
      {author} {\bibfnamefont {C.~J.~B.}\ \bibnamefont {Ford}},\ and\ \bibinfo
      {author} {\bibfnamefont {M.~F.}\ \bibnamefont {Gonzalez-Zalba}},\ }\href
      {https://doi.org/10.22331/q-2024-03-21-1294} {\bibfield  {journal} {\bibinfo
      {journal} {{Quantum}}\ }\textbf {\bibinfo {volume} {8}},\ \bibinfo {pages}
      {1294} (\bibinfo {year} {2024}{\natexlab{a}})}\BibitemShut {NoStop}%
    \bibitem [{\citenamefont {Bashir}\ \emph {et~al.}(2019)\citenamefont {Bashir},
      \citenamefont {Giounanlis}, \citenamefont {Blokhina}, \citenamefont
      {Leipold}, \citenamefont {Pomorski},\ and\ \citenamefont
      {Staszewski}}]{Bashir_2019}%
      \BibitemOpen
      \bibfield  {author} {\bibinfo {author} {\bibfnamefont {I.}~\bibnamefont
      {Bashir}}, \bibinfo {author} {\bibfnamefont {P.}~\bibnamefont {Giounanlis}},
      \bibinfo {author} {\bibfnamefont {E.}~\bibnamefont {Blokhina}}, \bibinfo
      {author} {\bibfnamefont {D.}~\bibnamefont {Leipold}}, \bibinfo {author}
      {\bibfnamefont {K.}~\bibnamefont {Pomorski}},\ and\ \bibinfo {author}
      {\bibfnamefont {R.~B.}\ \bibnamefont {Staszewski}},\ }in\ \href
      {https://doi.org/10.1109/NEWCAS44328.2019.8961307} {\emph {\bibinfo
      {booktitle} {2019 17th IEEE International New Circuits and Systems Conference
      (NEWCAS)}}}\ (\bibinfo {year} {2019})\ p.\ \bibinfo {pages}
      {1–4}\BibitemShut {NoStop}%
    \bibitem [{\citenamefont {Vigneau}\ \emph {et~al.}(2023)\citenamefont
      {Vigneau}, \citenamefont {Fedele}, \citenamefont {Chatterjee}, \citenamefont
      {Reilly}, \citenamefont {Kuemmeth}, \citenamefont {Gonzalez-Zalba},
      \citenamefont {Laird},\ and\ \citenamefont {Ares}}]{Vigneau_2023}%
      \BibitemOpen
      \bibfield  {author} {\bibinfo {author} {\bibfnamefont {F.}~\bibnamefont
      {Vigneau}}, \bibinfo {author} {\bibfnamefont {F.}~\bibnamefont {Fedele}},
      \bibinfo {author} {\bibfnamefont {A.}~\bibnamefont {Chatterjee}}, \bibinfo
      {author} {\bibfnamefont {D.}~\bibnamefont {Reilly}}, \bibinfo {author}
      {\bibfnamefont {F.}~\bibnamefont {Kuemmeth}}, \bibinfo {author}
      {\bibfnamefont {M.~F.}\ \bibnamefont {Gonzalez-Zalba}}, \bibinfo {author}
      {\bibfnamefont {E.}~\bibnamefont {Laird}},\ and\ \bibinfo {author}
      {\bibfnamefont {N.}~\bibnamefont {Ares}},\ }\href
      {https://doi.org/10.1063/5.0088229} {\bibfield  {journal} {\bibinfo
      {journal} {Applied Physics Reviews}\ }\textbf {\bibinfo {volume} {10}},\
      \bibinfo {pages} {021305} (\bibinfo {year} {2023})}\BibitemShut {NoStop}%
    \bibitem [{\citenamefont {Peri}\ \emph
      {et~al.}(2024{\natexlab{b}})\citenamefont {Peri}, \citenamefont {Benito},
      \citenamefont {Ford},\ and\ \citenamefont {Gonzalez-Zalba}}]{Peri_2023}%
      \BibitemOpen
      \bibfield  {author} {\bibinfo {author} {\bibfnamefont {L.}~\bibnamefont
      {Peri}}, \bibinfo {author} {\bibfnamefont {M.}~\bibnamefont {Benito}},
      \bibinfo {author} {\bibfnamefont {C.~J.~B.}\ \bibnamefont {Ford}},\ and\
      \bibinfo {author} {\bibfnamefont {M.~F.}\ \bibnamefont {Gonzalez-Zalba}},\
      }\href {https://doi.org/10.1038/s41534-024-00907-9} {\bibfield  {journal}
      {\bibinfo  {journal} {npj Quantum Information}\ }\textbf {\bibinfo {volume}
      {10}},\ \bibinfo {pages} {1–14} (\bibinfo {year}
      {2024}{\natexlab{b}})}\BibitemShut {NoStop}%
    \bibitem [{\citenamefont {Mizuta}\ \emph {et~al.}(2017)\citenamefont {Mizuta},
      \citenamefont {Otxoa}, \citenamefont {Betz},\ and\ \citenamefont
      {Gonzalez-Zalba}}]{Mizuta2017}%
      \BibitemOpen
      \bibfield  {author} {\bibinfo {author} {\bibfnamefont {R.}~\bibnamefont
      {Mizuta}}, \bibinfo {author} {\bibfnamefont {R.~M.}\ \bibnamefont {Otxoa}},
      \bibinfo {author} {\bibfnamefont {A.~C.}\ \bibnamefont {Betz}},\ and\
      \bibinfo {author} {\bibfnamefont {M.~F.}\ \bibnamefont {Gonzalez-Zalba}},\
      }\href {https://doi.org/10.1103/PhysRevB.95.045414} {\bibfield  {journal}
      {\bibinfo  {journal} {Phys. Rev. B}\ }\textbf {\bibinfo {volume} {95}},\
      \bibinfo {pages} {045414} (\bibinfo {year} {2017})}\BibitemShut {NoStop}%
    \bibitem [{\citenamefont {Esterli}\ \emph {et~al.}(2019)\citenamefont
      {Esterli}, \citenamefont {Otxoa},\ and\ \citenamefont
      {Gonzalez-Zalba}}]{Esterli_Otxoa_Gonzalez-Zalba_2019}%
      \BibitemOpen
      \bibfield  {author} {\bibinfo {author} {\bibfnamefont {M.}~\bibnamefont
      {Esterli}}, \bibinfo {author} {\bibfnamefont {R.~M.}\ \bibnamefont {Otxoa}},\
      and\ \bibinfo {author} {\bibfnamefont {M.~F.}\ \bibnamefont
      {Gonzalez-Zalba}},\ }\href {https://doi.org/10.1063/1.5098889} {\bibfield
      {journal} {\bibinfo  {journal} {Applied Physics Letters}\ }\textbf {\bibinfo
      {volume} {114}},\ \bibinfo {pages} {253505} (\bibinfo {year}
      {2019})}\BibitemShut {NoStop}%
    \bibitem [{\citenamefont {Gonzalez-Zalba}\ \emph {et~al.}(2015)\citenamefont
      {Gonzalez-Zalba}, \citenamefont {Barraud}, \citenamefont {Ferguson},\ and\
      \citenamefont {Betz}}]{Gonzalez_Zalba_2015}%
      \BibitemOpen
      \bibfield  {author} {\bibinfo {author} {\bibfnamefont {M.~F.}\ \bibnamefont
      {Gonzalez-Zalba}}, \bibinfo {author} {\bibfnamefont {S.}~\bibnamefont
      {Barraud}}, \bibinfo {author} {\bibfnamefont {A.~J.}\ \bibnamefont
      {Ferguson}},\ and\ \bibinfo {author} {\bibfnamefont {A.~C.}\ \bibnamefont
      {Betz}},\ }\href {https://doi.org/10.1038/ncomms7084} {\bibfield  {journal}
      {\bibinfo  {journal} {Nature Communications}\ }\textbf {\bibinfo {volume}
      {6}},\ \bibinfo {pages} {6084} (\bibinfo {year} {2015})}\BibitemShut
      {NoStop}%
    \bibitem [{\citenamefont {von Horstig}\ \emph
      {et~al.}(2024{\natexlab{a}})\citenamefont {von Horstig}, \citenamefont
      {Peri}, \citenamefont {Barraud}, \citenamefont {Shevchenko}, \citenamefont
      {Ford},\ and\ \citenamefont {Gonzalez-Zalba}}]{vonHorstig_floquet_2024}%
      \BibitemOpen
      \bibfield  {author} {\bibinfo {author} {\bibfnamefont {F.-E.}\ \bibnamefont
      {von Horstig}}, \bibinfo {author} {\bibfnamefont {L.}~\bibnamefont {Peri}},
      \bibinfo {author} {\bibfnamefont {S.}~\bibnamefont {Barraud}}, \bibinfo
      {author} {\bibfnamefont {S.~N.}\ \bibnamefont {Shevchenko}}, \bibinfo
      {author} {\bibfnamefont {C.~J.~B.}\ \bibnamefont {Ford}},\ and\ \bibinfo
      {author} {\bibfnamefont {M.~F.}\ \bibnamefont {Gonzalez-Zalba}},\ }\href
      {http://arxiv.org/abs/2407.14241} {\bibinfo {title} {Floquet interferometry
      of a dressed semiconductor quantum dot}} (\bibinfo {year}
      {2024}{\natexlab{a}}),\ \bibinfo {note} {arXiv:2407.14241}\BibitemShut {NoStop}%
    \bibitem [{\citenamefont {Oakes}\ \emph
      {et~al.}(2023{\natexlab{b}})\citenamefont {Oakes}, \citenamefont
      {Ciriano-Tejel}, \citenamefont {Wise}, \citenamefont {Fogarty}, \citenamefont
      {Lundberg}, \citenamefont {{Lainé}}, \citenamefont {Schaal}, \citenamefont
      {Martins}, \citenamefont {Ibberson}, \citenamefont {Hutin}, \citenamefont
      {Bertrand}, \citenamefont {Stelmashenko}, \citenamefont {Robinson},
      \citenamefont {Ibberson}, \citenamefont {Hashim}, \citenamefont {Siddiqi},
      \citenamefont {Lee}, \citenamefont {Vinet}, \citenamefont {Smith},
      \citenamefont {Morton},\ and\ \citenamefont {Gonzalez-Zalba}}]{Oakes_2023}%
      \BibitemOpen
      \bibfield  {author} {\bibinfo {author} {\bibfnamefont {G.}~\bibnamefont
      {Oakes}}, \bibinfo {author} {\bibfnamefont {V.}~\bibnamefont
      {Ciriano-Tejel}}, \bibinfo {author} {\bibfnamefont {D.}~\bibnamefont {Wise}},
      \bibinfo {author} {\bibfnamefont {M.}~\bibnamefont {Fogarty}}, \bibinfo
      {author} {\bibfnamefont {T.}~\bibnamefont {Lundberg}}, \bibinfo {author}
      {\bibfnamefont {C.}~\bibnamefont {{Lainé}}}, \bibinfo {author}
      {\bibfnamefont {S.}~\bibnamefont {Schaal}}, \bibinfo {author} {\bibfnamefont
      {F.}~\bibnamefont {Martins}}, \bibinfo {author} {\bibfnamefont
      {D.}~\bibnamefont {Ibberson}}, \bibinfo {author} {\bibfnamefont
      {L.}~\bibnamefont {Hutin}}, \bibinfo {author} {\bibfnamefont
      {B.}~\bibnamefont {Bertrand}}, \bibinfo {author} {\bibfnamefont
      {N.}~\bibnamefont {Stelmashenko}}, \bibinfo {author} {\bibfnamefont
      {J.}~\bibnamefont {Robinson}}, \bibinfo {author} {\bibfnamefont
      {L.}~\bibnamefont {Ibberson}}, \bibinfo {author} {\bibfnamefont
      {A.}~\bibnamefont {Hashim}}, \bibinfo {author} {\bibfnamefont
      {I.}~\bibnamefont {Siddiqi}}, \bibinfo {author} {\bibfnamefont
      {A.}~\bibnamefont {Lee}}, \bibinfo {author} {\bibfnamefont {M.}~\bibnamefont
      {Vinet}}, \bibinfo {author} {\bibfnamefont {C.}~\bibnamefont {Smith}},
      \bibinfo {author} {\bibfnamefont {J.}~\bibnamefont {Morton}},\ and\ \bibinfo
      {author} {\bibfnamefont {M.}~\bibnamefont {Gonzalez-Zalba}},\ }\href
      {https://doi.org/10.1103/PhysRevX.13.011023} {\bibfield  {journal} {\bibinfo
      {journal} {Physical Review X}\ }\textbf {\bibinfo {volume} {13}},\ \bibinfo
      {pages} {011023} (\bibinfo {year} {2023}{\natexlab{b}})}\BibitemShut
      {NoStop}%
    \bibitem [{\citenamefont {Yamaguchi}\ \emph {et~al.}(2017)\citenamefont
      {Yamaguchi}, \citenamefont {Yuge},\ and\ \citenamefont
      {Ogawa}}]{Yamaguchi_Yuge_Ogawa_2017}%
      \BibitemOpen
      \bibfield  {author} {\bibinfo {author} {\bibfnamefont {M.}~\bibnamefont
      {Yamaguchi}}, \bibinfo {author} {\bibfnamefont {T.}~\bibnamefont {Yuge}},\
      and\ \bibinfo {author} {\bibfnamefont {T.}~\bibnamefont {Ogawa}},\ }\href
      {https://doi.org/10.1103/PhysRevE.95.012136} {\bibfield  {journal} {\bibinfo
      {journal} {Physical Review E}\ }\textbf {\bibinfo {volume} {95}},\ \bibinfo
      {pages} {012136} (\bibinfo {year} {2017})}\BibitemShut {NoStop}%
    \bibitem [{\citenamefont {Cattaneo}\ \emph {et~al.}(2019)\citenamefont
      {Cattaneo}, \citenamefont {Giorgi}, \citenamefont {Maniscalco},\ and\
      \citenamefont {Zambrini}}]{Cattaneo_2019}%
      \BibitemOpen
      \bibfield  {author} {\bibinfo {author} {\bibfnamefont {M.}~\bibnamefont
      {Cattaneo}}, \bibinfo {author} {\bibfnamefont {G.~L.}\ \bibnamefont
      {Giorgi}}, \bibinfo {author} {\bibfnamefont {S.}~\bibnamefont {Maniscalco}},\
      and\ \bibinfo {author} {\bibfnamefont {R.}~\bibnamefont {Zambrini}},\ }\href
      {https://doi.org/10.1088/1367-2630/ab54ac} {\bibfield  {journal} {\bibinfo
      {journal} {New Journal of Physics}\ }\textbf {\bibinfo {volume} {21}},\
      \bibinfo {pages} {113045} (\bibinfo {year} {2019})}\BibitemShut {NoStop}%
    \bibitem [{\citenamefont {Mori}(2023)}]{Mori_2023}%
      \BibitemOpen
      \bibfield  {author} {\bibinfo {author} {\bibfnamefont {T.}~\bibnamefont
      {Mori}},\ }\href {https://doi.org/10.1146/annurev-conmatphys-040721-015537}
      {\bibfield  {journal} {\bibinfo  {journal} {Annual Review of Condensed Matter
      Physics}\ }\textbf {\bibinfo {volume} {14}},\ \bibinfo {pages} {35} (\bibinfo
      {year} {2023})}\BibitemShut {NoStop}%
    \bibitem [{\citenamefont {Ikeda}\ \emph {et~al.}(2021)\citenamefont {Ikeda},
      \citenamefont {Chinzei},\ and\ \citenamefont
      {Sato}}]{Ikeda_Chinzei_Sato_2021}%
      \BibitemOpen
      \bibfield  {author} {\bibinfo {author} {\bibfnamefont {T.}~\bibnamefont
      {Ikeda}}, \bibinfo {author} {\bibfnamefont {K.}~\bibnamefont {Chinzei}},\
      and\ \bibinfo {author} {\bibfnamefont {M.}~\bibnamefont {Sato}},\ }\href
      {https://doi.org/10.21468/SciPostPhysCore.4.4.033} {\bibfield  {journal}
      {\bibinfo  {journal} {SciPost Physics Core}\ }\textbf {\bibinfo {volume}
      {4}},\ \bibinfo {pages} {033} (\bibinfo {year} {2021})}\BibitemShut {NoStop}%
    \bibitem [{\citenamefont {Kohler}(2017)}]{kohler_dispersive_2017}%
      \BibitemOpen
      \bibfield  {author} {\bibinfo {author} {\bibfnamefont {S.}~\bibnamefont
      {Kohler}},\ }\href {https://doi.org/10.1103/PhysRevLett.119.196802}
      {\bibfield  {journal} {\bibinfo  {journal} {Physical Review Letters}\
      }\textbf {\bibinfo {volume} {119}},\ \bibinfo {pages} {196802} (\bibinfo
      {year} {2017})},\ \bibinfo {note} {publisher: American Physical
      Society}\BibitemShut {NoStop}%
    \bibitem [{\citenamefont {Benito}\ \emph {et~al.}(2017)\citenamefont {Benito},
      \citenamefont {Mi}, \citenamefont {Taylor}, \citenamefont {Petta},\ and\
      \citenamefont {Burkard}}]{Benito_Mi_2017}%
      \BibitemOpen
      \bibfield  {author} {\bibinfo {author} {\bibfnamefont {M.}~\bibnamefont
      {Benito}}, \bibinfo {author} {\bibfnamefont {X.}~\bibnamefont {Mi}}, \bibinfo
      {author} {\bibfnamefont {J.~M.}\ \bibnamefont {Taylor}}, \bibinfo {author}
      {\bibfnamefont {J.~R.}\ \bibnamefont {Petta}},\ and\ \bibinfo {author}
      {\bibfnamefont {G.}~\bibnamefont {Burkard}},\ }\href
      {https://doi.org/10.1103/PhysRevB.96.235434} {\bibfield  {journal} {\bibinfo
      {journal} {Physical Review B}\ }\textbf {\bibinfo {volume} {96}},\ \bibinfo
      {pages} {235434} (\bibinfo {year} {2017})}\BibitemShut {NoStop}%
    \bibitem [{\citenamefont {von Horstig}\ \emph
      {et~al.}(2024{\natexlab{b}})\citenamefont {von Horstig}, \citenamefont
      {Peri}, \citenamefont {Barraud}, \citenamefont {Robinson}, \citenamefont
      {Benito}, \citenamefont {Martins},\ and\ \citenamefont
      {Gonzalez-Zalba}}]{vonhorstig2024electrical}%
      \BibitemOpen
      \bibfield  {author} {\bibinfo {author} {\bibfnamefont {F.-E.}\ \bibnamefont
      {von Horstig}}, \bibinfo {author} {\bibfnamefont {L.}~\bibnamefont {Peri}},
      \bibinfo {author} {\bibfnamefont {S.}~\bibnamefont {Barraud}}, \bibinfo
      {author} {\bibfnamefont {J.~A.~W.}\ \bibnamefont {Robinson}}, \bibinfo
      {author} {\bibfnamefont {M.}~\bibnamefont {Benito}}, \bibinfo {author}
      {\bibfnamefont {F.}~\bibnamefont {Martins}},\ and\ \bibinfo {author}
      {\bibfnamefont {M.~F.}\ \bibnamefont {Gonzalez-Zalba}},\ }\href@noop {}
      {\bibinfo {title} {Electrical readout of spins in the absence of spin
      blockade}} (\bibinfo {year} {2024}{\natexlab{b}}),\ \Eprint
      {https://arxiv.org/abs/2403.12888} {arXiv:2403.12888}
      \BibitemShut {NoStop}%
    \bibitem [{\citenamefont {Mi}\ \emph {et~al.}(2018)\citenamefont {Mi},
      \citenamefont {Benito}, \citenamefont {Putz}, \citenamefont {Zajac},
      \citenamefont {Taylor}, \citenamefont {Burkard},\ and\ \citenamefont
      {Petta}}]{Mi_2018}%
      \BibitemOpen
      \bibfield  {author} {\bibinfo {author} {\bibfnamefont {X.}~\bibnamefont
      {Mi}}, \bibinfo {author} {\bibfnamefont {M.}~\bibnamefont {Benito}}, \bibinfo
      {author} {\bibfnamefont {S.}~\bibnamefont {Putz}}, \bibinfo {author}
      {\bibfnamefont {D.~M.}\ \bibnamefont {Zajac}}, \bibinfo {author}
      {\bibfnamefont {J.~M.}\ \bibnamefont {Taylor}}, \bibinfo {author}
      {\bibfnamefont {G.}~\bibnamefont {Burkard}},\ and\ \bibinfo {author}
      {\bibfnamefont {J.~R.}\ \bibnamefont {Petta}},\ }\href
      {https://doi.org/10.1038/nature25769} {\bibfield  {journal} {\bibinfo
      {journal} {Nature}\ }\textbf {\bibinfo {volume} {555}},\ \bibinfo {pages}
      {599–603} (\bibinfo {year} {2018})}\BibitemShut {NoStop}%
    \bibitem [{\citenamefont {Peri}\ \emph
      {et~al.}(2024{\natexlab{c}})\citenamefont {Peri}, \citenamefont {von
      Horstig}, \citenamefont {Barraud}, \citenamefont {Ford}, \citenamefont
      {Benito},\ and\ \citenamefont {Gonzalez-Zalba}}]{Peri_vonHorstig_2024}%
      \BibitemOpen
      \bibfield  {author} {\bibinfo {author} {\bibfnamefont {L.}~\bibnamefont
      {Peri}}, \bibinfo {author} {\bibfnamefont {F.-E.}\ \bibnamefont {von
      Horstig}}, \bibinfo {author} {\bibfnamefont {S.}~\bibnamefont {Barraud}},
      \bibinfo {author} {\bibfnamefont {C.~J.~B.}\ \bibnamefont {Ford}}, \bibinfo
      {author} {\bibfnamefont {M.}~\bibnamefont {Benito}},\ and\ \bibinfo {author}
      {\bibfnamefont {M.~F.}\ \bibnamefont {Gonzalez-Zalba}},\ }\href
      {https://doi.org/10.48550/arXiv.2410.17867} {\bibinfo {title} {Polarimetry
      with spins in the solid state}} (\bibinfo {year} {2024}{\natexlab{c}}),\
      \bibinfo {note} {arXiv:2410.17867}\BibitemShut
      {NoStop}%
    \bibitem [{\citenamefont {Petta}\ \emph {et~al.}(2010)\citenamefont {Petta},
      \citenamefont {Lu},\ and\ \citenamefont {Gossard}}]{Petta_Lu_Gossard_2010}%
      \BibitemOpen
      \bibfield  {author} {\bibinfo {author} {\bibfnamefont {J.~R.}\ \bibnamefont
      {Petta}}, \bibinfo {author} {\bibfnamefont {H.}~\bibnamefont {Lu}},\ and\
      \bibinfo {author} {\bibfnamefont {A.~C.}\ \bibnamefont {Gossard}},\ }\href
      {https://doi.org/10.1126/science.1183628} {\bibfield  {journal} {\bibinfo
      {journal} {Science}\ }\textbf {\bibinfo {volume} {327}},\ \bibinfo {pages}
      {669–672} (\bibinfo {year} {2010})}\BibitemShut {NoStop}%
    \bibitem [{\citenamefont {Glasbrenner}\ and\ \citenamefont
      {Schleich}(2023)}]{Glasbrenner_Schleich_2023}%
      \BibitemOpen
      \bibfield  {author} {\bibinfo {author} {\bibfnamefont {E.~P.}\ \bibnamefont
      {Glasbrenner}}\ and\ \bibinfo {author} {\bibfnamefont {W.~P.}\ \bibnamefont
      {Schleich}},\ }\href {https://doi.org/10.1088/1361-6455/acc774} {\bibfield
      {journal} {\bibinfo  {journal} {Journal of Physics B: Atomic, Molecular and
      Optical Physics}\ }\textbf {\bibinfo {volume} {56}},\ \bibinfo {pages}
      {104001} (\bibinfo {year} {2023})}\BibitemShut {NoStop}%
    \bibitem [{\citenamefont {Shevchenko}\ \emph {et~al.}(2010)\citenamefont
      {Shevchenko}, \citenamefont {Ashhab},\ and\ \citenamefont
      {Nori}}]{Shevchenko_2010}%
      \BibitemOpen
      \bibfield  {author} {\bibinfo {author} {\bibfnamefont {S.~N.}\ \bibnamefont
      {Shevchenko}}, \bibinfo {author} {\bibfnamefont {S.}~\bibnamefont {Ashhab}},\
      and\ \bibinfo {author} {\bibfnamefont {F.}~\bibnamefont {Nori}},\ }\href
      {https://doi.org/10.1016/j.physrep.2010.03.002} {\bibfield  {journal}
      {\bibinfo  {journal} {Physics Reports}\ }\textbf {\bibinfo {volume} {492}},\
      \bibinfo {pages} {1–30} (\bibinfo {year} {2010})},\ \bibinfo {note}
      {arXiv:0911.1917}\BibitemShut
      {NoStop}%
    \bibitem [{\citenamefont {Ivakhnenko}\ \emph {et~al.}(2023)\citenamefont
      {Ivakhnenko}, \citenamefont {Shevchenko},\ and\ \citenamefont
      {Nori}}]{Ivakhnenko_2023}%
      \BibitemOpen
      \bibfield  {author} {\bibinfo {author} {\bibfnamefont {O.~V.}\ \bibnamefont
      {Ivakhnenko}}, \bibinfo {author} {\bibfnamefont {S.~N.}\ \bibnamefont
      {Shevchenko}},\ and\ \bibinfo {author} {\bibfnamefont {F.}~\bibnamefont
      {Nori}},\ }\href {https://doi.org/10.1016/j.physrep.2022.10.002} {\bibfield
      {journal} {\bibinfo  {journal} {Physics Reports}\ }\bibinfo {series}
      {Nonadiabatic Landau-Zener-Stückelberg-Majorana transitions, dynamics, and
      interference},\ \textbf {\bibinfo {volume} {995}},\ \bibinfo {pages} {1–89}
      (\bibinfo {year} {2023})}\BibitemShut {NoStop}%
    \bibitem [{\citenamefont {Gonzalez-Zalba}\ \emph {et~al.}(2016)\citenamefont
      {Gonzalez-Zalba}, \citenamefont {Shevchenko}, \citenamefont {Barraud},
      \citenamefont {Johansson}, \citenamefont {Ferguson}, \citenamefont {Nori},\
      and\ \citenamefont {Betz}}]{GonzalezZalba_2016}%
      \BibitemOpen
      \bibfield  {author} {\bibinfo {author} {\bibfnamefont {M.~F.}\ \bibnamefont
      {Gonzalez-Zalba}}, \bibinfo {author} {\bibfnamefont {S.~N.}\ \bibnamefont
      {Shevchenko}}, \bibinfo {author} {\bibfnamefont {S.}~\bibnamefont {Barraud}},
      \bibinfo {author} {\bibfnamefont {J.~R.}\ \bibnamefont {Johansson}}, \bibinfo
      {author} {\bibfnamefont {A.~J.}\ \bibnamefont {Ferguson}}, \bibinfo {author}
      {\bibfnamefont {F.}~\bibnamefont {Nori}},\ and\ \bibinfo {author}
      {\bibfnamefont {A.~C.}\ \bibnamefont {Betz}},\ }\href
      {https://doi.org/10.1021/acs.nanolett.5b04356} {\bibfield  {journal}
      {\bibinfo  {journal} {Nano Letters}\ }\textbf {\bibinfo {volume} {16}},\
      \bibinfo {pages} {1614–1619} (\bibinfo {year} {2016})}\BibitemShut
      {NoStop}%
    \bibitem [{\citenamefont {Yoneda}\ \emph {et~al.}(2014)\citenamefont {Yoneda},
      \citenamefont {Otsuka}, \citenamefont {Nakajima}, \citenamefont {Takakura},
      \citenamefont {Obata}, \citenamefont {Pioro-{Ladrière}}, \citenamefont {Lu},
      \citenamefont {Palmstrøm}, \citenamefont {Gossard},\ and\ \citenamefont
      {Tarucha}}]{Yoneda_2014}%
      \BibitemOpen
      \bibfield  {author} {\bibinfo {author} {\bibfnamefont {J.}~\bibnamefont
      {Yoneda}}, \bibinfo {author} {\bibfnamefont {T.}~\bibnamefont {Otsuka}},
      \bibinfo {author} {\bibfnamefont {T.}~\bibnamefont {Nakajima}}, \bibinfo
      {author} {\bibfnamefont {T.}~\bibnamefont {Takakura}}, \bibinfo {author}
      {\bibfnamefont {T.}~\bibnamefont {Obata}}, \bibinfo {author} {\bibfnamefont
      {M.}~\bibnamefont {Pioro-{Ladrière}}}, \bibinfo {author} {\bibfnamefont
      {H.}~\bibnamefont {Lu}}, \bibinfo {author} {\bibfnamefont {C.~J.}\
      \bibnamefont {Palmstrøm}}, \bibinfo {author} {\bibfnamefont {A.~C.}\
      \bibnamefont {Gossard}},\ and\ \bibinfo {author} {\bibfnamefont
      {S.}~\bibnamefont {Tarucha}},\ }\href
      {https://doi.org/10.1103/PhysRevLett.113.267601} {\bibfield  {journal}
      {\bibinfo  {journal} {Physical Review Letters}\ }\textbf {\bibinfo {volume}
      {113}},\ \bibinfo {pages} {267601} (\bibinfo {year} {2014})}\BibitemShut
      {NoStop}%
    \bibitem [{\citenamefont {Suarez}\ and\ \citenamefont
      {Melville}(2006)}]{Suarez_2006}%
      \BibitemOpen
      \bibfield  {author} {\bibinfo {author} {\bibfnamefont {A.}~\bibnamefont
      {Suarez}}\ and\ \bibinfo {author} {\bibfnamefont {R.}~\bibnamefont
      {Melville}},\ }\href {https://doi.org/10.1109/TMTT.2005.864108} {\bibfield
      {journal} {\bibinfo  {journal} {IEEE Transactions on Microwave Theory and
      Techniques}\ }\textbf {\bibinfo {volume} {54}},\ \bibinfo {pages}
      {1166–1179} (\bibinfo {year} {2006})}\BibitemShut {NoStop}%
    \bibitem [{\citenamefont {Bava}\ \emph {et~al.}(1979)\citenamefont {Bava},
      \citenamefont {Bava}, \citenamefont {Godone},\ and\ \citenamefont
      {Rietto}}]{Bava_1979}%
      \BibitemOpen
      \bibfield  {author} {\bibinfo {author} {\bibfnamefont {E.}~\bibnamefont
      {Bava}}, \bibinfo {author} {\bibfnamefont {G.}~\bibnamefont {Bava}}, \bibinfo
      {author} {\bibfnamefont {A.}~\bibnamefont {Godone}},\ and\ \bibinfo {author}
      {\bibfnamefont {G.}~\bibnamefont {Rietto}},\ }\href
      {https://doi.org/10.1109/TMTT.1979.1129575} {\bibfield  {journal} {\bibinfo
      {journal} {IEEE Transactions on Microwave Theory and Techniques}\ }\textbf
      {\bibinfo {volume} {27}},\ \bibinfo {pages} {141–147} (\bibinfo {year}
      {1979})}\BibitemShut {NoStop}%
    \bibitem [{\citenamefont {Diamond}(1963)}]{Diamond_1963}%
      \BibitemOpen
      \bibfield  {author} {\bibinfo {author} {\bibfnamefont {B.}~\bibnamefont
      {Diamond}},\ }\href {https://doi.org/10.1109/TCT.1963.1082066} {\bibfield
      {journal} {\bibinfo  {journal} {IEEE Transactions on Circuit Theory}\
      }\textbf {\bibinfo {volume} {10}},\ \bibinfo {pages} {35–44} (\bibinfo
      {year} {1963})}\BibitemShut {NoStop}%
    \bibitem [{\citenamefont {Cochrane}\ \emph {et~al.}(2024)\citenamefont
      {Cochrane}, \citenamefont {Seshia},\ and\ \citenamefont
      {Gonzalez-Zalba}}]{Cochrane_2024}%
      \BibitemOpen
      \bibfield  {author} {\bibinfo {author} {\bibfnamefont {L.}~\bibnamefont
      {Cochrane}}, \bibinfo {author} {\bibfnamefont {A.~A.}\ \bibnamefont
      {Seshia}},\ and\ \bibinfo {author} {\bibfnamefont {M.~F.}\ \bibnamefont
      {Gonzalez-Zalba}},\ }\href {https://doi.org/10.1103/PhysRevApplied.21.064066}
      {\bibfield  {journal} {\bibinfo  {journal} {Physical Review Applied}\
      }\textbf {\bibinfo {volume} {21}},\ \bibinfo {pages} {064066} (\bibinfo
      {year} {2024})}\BibitemShut {NoStop}%
    \bibitem [{\citenamefont {Ibberson}\ \emph {et~al.}(2021)\citenamefont
      {Ibberson}, \citenamefont {Lundberg}, \citenamefont {Haigh}, \citenamefont
      {Hutin}, \citenamefont {Bertrand}, \citenamefont {Barraud}, \citenamefont
      {Lee}, \citenamefont {Stelmashenko}, \citenamefont {Oakes}, \citenamefont
      {Cochrane}, \citenamefont {Robinson}, \citenamefont {Vinet}, \citenamefont
      {Gonzalez-Zalba},\ and\ \citenamefont {Ibberson}}]{Ibberson_2021}%
      \BibitemOpen
      \bibfield  {author} {\bibinfo {author} {\bibfnamefont {D.~J.}\ \bibnamefont
      {Ibberson}}, \bibinfo {author} {\bibfnamefont {T.}~\bibnamefont {Lundberg}},
      \bibinfo {author} {\bibfnamefont {J.~A.}\ \bibnamefont {Haigh}}, \bibinfo
      {author} {\bibfnamefont {L.}~\bibnamefont {Hutin}}, \bibinfo {author}
      {\bibfnamefont {B.}~\bibnamefont {Bertrand}}, \bibinfo {author}
      {\bibfnamefont {S.}~\bibnamefont {Barraud}}, \bibinfo {author} {\bibfnamefont
      {C.-M.}\ \bibnamefont {Lee}}, \bibinfo {author} {\bibfnamefont {N.~A.}\
      \bibnamefont {Stelmashenko}}, \bibinfo {author} {\bibfnamefont {G.~A.}\
      \bibnamefont {Oakes}}, \bibinfo {author} {\bibfnamefont {L.}~\bibnamefont
      {Cochrane}}, \bibinfo {author} {\bibfnamefont {J.~W.}\ \bibnamefont
      {Robinson}}, \bibinfo {author} {\bibfnamefont {M.}~\bibnamefont {Vinet}},
      \bibinfo {author} {\bibfnamefont {M.~F.}\ \bibnamefont {Gonzalez-Zalba}},\
      and\ \bibinfo {author} {\bibfnamefont {L.~A.}\ \bibnamefont {Ibberson}},\
      }\href {https://doi.org/10.1103/PRXQuantum.2.020315} {\bibfield  {journal}
      {\bibinfo  {journal} {PRX Quantum}\ }\textbf {\bibinfo {volume} {2}},\
      \bibinfo {pages} {020315} (\bibinfo {year} {2021})}\BibitemShut {NoStop}%
    \end{thebibliography}
%

\end{document}